\documentclass[a4paper,12pt,twoside]{article}
\usepackage{url}
\usepackage[pdftex,colorlinks,bookmarksnumbered,bookmarksopen,citecolor=red,urlcolor=red]{hyperref}
\usepackage{graphicx}\usepackage{a4wide}
\usepackage{amssymb,amsmath,amsthm}
\usepackage{siunitx}
\usepackage{subfigure,color,colordvi,graphicx,xcolor}
\usepackage[only,llbracket,rrbracket,interleave]{stmaryrd}
\usepackage[compress]{cite}
\usepackage{enumitem}
\usepackage{textcomp}
\usepackage{upquote}

\newcommand{\bm}[1]{\text{\boldmath $#1$\unboldmath}}
\newcommand{\grad}{\bm{\nabla}}
\newcommand{\bx}{\bm{x}}
\newcommand{\bn}{\bm{n}}
\newcommand{\bt}{\bm{t}}
\newcommand{\bXi}{\bm{\xi}}
\newcommand{\bEta}{\bm{\eta}}
\newcommand{\Ga}[1][1]{\ensuremath{\Gamma_{_{\!\! #1}}}}
\newcommand{\uD}{u_{\!_D}}
\newcommand{\buD}{\bu_{\!_D}}
\newcommand{\bq}{\bm{q}}
\newcommand{\bw}{\bm{w}}
\newcommand{\hu}{\hat{u}}
\newcommand{\hv}{\hat{v}}
\newcommand{\uS}{u_{\star}}
\newcommand{\vS}{v_{\star}}
\newcommand{\bu}{\bm{u}}
\newcommand{\bv}{\bm{v}}
\newcommand{\bhu}{\widehat{\bu}}
\newcommand{\bhv}{\widehat{\bv}}
\newcommand{\buS}{\bu_{\star}}
\newcommand{\bvS}{\bv_{\star}}
\newcommand{\bL}{\bm{L}}
\newcommand{\bW}{\bm{W}}
\newcommand{\bD}{\bm{D}}
\newcommand{\bE}{\bm{E}}
\newcommand{\Insd}{\mat{I}_{\nsd}}
\newcommand{\Imsd}{\mat{I}_{\msd}}
\newcommand{\numel}{\ensuremath{\texttt{n}_{\texttt{el}}}}
\newcommand{\nsd}{\ensuremath{\texttt{n}_{\texttt{sd}}}}
\newcommand{\msd}{\ensuremath{\texttt{m}_{\texttt{sd}}}}
\newcommand{\RR}{\mathbb{R}}

\newcommand{\eltwo}{\ensuremath{\mathcal{L}_2}}
\newcommand{\Hdiv}{H(\operatorname{div})}
\newcommand{\Vh}{\ensuremath{\mathcal{V}^h}}
\newcommand{\HVh}{\ensuremath{\reallywidehat{\mathcal{V}}^h}}
\newcommand{\VhS}{\ensuremath{\mathcal{V}_{\star}^h}}
\newcommand{\Poly}[1][1]{\ensuremath{\mathcal{P}^{#1}}}
\newcommand{\jump}[1]{\llbracket #1\rrbracket}
\newcommand{\abs}[1]{\lvert#1\rvert}

\newcommand{\mat}[1]{\mathbf{#1}}
\newcommand{\vect}[1]{\mathbf{#1}}
\newcommand{\node}[1]{\mathrm{#1}}
\newcommand{\vecF}[1][1]{\ensuremath{\vect{f}_{#1}}}
\newcommand{\vecHF}[1][1]{\ensuremath{\vect{\hat{f}}_{#1}}}
\newcommand{\Assem}{\mbox{\textsf{\textbf{\Large A}}}}
\newcommand{\hN}{\hat{N}}
\newcommand{\NmatSub}[1][]{\ensuremath{\bm{\mathcal{N}}_{\!\! #1}}}
\newcommand{\NmatHatSub}[1][]{\ensuremath{\bm{\widehat{\mathcal{N}}}_{\!\! #1}}}
\newcommand{\NmatStarSub}[1][]{\ensuremath{\bm{\mathcal{N}^{\star}}_{\!\!\!\!\!\! #1}}}
\newcommand{\QmatSub}[1][]{\ensuremath{\bm{\mathcal{Q}}_{ #1}}}
\newcommand{\QmatStarSub}[1][]{\ensuremath{\bm{\mathcal{Q}^{\star}}_{\!\!\!\! #1}}}
\newcommand{\Nmat}{\bm{\mathcal{N}}}
\newcommand{\NmatHat}{\bm{\widehat{\mathcal{N}}}}
\newcommand{\NmatStar}{\bm{\mathcal{N}^{\star}}}
\newcommand{\Qmat}{\bm{\mathcal{Q}}}
\newcommand{\QmatStar}{\bm{\mathcal{Q}^{\star}}}
\newcommand{\nen}{\ensuremath{\texttt{n}_{\texttt{en}}}}
\newcommand{\nfn}{\ensuremath{\texttt{n}_{\texttt{fn}}}}

\newcommand{\numfaE}{\ensuremath{\texttt{n}_{\texttt{fa}}^{\texttt{e}}}}
\newcommand{\sumfa}{\sum_{\texttt{f}=1}^{\numfaE}}
\newcommand{\tauF}{\tau_{\texttt{f}}}
\newcommand{\chiF}[1][1]{\ensuremath{\chi_{\!_{#1}}^{\texttt{f}}}}
\newcommand{\bJ}{\mat{J}}
\newcommand{\nipe}{\texttt{n}_{\texttt{ip}}^{\texttt{e}}}
\newcommand{\sumge}{\sum_{\texttt{g}=1}^{\nipe}}
\newcommand{\bXige}{\bXi^\texttt{e}_\texttt{g}}
\newcommand{\wge}{w^\texttt{e}_\texttt{g}}
\newcommand{\nipf}{\texttt{n}_{\texttt{ip}}^{\texttt{f}}}
\newcommand{\sumgf}{\sum_{\texttt{g}=1}^{\nipf}}
\newcommand{\bXigf}{\bXi^\texttt{f}_\texttt{g}}
\newcommand{\wgf}{w^\texttt{f}_\texttt{g}}
\newcommand{\bJs}{\mat{J}_{\! \star}}
\newcommand{\nenS}{\ensuremath{\texttt{n}_{\texttt{en}}^{\star}}}
\newcommand{\nipS}{\texttt{n}_{\texttt{ip}}^{\star}}
\newcommand{\sumgS}{\sum_{\texttt{g}=1}^{\nipS}}
\newcommand{\bXigS}{\bXi^{\star}_\texttt{g}}
\newcommand{\wgS}{w^{\star}_\texttt{g}}
\newcommand{\sumJ}{\sum_{j {=} 1}^{\nsd}}
\renewcommand{\deg}{\texttt{p}}
\newcommand{\Rint} {R_{\text{int}}}
\newcommand{\Rext}{R_{\text{ext}}}
\newcommand{\wInt} {\omega_{\text{int}}}
\newcommand{\wExt}{\omega_{\text{ext}}}
\newcommand{\Ball}{\mathcal{B}_{\bm{0},1}}

\usepackage{xcolor}

\renewcommand{\emph}[1]{\textit{#1}}
\newenvironment{keywords}{\begin{quote}\emph{\textbf{Keywords:}}}{\end{quote}}

\theoremstyle{definition}

\newtheorem{remark}{Remark}

\RequirePackage{algorithm}[0.1] 
\usepackage{algorithmic} 

\usepackage{scalerel}
\usepackage{stackengine}
\stackMath
\def\hatgap{0pt}
\def\subdown{-2pt}
\newcommand\reallywidehat[2][]{
	\renewcommand\stackalignment{l}
	\stackon[\hatgap]{#2}{
		\stretchto{\scalerel*[\widthof{$#2$}]{\kern-.6pt\bigwedge\kern-.6pt}{\rule[-\textheight/2]{1ex}{\textheight}}}{0.5ex}_{\smash{\belowbaseline[\subdown]{\scriptstyle#1}}}}}
		
\makeatletter
\newcommand*\wt[2][0.2ex]{%
        \begingroup
        \mathchoice{\wt@helper{#1}{#2}{\displaystyle}{\textfont}}
                   {\wt@helper{#1}{#2}{\textstyle}{\textfont}}
                   {\wt@helper{#1}{#2}{\scriptstyle}{\scriptfont}}
                   {\wt@helper{#1}{#2}{\scriptscriptstyle}{\scriptscriptfont}}%
        \endgroup
        #2%
}
\newcommand*\wt@helper[4]{%
        \def\currentfont{\the#41}%
        \def\currentskewchar{\char\the\skewchar\currentfont}%
        \setbox\tw@\hbox{\currentfont#2\currentskewchar}%
        \dimen@ii\wd\tw@
        \setbox\tw@\hbox{\currentfont#2{}\currentskewchar}%
        \advance\dimen@ii-\wd\tw@
        \rlap{\raisebox{-#1}{$\m@th#3\kern\dimen@ii\widetilde{\phantom{#2}}$}}%
}
\makeatother

\graphicspath{{figsPDF/}{figsPNG/}{figs/}}

\begin{document}
\title{\texttt{HDGlab}: An open-source implementation of the hybridisable discontinuous Galerkin method in MATLAB}

\author{
\renewcommand{\thefootnote}{\arabic{footnote}}
			  M. Giacomini\footnotemark[1]\textsuperscript{ \ ,}\footnotemark[2] , 
			  R. Sevilla\footnotemark[3]\textsuperscript{ \ ,}* \ and
             A. Huerta\footnotemark[1]\textsuperscript{ \ ,}\footnotemark[2]
}

\date{\today}
\maketitle

\renewcommand{\thefootnote}{\arabic{footnote}}

\footnotetext[1]{Laboratori de C\`alcul Num\`eric (LaC\`aN), ETS de Ingenieros de Caminos, Canales y Puertos, Universitat Polit\`ecnica de Catalunya, Barcelona, Spain}
\footnotetext[2]{International Centre for Numerical Methods in Engineering (CIMNE), Barcelona, Spain.}
\footnotetext[3]{Zienkiewicz Centre for Computational Engineering, College of Engineering, Swansea University, Wales, UK
\vspace{5pt}\\
* Corresponding author: Ruben Sevilla. \textit{E-mail:} \texttt{r.sevilla@swansea.ac.uk}
}

\begin{abstract}
This paper presents \texttt{HDGlab}, an open source MATLAB implementation of the hybridisable discontinuous Galerkin (HDG) method. The main goal is to provide a detailed description of both the HDG method for elliptic problems and its implementation available in \texttt{HDGlab}. Ultimately, this is expected to make this relatively new advanced discretisation method more accessible to the computational engineering community. 
\texttt{HDGlab} presents some features not available in other implementations of the HDG method that can be found in the free domain. First, it implements high-order polynomial shape functions up to degree nine, with both equally-spaced and Fekete nodal distributions. Second, it supports curved isoparametric simplicial elements in two and three dimensions. Third, it supports non-uniform degree polynomial approximations and it provides a flexible structure to devise degree adaptivity strategies. Finally, an interface with the open-source high-order mesh generator \texttt{Gmsh} is provided to facilitate its application to practical engineering problems. 
\end{abstract}

\begin{keywords}
hybridizable discontinuous Galerkin, high-order, elliptic problems, MATLAB, open-source
\end{keywords}

\section{Introduction} \label{sc:intro}

In recent years, hybrid discretisation methods have received increasing attention by the applied mathematics and computational engineering community. The main interest in these methodologies is due to their reduced computational cost with respect to classical discontinuous Galerkin (DG) methods, see~\cite{Cockburn-KSC:11,AA-HARP:13,May-WBMS-14,Fidkowski-19}, from which they inherit appealing stability and convergence properties as well as the flexibility to devise high-order, non-uniform degree and adaptive discretisations and the capability to efficiently exploit parallel computing architectures~\cite{Karniadakis-CKS-00,Hesthaven-HW-02,Riviere2008,ErnBook,Cangiani2017}.

The purpose of the present contribution is two-fold: to present a review on the state-of-the-art of hybrid discretisation methods including both fundamental and applied contributions; to provide an educational implementation of the hybridisable discontinuous Galerkin (HDG) method in MATLAB, the so-called \texttt{HDGlab} library, and describe its structure, capabilities and functioning. \texttt{HDGlab} is an open-source library released under GNU GPL licence and designed for rapid prototyping and testing. It supports simplicial meshes and it provides a seamless 2D and 3D implementation with vectorised loops on the integration points. In addition, \texttt{HDGlab} presents four specific features, currently not available in existing open-source HDG implementations in MATLAB:
\begin{enumerate}
	\item Availability of high-order polynomial shape functions up to degree 9, with both equally-spaced and Fekete nodal distributions.
	\item Support of curved isoparametric simplicial elements in 2D and 3D.
	\item Support of non-uniform degree polynomial approximations and flexibility to devise degree adaptivity strategy.
	\item Interface with the open-source high-order mesh generator \texttt{Gmsh}.
\end{enumerate}

The remainder of this paper is organised as follows. First, a review of the state-of-the-art on hybrid discretisation methods is presented in section~\ref{sc:review}. The formulation of the HDG method for the Poisson and Stokes problems is briefly recalled in sections~\ref{sc:PoissonHDG} and~\ref{sc:StokesHDG}, respectively. Section~\ref{sc:repo} provides a description of the structure of the \texttt{HDGlab} library and the url of the repository available under GNU GPL licence. The data structures for the storage of the mesh information, the reference element and the reference face are presented in section~\ref{sc:data}. Section~\ref{sc:preprocess} is devoted to the preprocessing operations, whereas the core of the \texttt{HDGlab} solver for the scalar Poisson equation is described in section~\ref{sc:solver}. Its extension to vectorial problems involving incompressible Stokes flows is discussed in section~\ref{sc:solverStokes}. The visualisation library is introduced in section~\ref{sc:visualisation}. Section~\ref{sc:examples} is devoted to numerical examples, in 2D and 3D, validating the optimal convergence properties of the HDG method and showing the potentialities of the \texttt{HDGlab} implementation. Finally, section~\ref{sc:conclusions} summarises the capabilities of the presented library and three appendices provide implementation details for the Poisson (Appendix~\ref{app:Poisson}) and Stokes (Appendix~\ref{app:Stokes}) solvers and for the interface with the mesh generator \texttt{Gmsh} (Appendix~\ref{app:Gmsh}).

\section{Literature review} \label{sc:review}

The common idea of all hybrid discretisation methods stems from the seminal works of Guyan on static condensation of primal formulations~\cite{Guyan-65} and of Fraeijs de Veubeke on hybridisation of mixed formulations~\cite{Fraeijs-65} of the finite element method. In the context of element-by-element discontinuous approximations, these techniques allow to remedy the drawback of node duplication in DG methods by considering only the unknowns on the mesh faces (edges in 2D) as globally-coupled degrees of freedom. More precisely, the unknowns in each element are expressed as a function of the degrees of freedom on the element faces by solving a local boundary value problem with purely Dirichlet data, whereas appropriate transmission conditions are imposed to guarantee the interelement continuity of the solution and the fluxes, see~\cite{Cockburn-16}.

Three families of hybrid numerical schemes lay within this description, namely, (i) hybrid/hybridised DG, (ii) hybridisable DG, henceforth referred to as HDG, and (iii) hybrid high order (HHO) methods. Stemming from classical DG primal formulations, the hybrid or hybridised DG method reduces the number of globally coupled degrees of freedom by performing static condensation~\cite{Egger-ES-09,Egger-EW-12,Egger-EW-12b}. In addition, improved efficiency can be achieved using polynomial spaces of degree $\deg {+} 1$ and $\deg$ for the primal and hybrid variables, respectively and resorting to the reduced stabilisation approach~\cite{oikawa2015hybridized,oikawa2016analysis}. The hybridisable DG method, henceforth named HDG, is derived from the mixed formulation of the local DG method~\cite{Cockburn-CS-98,cockburn2004characterization,Cockburn-CDG:08} with hybridisation. The main advantage of HDG with respect to other hybridised DG methods relies in the introduction of a mixed variable approximating the gradient of the primal unknown~\cite{Jay-CG:09,Jay-CGL:09}. This approach is of special interest in the context of engineering problems where quantities of interest often depend on the flux of the solution or on the stress. Finally, HHO bridges the two approaches above by utilising a primal formulation and introducing a local reconstruction operator for the gradient of the solution and an appropriate stabilisation term in the static condensation problem~\cite{DiPietro-DPEL-14,Ern-DPE-15}. It is worth noting that many hybrid discretisation schemes can be interpreted in a unique framework as HDG-type methods via appropriate definitions of the stabilisation term, see e.g.~\cite{Chung-CCF-14,Chung-CCF-16} for the staggered DG method and~\cite{Ern2016-CDPE} for HHO.

Unified presentations of hybrid discretisation techniques and their relationship with other known numerical methods are available in~\cite{Jay-CGL:09,BuiThanh-15,DiPietro-DDM-18}. Interested readers are also referred to the review papers~\cite{Cockburn-16,MG-GS:19} and to the recent monograph~\cite{HHO-book}. In the following subsections, an overview of the contributions on hybrid discretisation methods according to the authors' vision is presented.

\subsection{From linear to nonlinear scalar equations}

Second-order scalar elliptic problems have been extensively studied using HDG~\cite{Jay-CGL:09}, HHO~\cite{DiPietro-DPEL-14,Ern-DE-15} and the hybridised DG method~\cite{oikawa2015hybridized}, whereas their extension to linear convection-diffusion problems is discussed in~\cite{Sacco-CDGRS-09,Nguyen-NPC:09,Ern-DDE-15,Egger-ES-09}. Cases of higher-order partial differential equations (PDEs) are presented in~\cite{Cockburn-CDG-09} and~\cite{Chen-CCD-16} for HDG discretisations of biharmonic and third-order equations, respectively, whereas an HHO approximation of the Cahn-Hilliard equation is proposed in~\cite{DiPietro-CDMP-16}. In addition, time-fractional diffusion problems are discussed in~\cite{Cockburn-CM-15,Cockburn-MNC-16}.

More recently, there has been growing interest towards the analysis and simulation of quasilinear and semilinear problems, including the quasilinear $p$-Laplace operator~\cite{Cockburn-CS-16,Shi-QS-19} and the semilinear Grad-Shafranov equation~\cite{SanchezVizuet-SS-19,SanchezVizuet-SSC-20}. To reduce the computational cost of semilinear problems, the interpolatory HDG method was recently devised introducing an interpolation procedure for the efficient and accurate approximation of nonlinear terms~\cite{Cockburn-CSZ-19,Cockburn-CCSZ-19}.

Concerning nonlinear problems, HDG discretisations were proposed for nonlinear convection diffusion~\cite{Nguyen-NPC:09b} and nonlinear Schr\"{o}dinger~\cite{Castillo-CG-20} equations, whereas an HHO formulation of the nonlinear Leray-Lions equation is presented in~\cite{DiPietro-DD-17}. Recent applications involving HDG approximations of nonlinear scalar equations focus on the optoelectronic simulation of photovoltaic solar cells. This problem couples a high-order HDG method for the drift-diffusion electronic model in the semiconductor layer of the solar cells with an efficient approximation of the time-harmonic Maxwell's equations~\cite{Monk-CMZ-19,Monk-ACML-20}.

\subsection{Incompressible flows}

In the context of incompressible flows, HDG formulations of the Stokes equations were devised and analysed in~\cite{Nguyen-NPC:10,Nguyen-CNP:10,Jay-CGNPS-11,Cockburn-CS:14}. The corresponding analysis of the HDG method for Oseen flow is presented in~\cite{Cesmelioglu-CCNP-13}. In~\cite{Nguyen-CNP:10}, it was observed that the HDG method based on Cauchy stress tensor formulation experiences suboptimal convergence of the mixed variable and loss of superconvergence of the postprocessed velocity, when low-order polynomial approximations are considered. The $M$-decomposition approach~\cite{Cockburn-CFQ-17} remedies this issue by appropriately enriching the discrete local spaces of approximation. An alternative strategy imposing the symmetry of the mixed variable pointwise via Voigt notation is discussed in~\cite{MG-GKSH-18} along with a postprocessing procedure to handle translational and rotational rigid body modes. Divergence-conforming HDG~\cite{Cockburn-CS-14}, hybridised DG~\cite{Egger-EW-12} and embedded-hybridised DG (EHDG)~\cite{Rhebergen-RW-20} discretisations were also studied for the incompressible Stokes equations, whereas a pressure-robust HHO method for viscosity-dependent Stokes flows is proposed in~\cite{Ern-DELS-16}.

HDG formulations for the nonlinear incompressible Navier-Stokes equations using equal order and different order of polynomial approximations for the primal, mixed and hybrid variables are described in~\cite{Nguyen-NPC:11,Cesmelioglu-CCQ-17} and~\cite{Shi-QS-16}, respectively. The former approach is also employed in~\cite{Giorgiani-GFH-13} to devise a degree adaptive strategy relying on the local superconvergence of the postprocessed velocity. Stemming from the work in~\cite{DiPietro-DK-18}, different HHO formulations of the incompressible Navier-Stokes equations were proposed, incorporating a skew-symmetric form of the convection term~\cite{DiPietro-BDD-19} and a globally divergence-free velocity approximation to achieve robustness in presence of large irrotational body forces \cite{DiPietro-CD-20}. Moreover, special attention was devoted in recent years to the development of hybridised DG schemes \cite{Egger-EW-12b} with pointwise divergence-free velocity~\cite{Schoberl-LS-16,Wells-RW-18b,Rhebergen-KR-19} and with relaxed $\Hdiv$-conformity~\cite{Lederer-LLS-18,Lederer-LLS-19}, as well as divergence-conforming hybrid DG discretisations for incompressible flows on surfaces~\cite{Lederer-LLS-20}. It is worth noting that all the above mentioned references focus on viscous laminar flows and preliminary promising results on the incompressible Reynolds averaged Navier-Stokes (RANS) equations coupled with the Spalart Allmaras turbulence model were recently presented in~\cite{Evans-PE-19}.

Besides classical approaches to steady and unsteady Navier-Stokes equations, HDG-based space-time formulations were studied for their ability to effectively handle moving and deforming domains. More precisely, stemming from the HDG formulation introduced in~\cite{Cockburn-RC-12}, $\Hdiv$-conforming hybridised DG~\cite{Rhebergen-HR-19} and EHDG~\cite{Rhebergen-HR-20} methods were proposed. Hybridised DG and HDG methods with arbitrary Lagrangian Eulerian (ALE) formulations were thus presented in~\cite{Fidkowski-16,Fu-20} and the resulting HDG-ALE framework was applied to fluid-structure interaction (FSI) problems involving incompressible~\cite{Pitt-SMP-16}  and weakly-compressible flows~\cite{LaSpina-SKGWH-20}.

Among the applications of hybrid discretisation methods to incompressible flows, it is also worth mentioning the recent attempts to simulate quasi-Newtonian fluids~\cite{Gatica-GS-15} and viscoplastic materials~\cite{Ern-CBCE-18}.

\subsection{Two-phase flows and heterogeneous porous media}

HDG simulations of immiscible incompressible two-phase flows in heterogeneous porous media were first proposed in~\cite{Riviere-FKR-18} and coupled with high-order diagonally implicit Runge-Kutta (DIRK) time integrators in~\cite{CostaSole-CRS-19}. Moreover, in~\cite{BuiThanh-KBA-19}  a linear degenerate elliptic problem modelling two-phase mixture is approximated using a hybridised DG approach. Darcy flow and two-phase flow simulations in highly heterogeneous media are performed in~\cite{Shi-YSF-19} via the so-called generalised multiscale HDG (GMsHDG) method which is connected to the mortar mixed finite element method described in~\cite{Wheeler-APWY-07}. GMsHDG was also employed for multiscale simulations of elliptic PDEs in heterogeneous media~\cite{Shi-ELMS-15,Efendiev-CEL-19} and perforated domains~\cite{Moon-CM-20} and of parabolic PDEs in heterogeneous media~\cite{Lazarov-MLJ-19}.

An alternative to GMsHDG is the HHO frameowrk for highly oscillatory elliptic problems introduced in~\cite{Ern-CEL-18}. Moreover, in the context of coupled problems involving porous media, HHO simulations of passive transport of a solute in a fractured medium are presented in~\cite{DiPietro-CDF-19}, whereas nonlinear poroelastic phenomena in a saturated porous medium with a slightly compressible fluid are described in~\cite{DiPietro-BDS-20,DiPietro-BDLS-20}. 

Extensive research has been also devoted to coupled Stokes/Darcy and Brinkman models. In~\cite{Rhebergen-CRW-20}, an EHDG formulation of the Stokes/Darcy system is described. Concerning the Brinkman model, an analysis of its HDG approximation is presented in~\cite{Solano-ASV-19}, its simulation in the context of heterogeneous media with high-contrast is discussed in~\cite{Shi-LS-18} and an $\Hdiv$-conforming discretisation is proposed in~\cite{Fu-FJQ-19}. In~\cite{DiPietro-BDD-18}, an HHO formulation with divergence-conforming Darcy velocity and higher-order Stokes velocity is devised.

\subsection{Compressible flows and gas kinetics equations}

Hybrid formulations for inviscid Euler and laminar compressible Navier-Stokes equations are proposed in~\cite{peraire2010hybridizable} in the context of HDG and in~\cite{Peraire-NPC-15} for the embedded DG (EDG) method. Extension to viscous turbulent compressible flows using RANS equations with Spalart-Allmaras turbulence model is presented in~\cite{Peraire-MNP-11}, whereas a large-eddy simulation framework is introduced in~\cite{Peraire-FNP-17}. In addition, an entropy stable space-time discretisation was proposed for the compressible Navier-Stokes equations using an HDG approach in space and a discontinuous approximation in time~\cite{Williams-18}. More recently, special attention was dedicated to the development of positivity-preserving Riemann solvers in the context of hybridised DG methods~\cite{Vila-VGSH-review}. For a complete review on HDG methods for compressible flows, interested readers are referred to~\cite{Vila-VGSH-review}, whereas the application to gas kinetics modelled by means of the linearised Bhatnagar-Gross-Krook equation is discussed in~\cite{Su-SWZW-19}.

\subsection{Plasma physics and magnetohydrodynamics}

Computational physics community is showing increasing interest towards the application of hybrid discretisation methods to the simulation of magnetic plasma physics. Promising preliminary results concerning the HDG approximation of the Grad-Shafranov equation in axisymmetric confinement devices modelling fusion reactors are described in~\cite{SanchezVizuet-SS-19,SanchezVizuet-SSC-20}. In the context of magnetohydrodynamics (MHD), an HDG method for steady-state linearised incompressible MHD equations is proposed in~\cite{BuiThanh-LSBS-19}. Approximation strategies for the unsteady compressible MHD equations using HDG, EDG and the interior embedded DG (IEDG) methods with DIRK time integrators are explored in~\cite{Peraire-CFCNP-20}.

\subsection{Shallow water equations}

The shallow water equations have been extensively studied in the context of hybrid DG methods, starting from the linearised shallow water system in~\cite{BuiThanh-16} to the nonlinear Korteweg-de Vries equation in~\cite{Dawson-SPMD-16,Chen-CDJ-18}. In both the above mentioned works, time integration is performed implicitly using a backward Euler method. Extension to high-order backward differentiation formulas is discussed in~\cite{Fabien-20-shallow} in the context of the Benjamin-Bona-Mahony equation. To reduce the computational cost of fully-implicit procedures, in~\cite{Dawson-SD-18} an operator splitting is applied to the Green-Naghdi equation and the nonlinear hyperbolic subproblem is solved using an explicit approach, whereas the implicit time integrator is only applied to the linear dispersive subproblem. A similar idea is presented in~\cite{BuiThanh-KGB-20} to devise an implicit-explicit (IMEX) HDG-DG scheme in which the linear part of the problem is solved using a hybridised DG method and a singly diagonally implicit Runge-Kutta (SDIRK) scheme and the nonlinear one is approximated by means of an explicit Runge-Kutta (RK) DG discretisation. A detailed comparison of explicit and implicit approaches to the nonlinear shallow water equations is provided in~\cite{Dawson-SKMD-19}.


\subsection{Wave propagation phenomena}

The benefits of high-order methods in the simulation of wave propagation prompted extensive research on hybrid discretisation methods in the fields of electromagnetics, elastodynamics and acoustics. A detailed review on HDG and EDG approaches for these problems is available in~\cite{Peraire-FCTNP-18}.

Starting from the work in~\cite{nguyen2011hybridizable}, research on time-harmonic Maxwell's equations tackled the analysis and development of HDG formulations~\cite{Lanteri-LLP-15}, including methods suitable for simulations at large wave numbers~\cite{Qiu-LCQ-17} and Schwarz-type domain decomposition (DD) strategies designed for HDG~\cite{Lanteri-LLP-14,Lanteri-AGGKLM-20}. Recent applications of HDG to time-harmonic Maxwell's equations focus on wave propagation in heterogeneous media modelling photovoltaic cells~\cite{Solano-CLOS-20}, coupling with nonlocal hydrodynamic Drude and generalised nonlocal optical response models~\cite{Lanteri-LLMW-17} and with hydrodynamic models for metals~\cite{Peraire-VCNOP-18,Peraire-VNP-18,Peraire-YVCNSPO-19,Peraire-VMCYNOP-20} to simulate plasmonic nanostructures. In the context of time-domain Maxwell's equations, HDG methods are presented and analysed in~\cite{Shi-CQSS-17,Chen-CCX-19,Sayas-DS-20}, whereas implicit hybridised DG discretisations are proposed in~\cite{Lanteri-CDL-18}.

In the framework of elastodynamics, HDG with DIRK time integrators were introduced in~\cite{nguyen2011high}, whereas in~\cite{Terrana-TVG-17} an HDG spectral element method (HDG-SEM) is utilised to simulate wave propagation in coupled elastic-acoustic media. In the frequency-domain, HDG methods for elastodynamics are analysed and presented in~\cite{Sayas-HPS-17,Lanteri-BCDL-17}.

The first HDG solver for acoustics, introduced in \cite{nguyen2011high}, relied on a fully-implicit approach based on DIRK time integrators. Since then, explicit HDG formulations utilising strong stability-preserving RK (SSPRK) and explicit RK integrators were proposed in~\cite{Peraire-SNPC-16}, wheras an explicit arbitrary derivative (ADER) approach is discussed in~\cite{Kronbichler-SKW-18}. In addition, a comparison of implicit and explicit HDG schemes for acoustic wave propagation is performed in~\cite{Kronbichler-KSMW-16}. More recently, an HDG-based cut finite element startegy with local time stepping was presented in~\cite{Kronbichler-SSKK-20}. It is worth recalling that devising a conservative numerical scheme is a critical aspect for the accurate simulation of acoustic wave propagation. To correct the dissipative nature of the method analysed in~\cite{Cockburn-CQ-14}, an energy-conservative HDG formulation with a two-step Stormer-Numerov time-marching is proposed in~\cite{Cockburn-CFHJSS-18}. Moreover, symplectic~\cite{Peraire-SCNPC-17}  and multisymplectic~\cite{Stern-MS-20} HDG schemes preserving the Hamiltonian structure of the PDEs under analysis are developed to achieve energy conservation.

Among the applications of HDG to wave propagation phenomena, it is also worth mentioning the degree adaptive approximation of the mild slope equation to perform harbour simulations~\cite{Giorgiani-GFH-13} and the cardiac electrophysiology simulations of the monodomain model~\cite{Kronbichler-HBKPCW-18,Loula-RDIL-20}.


\subsection{Linear and nonlinear elasticity}

In linear elasticity, the imposition of the symmetry of the stress tensor using HDG methods based on mixed formulations has been extensively studied in the literature. Indeed, the first formulations introduced in~\cite{soon2009hybridizable,Fu-FCS-15} experienced suboptimal convergence of the mixed variable and a loss of superconvergence of the postprocessed displacement field. To remedy this issue, a formulation considering a weakly symmetric stress tensor was presented in~\cite{Shi-CS-13}. The strong imposition of the symmetry can be achieved via several strategies: in~\cite{Shi-QSS-18}, different degrees of polynomial approximation are considered for the primal and hybrid variables; the $M$-decomposition framework~\cite{Cockburn-CF-17-I,Cockburn-CF-17-II,Cockburn-CF-17-III} is applied to the linear elastic problem~\cite{Cockburn-CF-17} to enrich the discrete spaces of approximation utilised in the local problem; an alternative formulation imposing the symmetry of the mixed variable pointwise via Voigt notation is proposed for high-order and the lowest-order HDG discretisations in~\cite{RS-SGKH-18} and~\cite{RS-SGH-19}, respectively. In the context of the high-order discretisation, a novel postprocessing strategy accounting for rigid translation and rotation is also devised. It is worth noting that hybrid methods relying on primal formulations do not suffer from these issues, see e.g. HHO~\cite{Ern-DPE-15}.

Timoshenko beams are discussed in~\cite{Celiker-CCS-10,Celiker-CCS-11}, whereas the case of Kirchhoff plates is considered in~\cite{Huang-HH-19} using HDG and in~\cite{DiPietro-BDGK-18} using HHO methods.

In the context of nonlinear elasticity, the first hybrid discretisation formulation was presented in~\cite{Lew-KLC-15}. In this work, it was observed that the method may not converge to the exact solution if the interelement jumps are not appropriately penalised and a detailed numerical study on the choice of the HDG stabilisation is discussed in~\cite{Cockburn-CS-19}. More recently, a locking-free HDG formulation for nonlinear elasticity of thin structures subject to large deformations was proposed~\cite{Peraire-TNBP-19}. In addition, HHO discretisations of hyperelastic materials in small and finite deformations were presented in~\cite{DiPietro-BDS-17} and~\cite{Ern-AEP-18}, respectively. HHO discretisations for problems involving plastic and elastoplatic simulations are discussed in~\cite{Ern-AEP-19,Ern-AEP-19b}, whereas contact phenomena are addressed in~\cite{Ern-CEP-20}.

\subsection{Interface problems and immersed discretisations}

The first attempt to solve interface problems using hybrid discretisation techniques was proposed in~\cite{Peraire-HNPK-13} using a body-fitted mesh. In this context, a superparametric HDG formulation was considered to limit the geometric error due to the polygonal approximation of curved interfaces. 

Recently, immersed methods have received special attention, both in the context of HHO and HDG formulations. More precisely, unfitted HHO methods relying on a cell agglomeration procedure to remedy small cut instabilities are analysed for scalar and vectorial second-order elliptic problems in~\cite{Burman-BE-18,Burman-BDE-20}. In the framework of HDG, Poisson interface problems are treated in~\cite{Dong-DWXW-16} by means of an unfitted method introducing appropriately defined ansatz functions in the vicinity of the interface. An alternative approach to handle curved interfaces is proposed in~\cite{Solano-QSV-16} where a fictitious domain strategy is developed coupling a mesh of planar faces and a transferring function for the imposition of the transmission conditions on the fictitious subdomain. Inspired by the cut finite element method, in~\cite{Kronbichler-SSKK-20}, a high-order HDG strategy employing a level-set function to describe the immersed interfaces and a cell agglomeration procedure is described for the wave equation. Similarly, the extended HDG (X-HDG) method introduces a framework in which the HDG local problem is modified only in the elements cut by the interface. In this context, cut instabilities are handled by displacing the mesh nodes responsible for the bad cuts~\cite{Gurkan-GSKF-16,Gurkan-GKF-17,Gurkan-GKF-19}. Finally, an HDG-based phase-field model for brittle fracture was recently proposed in~\cite{Muixi-MRF-20}.

\subsection{High-order and exact geometry representations} 

Geometry representation plays a crucial role in the capability of high-order methods to achieve optimal accuracy. In the context of HDG, high-order isoparametric approaches in presence of curved meshes are utilised in many references, see e.g.~\cite{Peraire-MNP-11,MG-GKSH-18}, whereas this technique is addressed for HHO in~\cite{DiPietro-BD-18}. An alternative approach relying on meshes with planar faces and the extension to a fictitious subdomain is discussed in~\cite{Solano-CS-12,Solano-CS-14,Solano-SV-19} for several linear problems and was recently extended to the semilinear Grad-Shafranov equation~\cite{SanchezVizuet-SS-19,SanchezVizuet-SSC-20}. It is worth noting that all the techniques mentioned above introduce geometric errors due to the polynomial approximation of the boundaries. In order to exploit the exact CAD representation of the boundaries, the NURBS-enhanced finite element method (NEFEM) \cite{Sevilla-SFH-08,Sevilla-SFH-11} is employed in~\cite{RS-SH-18,RS-19,MG-GS:19} to devise HDG formulations with exact geometry for Stokes, linear elastic and electrostatics problems, respectively.

\subsection{Lowest-order hybrid discretisations}

Hybrid discretisation methods have been traditionally developed in the context of high-order approximations. Nonetheless, it is well-known that lowest-order discretisations, e.g. the finite volume (FV) method, are more robust than high-order techniques. In this framework, a new class of lowest-order hybrid discretisations was developed, with unknowns approximated by means of constant functions on the mesh faces. The recently proposed face-centred finite volume (FCFV) for Poisson, Stokes~\cite{RS-SGH-18} and linear elasticity~\cite{RS-SGH-19} can be interpreted as an HDG method of degree zero. Variants of this approach achieving optimal second-order convergence of the primal variable are discussed in~\cite{Vieira-VGSH-20,MG-GS-20}. Stemming from HHO, lowest-order nonconforming discretisations are proposed in~\cite{DiPietro-BDG-19} for linear elasticity and in~\cite{Ern-CEG-20} for elliptic obstacle problems. As their high-order counterparts, the above mentioned methodologies allow the use of generic polygonal and polyhedral elements and provide a workaround to the sensitivity issues of FV methods to mesh distortion and stretching~\cite{diskin2010comparison,diskin2011comparison}.

\subsection{Iterative solvers and preconditioning}

Although hybrid discretisation methods are responsible for a substantial reduction of degrees of freedom with respect to classical DG methods, their applicability to realistic problems of engineering interest still rely on the development of efficient solution strategies for large scale systems. 

In~\cite{Dolean-BBDNT-18}, a DD strategy based on restricted additive Schwarz methods is proposed for hybridised DG approximations, whereas an optimised Schwarz DD approach suitable to handle the many-subdomain case is discussed in~\cite{Gander-GH-18}. Starting from~\cite{Jay-CDGT-14}, several works also explored the capabilities of multigrid solvers for HDG formulations, including hierarchical scale separation~\cite{Schutz-SA-17}, geometric multigrid~\cite{BuiThanh-WMB-19}, nested geometric multigrid on many-core processors~\cite{Riviere-FKMR-19}, $p$-multigrid in the context of second-order elliptic problems~\cite{Kronbichler-KW-18} and compressible Navier-Stokes flows~\cite{Crivellini-FFC-20} and GPU-accelerated $p$-multigrid for linear elasticity~\cite{Fabien-20}. Finally, iterative algorithms inspired by the Gauss-Seidel method were proposed in~\cite{BuiThanh-MTB-18} and tested on massively parallel architectures up to 16,384 cores. A block symmetric Gauss-Seidel type preconditioner was also introduced in~\cite{Wells-RW-18}, whereas a multilevel solver coupled with a block-Jacobi fine scale solver is proposed in~\cite{BuiThanh-MBS-20}.

\subsection{\emph{A posteriori} error estimates and adaptivity}

The quality of hybrid discretisation methods has been assessed in several works by means of \emph{a posteriori} estimates of the error in the primal, mixed and hybrid variables, as well as in quantities of interest.

Starting from the seminal works~\cite{Cockburn-CZ-12,Cockburn-CZ-13} establishing reliability and efficiency of error estimates for the HDG approximations of second-order elliptic equations, \emph{a posteriori} estimates were developed for steady and unsteady scalar convection-diffusion problems in~\cite{Qiu-CLQ-14} and~\cite{Chen-LC-20}, respectively, and for the vectorial case of incompressible Oseen~\cite{Solano-ASV-19b} and Brinkman~\cite{Solano-ASV-19} flows. In addition, constant-free computable \emph{a posteriori} error estimates are devised in~\cite{Fu-AF-18} for second-order elliptic problems using an equilibrated fluxes approach, whereas residual-based estimates are established for Maxwell's equations in~\cite{Shi-CQS-18}. 

In the context of adaptivity, on the one hand, the analysis of HDG approximations based on non-uniform polynomial degrees~\cite{Cockburn-CC-12,Cockburn-CC-14} and the superconvergence property of the postprocessed solution~\cite{Cockburn-CDG:08,Jay-CGL:09} prompted the development of degree adaptive procedures based on superparametric HDG methods~\cite{Giorgiani-GFH-13,Giorgiani-GFH-14} and on isoparametric HDG-NEFEM approaches~\cite{RS-SH-18,RS-19,MG-GS:19}. Degree adaptivity is also applied to the simulation of cardiac electrophysiology in~\cite{Kronbichler-HBKPCW-18}. On the other hand, mesh adaptivity procedures to capture localised abrupt changes in the solution were devised in~\cite{SanchezVizuet-SSC-20} and~\cite{Muixi-MRF-20} for the Grad-Shafranov equation and the phase-field model for brittle fracture, respectively. Octree-based mesh refinement is performed in~\cite{Dawson-SMD-16} for anisotropic inhomogeneous diffusion problems. Mesh adaptivity driven by local error indicators is also employed in the context of second-order FCFV approximations~\cite{Vieira-VGSH-20,MG-GS-20}. Concerning the error in quantities of interest, an adjoint-based method allowing to achieve superconvergent approximations of linear functionals is described in~\cite{Cockburn-CZ-17} and goal-oriented mesh adaptation strategies are proposed in~\cite{May-WMS-14,Fidkowski-FC-20}.

\subsection{Coupling HDG with other numerical methods}

The accuracy of high-order HDG approximations has been recently exploited to develop efficient algorithms coupling different numerical methodologies in different regions of the computational domain. 

In~\cite{Sevilla-KSH-20}, a strategy coupling HDG and a vertex-centred finite volume method is proposed to simulate transient inviscid flows using coarse meshes designed for steady-state problems. In addition, different couplings of HDG and continuous Galerkin (CG) discretisations were explored in the literature. A strategy inspired by a non-overlapping DD method is presented in~\cite{FernandezMendez-PTF-19} in the context of incompressible Navier-Stokes flows coupled with conjugate heat transfer phenomena. An alternative minimally-intrusive coupling based on a Nitsche's formulation of the CG method was first introduced in~\cite{MG-SpinaGH-20} for linear elastic problems involving nearly incompressible materials and was extended to FSI problems with weakly compressible flows in~\cite{LaSpina-SKGWH-20}.

\subsection{HDG-based reduced order models}

In recent years, the accuracy of the HDG method and its flexibility to devise high-order adaptive discretisation have been also employed to devise high-fidelity reduced and surrogate models. In~\cite{Peraire-VNGP-15,Peraire-VNGP-16}, a reduced order model to accelerate the Monte-Carlo simulation of stochastic elliptic PDEs is constructed coupling a high-order HDG method with a reduced basis and empirical interpolation approach. The combination of an HDG solver for time-harmonic Maxwell's equations and a proper orthogonal decomposition (POD) strategy to design parametrised plasmonic nanogap structures is proposed in~\cite{Peraire-VNP-18}. An HGD-POD reduced order model (ROM) is also discussed in~\cite{Shen-SSZ-19}  for the fast simulation of the unsteady heat equation. More recently, an \emph{a priori} ROM based on HDG and the proper generalised decomposition was proposed to simulate Stokes flows in geometrically parametrised domains~\cite{RS-SBGH-20,MG-GBSH-20}. 


\subsection{Availability of open-source implementations of hybrid discretisation methods}

The success of hybrid discretisation methods led to the development of targeted open-source libraries and to their implementation in  existing finite element libraries available open-source. To the best of the authors' knowledge, the hybridised DG method based on primal formulations is available in the following libraries:
\begin{itemize}
\item \texttt{MFEM} \cite{MFEM-19,MFEM-DOC}
\item \texttt{Netgen/NGSolve} \cite{NGSolve-14,NGSolve-DOC}
\end{itemize}
whereas the libraries
\begin{itemize}
\item \texttt{deal.II} \cite{deal-07,deal-DOC}
\item \texttt{Feel++} \cite{Feelpp-06,Feelpp-DOC}
\item \texttt{Firedrake} \cite{Firedrake-16,Firedrake-DOC}
\item \texttt{Nektar++} \cite{Nektar-15,Nektar-DOC}
\end{itemize}
provide implementations of the HDG method based on mixed formulations. Finally, the HHO method is available in
\begin{itemize}
\item \texttt{code\_aster} \cite{Aster-19,Aster-DOC}
\item \texttt{Code\_Saturne} \cite{Saturne-11,Saturne-DOC}
\item \texttt{DiSk++} \cite{disk-18,disk-DOC}
\item \texttt{GetFEM} \cite{GetFEM-20,GetFEM-DOC}
\item \texttt{HArDCore} \cite{HArDCore-DOC}
\end{itemize}
All above mentioned libraries rely on either Fortran or C/C++ implementations, whereas open-source libraries implementing HDG in MATLAB include:
\begin{itemize}
\item \texttt{HDG3D} \cite{HDG3D-15,HDG3D-DOC}
\item \texttt{FESTUNG} \cite{FESTUNG-18,FESTUNG-DOC}
\end{itemize}

\section{HDG formulation of the Poisson equation} \label{sc:PoissonHDG}

In this section, the formulation of the HDG method for the Poisson equation is briefly recalled. Special attention is devoted to the identification of the building blocks of the numerical scheme whose implementation will be detailed in section~\ref{sc:solver}. Interested readers are referred to~\cite{Jay-CGL:09} for a complete theoretical introduction to the HDG method for Poisson equation and to~\cite{RS-SH:16} for a tutorial on its derivation.

Let $\Omega \subset \RR^{\nsd}$ be an open bounded domain in $\nsd$ spatial dimensions such that its boundary is $\partial\Omega {=} \Ga[D] \cup \Ga[N]$ and $\Ga[D] \cap \Ga[N] {=} \emptyset$. The strong form of the Poisson equation is
\begin{equation}\label{eq:PoissonStrongForm}
\left\{
\begin{aligned}
-\grad {\cdot} (\kappa \grad u) &= s           &&\text{in $\Omega$,}   \\
u &= \uD                                     &&\text{on $\Ga[D]$,} \\
\bn {\cdot} \kappa \grad u &=g               &&\text{on $\Ga[N]$,}
\end{aligned}
\right.
\end{equation}
where the unknown $u$ represents the solution field, $\kappa$ denotes the material parameter (e.g. conductivity in a thermal problem) and $s$ is a volumetric source term. On the boundary, Dirichlet, $\uD$, and Neumann, $g$, data prescribe the values of the unknown and its flux on $\Ga[D]$ and $\Ga[N]$, respectively. The vector $\bn$ denotes the outward unit normal vector to the boundary.

\subsection{HDG local and global problems: strong form} 

Consider a partition of $\Omega$ in $\numel$ disjoint subdomains such that
$$
\Omega = \bigcup_{e=1}^{\numel} \Omega_e,
\quad \Omega_i \cap \Omega_j = \emptyset \text{ for } i \neq j 
$$
and define the \emph{mesh skeleton} as
$$
\Gamma := \left[ \bigcup_{e=1}^{\numel} \partial\Omega_e \right] \setminus \partial\Omega .
$$

Following the HDG rationale~\cite{Jay-CGL:09,Nguyen-NPC:09,Nguyen-NPC:09b,Nguyen-CNP:10,Nguyen-NPC:11,RS-SH:16}, a mixed variable $\bq {=} {-} \sqrt{\kappa} \grad u$ is introduced and problem~\eqref{eq:PoissonStrongForm} is rewritten as a system of first-order equations element-by-element, that is,
\begin{equation}\label{eq:PoissonHDGstrong}
\left\{
\begin{aligned}
\bq + \sqrt{\kappa} \grad u &= \bm{0}                            && \text{in $\Omega_e, \, e {=} 1,\ldots,\numel$,} \\
\grad {\cdot} (\sqrt{\kappa} \bq) &= s                               && \text{in $\Omega_e, \, e {=} 1,\ldots,\numel$,} \\
u &= \uD                                                   && \text{on $\Ga[D]$,} \\
\bn {\cdot} \sqrt{\kappa}\bq &= -g           && \text{on $\Ga[N]$,} \\
\jump{ u \bn } &= \bm{0}                         && \text{on $\Gamma$,}   \\
\jump{ \bn {\cdot} \sqrt{\kappa} \bq } &= 0                  &&\text{on $\Gamma$,}
\end{aligned}
\right.
\end{equation}
where the \emph{jump} operator $\jump{\cdot}$ is defined as
$$
\jump{ \odot } := \odot_i + \odot_j ,
$$
being $\odot_i$ and $\odot_j$ the evaluations of the quantity $\odot$ in two neighbouring elements $\Omega_i$ and $\Omega_j$ sharing a given interface~\cite{AdM-MFH:08}. The last two conditions in~\eqref{eq:PoissonHDGstrong}, known as \emph{transmission conditions}, enforce the continuity of the solution and of its normal flux across the internal mesh skeleton $\Gamma$.

The HDG algorithm solves equation~\eqref{eq:PoissonHDGstrong} in two stages. First, an independent hybrid variable $\hu$ is introduced to represent the trace of the solution on $\partial\Omega_e \setminus \Ga[D]$ and the primal and mixed variables $(u_e,\bq_e)$ in each element $\Omega_e, \, e {=} 1,\ldots,\numel$ are expressed as functions of the unknown $\hu$, namely
\begin{equation}\label{eq:PoissonHDGstrongLocal}
\left\{
\begin{aligned}
\bq_e + \sqrt{\kappa} \grad u_e &= \bm{0}                            && \text{in $\Omega_e, \, e {=} 1,\ldots,\numel$,} \\
\grad {\cdot} (\sqrt{\kappa} \bq_e ) &= s                               && \text{in $\Omega_e, \, e {=} 1,\ldots,\numel$,} \\
u_e &= \uD                                                   && \text{on $\partial\Omega_e \cap \Ga[D]$,} \\
u_e &= \hu                                                   && \text{on $\partial\Omega_e \setminus \Ga[D]$.}
\end{aligned}
\right.
\end{equation}
\begin{remark}
Equation~\eqref{eq:PoissonHDGstrongLocal} represents the $\numel$ HDG local problems. This stage corresponds to the \emph{hybridisation} of the mixed problem, see~\cite{Fraeijs-65}, and is equivalent to the \emph{static condensation} procedure in classical continuous Galerkin methods~\cite{Guyan-65}.
\end{remark}

Second, the hybrid variable is computed by solving the HDG global problem, which accounts for the transmission conditions on the mesh skeleton $\Gamma$ and the Neumann boundary condition on $\Ga[N]$, that is,
\begin{equation}\label{eq:PoissonHDGstrongGlobal}
\left\{
\begin{aligned}
\jump{ u \bn } &= \bm{0}                         && \text{on $\Gamma$,}   \\
\jump{ \bn {\cdot} \sqrt{\kappa} \bq } &= 0                  &&\text{on $\Gamma$,} \\
\bn {\cdot} \sqrt{\kappa} \bq &= -g                                && \text{on $\Ga[N]$.} \\
\end{aligned}
\right.
\end{equation}
\begin{remark}\label{rmrk:continuityU}
The first condition is automatically fulfilled owing to the Dirichlet boundary condition $u_e {=} \hu$ on $\partial\Omega_e \setminus \Ga[D]$ imposed in the local problem and to the uniqueness of the hybrid variable on each mesh face (respectively, edge in 2D).
\end{remark}

The solution $(u_e,\bq_e)$ in each element $\Omega_e, \, e {=} 1,\ldots,\numel$ is thus efficiently retrieved by solving $\numel$ independent problems, see equation~\eqref{eq:PoissonHDGstrongLocal}, element-by-element.

\subsection{HDG local and global problems: weak form} 

Following the rationale introduced in~\cite{RS-SH:16}, the discrete functional spaces 
\begin{subequations}\label{eq:HDGspaces}
\begin{align}
\Vh(\Omega) & {:=}
\lbrace 
v \in \eltwo(\Omega) : \, v\vert_{\Omega_e} \in \Poly[\deg](\Omega_e) \forall \Omega_e, \, e {=} 1,\ldots,\numel
\rbrace,
\\
\HVh(S) & {:=}
\lbrace
\hv \in \eltwo(S) : \, \hv\vert_{\Gamma_i} \in \Poly[\deg](\Gamma_i) \forall \Gamma_i \subset S \subseteq \Gamma \cup \partial\Omega
\rbrace ,
\end{align}
\end{subequations}
are defined for the approximation of the element-based and face-based variables, respectively. In~\eqref{eq:HDGspaces}, $\Poly[\deg](\Omega_e)$ and $\Poly[\deg](\Gamma_i)$ stand for the spaces of polynomial functions of complete degree at most $\deg$ in $\Omega_e$ and on $\Gamma_i$, respectively.

For $e {=} 1,\ldots,\numel$, the weak form of the HDG local problem is: given $\uD$ on $\Ga[D]$ and $\hu^h$ on $\Gamma\cup\Ga[N]$, find $(u_e^h,\bq_e^h) \in \Vh(\Omega_e) {\times} \left[\Vh(\Omega_e)\right]^{\nsd}$ that satisfy
\begin{subequations}\label{eq:PoissonHDGweakLocal}
\begin{align}
- (\bw, \bq_e^h)_{\Omega_e} {+} (\grad {\cdot} (\sqrt{\kappa} \bw), u_e^h)_{\Omega_e} 
&= \langle \bn {\cdot} \sqrt{\kappa}\bw, \uD \rangle_{\partial\Omega_e\cap\Ga[D]} {+} \langle \bn {\cdot} \sqrt{\kappa}\bw, \hu^h \rangle_{\partial\Omega_e\setminus\Ga[D]} ,
\\
(v , \grad {\cdot} (\sqrt{\kappa} \bq_e^h))_{\Omega_e} {+} \langle v, \tau u_e^h \rangle_{\partial\Omega_e} 
&= (v, s)_{\Omega_e} {+} \langle v, \tau \uD \rangle_{\partial\Omega_e\cap\Ga[D]} {+} \langle v, \tau \hu^h \rangle_{\partial\Omega_e\setminus\Ga[D]} ,
\end{align}
\end{subequations}
for all $(v,\bw) \in \Vh(\Omega_e) {\times} \left[\Vh(\Omega_e)\right]^{\nsd}$, where $(\cdot,\cdot)_D$ and $\langle \cdot,\cdot\rangle_S$ denote the $\eltwo$ inner products on a generic subdomain $D \subset \Omega$ and $S \subset \Gamma \cup \partial\Omega$, respectively.
\begin{remark}
In equation~\eqref{eq:PoissonHDGweakLocal}, $\tau$ represents a stabilisation parameter influencing accuracy, stability and convergence of the HDG method~\cite{Jay-CGL:09,Nguyen-NPC:09,Nguyen-NPC:09b,Nguyen-CNP:10,Nguyen-NPC:11}.
\end{remark}

Similarly, the weak form of the HDG global problem is: find $\hu^h \in \HVh(\Gamma\cup\Ga[N])$ that satisfies
\begin{equation}\label{eq:PoissonHDGweakGlobal}
\sum_{e=1}^{\numel} \Bigl\{ 
\langle \hv, \bn {\cdot} \sqrt{\kappa}\bq_e^h \rangle_{\partial\Omega_e\setminus\Ga[D]} {+} \langle \hv, \tau u_e^h \rangle_{\partial\Omega_e\setminus\Ga[D]} 
 {-} \langle \hv, \tau \hu^h \rangle_{\partial\Omega_e\setminus\Ga[D]} \Bigr\} = {-} \sum_{e=1}^{\numel} \langle \hv, g \rangle_{\partial\Omega_e\cap\Ga[N]} ,
\end{equation}
for all $\hv \in \HVh(\Gamma\cup\Ga[N])$.

\subsection{HDG local and global problems: discrete form} \label{sc:PoissonHDGdiscrete}

An isoparametric formulation is considered for the primal, mixed and hybrid variables in the discrete spaces~\eqref{eq:HDGspaces}, that is,
\begin{equation}\label{eq:isoParam}
u^h = \sum_{i=1}^{\nen} N_i \node{u}_i , \quad
\bq^h = \sum_{i=1}^{\nen} N_i \vect{q}_i , \quad
\hu^h = \sum_{i=1}^{\nfn} \hN_i \node{\hu}_i ,
\end{equation}
where $\node{u}_i$, $\vect{q}_i$ and $\node{\hu}_i$ are the nodal values of the unknowns, $N_i$ and $\hN_i$ are the polynomial shape functions of degree $\deg$ defined in a reference element and on a reference face, respectively and $\nen$ and $\nfn$ denote the number of nodes per element and per face, respectively.

Hence, for each element $\Omega_e, \ e=1,\ldots,\numel$, the local problem~\eqref{eq:PoissonHDGweakLocal} leads to the linear system of equations
\begin{equation}\label{eq:PoissonHDGdiscreteLocal}
  \begin{bmatrix}
   \mat{A}_{uu} & \mat{A}_{uq}  \\
   \mat{A}_{uq}^T & \mat{A}_{qq}
  \end{bmatrix}_{\! e}
  \hspace{-0.5ex}
\begin{Bmatrix}
   \vect{u} \\
   \vect{q} 
  \end{Bmatrix}_{\!\! e} 
  {=} 
  \begin{Bmatrix}
    \vecF[u]  \\
    \vecF[q]  
  \end{Bmatrix}_{\!\! e} 
  \hspace{-0.5ex}
  {+}
  \begin{bmatrix}
    \mat{A}_{u\hu} \\
    \mat{A}_{q\hu} 
  \end{bmatrix}_{\! e}
  \hspace{-0.5ex}
  \vect{\hu}_e ,
\end{equation}
from which the following solution is computed
\begin{subequations}\label{eq:PoissonHDGdiscreteLocalSolution}
\begin{equation}\label{eq:PoissonHDGdiscreteLocalPb}
\begin{Bmatrix}
   \vect{u} \\
   \vect{q} 
  \end{Bmatrix}_{\!\! e} 
  {=} 
    \begin{Bmatrix}
    \vect{z}_u^f  \\
    \vect{z}_q^f  
  \end{Bmatrix}_{\!\! e} 
  \hspace{-0.5ex}
  {+}
    \begin{bmatrix}
    \mat{Z}_{u\hu} \\
    \mat{Z}_{q\hu} 
  \end{bmatrix}_{\! e}
  \hspace{-0.5ex}
  \vect{\hu}_e ,
\end{equation}
with the matrices
\begin{equation}\label{eq:PoissonHDGdiscreteLocalMat}
  \begin{bmatrix}
    \mat{Z}_{u\hu} \\
    \mat{Z}_{q\hu} 
  \end{bmatrix}_{\! e}
  {:=}
  \begin{bmatrix}
   \mat{A}_{uu} & \mat{A}_{uq}  \\
   \mat{A}_{uq}^T & \mat{A}_{qq}
  \end{bmatrix}_{\! e}^{-1}
    \hspace{-0.5ex}
  \begin{bmatrix}
    \mat{A}_{u\hu} \\
    \mat{A}_{q\hu} 
  \end{bmatrix}_{\! e}
\end{equation}
and the vectors
\begin{equation}\label{eq:PoissonHDGdiscreteLocalVec}
    \begin{Bmatrix}
    \vect{z}_u^f  \\
    \vect{z}_q^f  
  \end{Bmatrix}_{\!\! e} 
  {:=}
    \begin{bmatrix}
   \mat{A}_{uu} & \mat{A}_{uq}  \\
   \mat{A}_{uq}^T & \mat{A}_{qq}
  \end{bmatrix}_{\! e}^{-1}
  \hspace{-0.5ex}
  \begin{Bmatrix}
    \vecF[u]  \\
    \vecF[q]  
  \end{Bmatrix}_{\!\! e} .
\end{equation}
\end{subequations}

The corresponding discretisation of the global problem~\eqref{eq:PoissonHDGweakGlobal} leads to
\begin{equation}\label{eq:PoissonHDGdiscreteGlobal}
  \sum_{e=1}^{\numel}\Big\{
    \begin{bmatrix} \mat{A}_{u \hu}^T & \mat{A}_{q \hu}^T \end{bmatrix}_{\! e}
    \begin{Bmatrix} \vect{u} \\ \vect{q} \end{Bmatrix}_{\!\! e} 
   +
    [\mat{A}_{\hu\hu}]_e \, \vect{\hu}_e \Big\}
  = 
  \sum_{e=1}^{\numel} [\vecF[\hu]]_e .
\end{equation}
By plugging the local elemental solution~\eqref{eq:PoissonHDGdiscreteLocalPb} into~\eqref{eq:PoissonHDGdiscreteGlobal}, the HDG problem 
$$
\mat{K} \vect{\hu} = \vecHF[]
$$
involving only the globally-coupled degrees of freedom is obtained, where the matrix and the right-hand side vector are obtained by assembling the elemental contributions
%
%
\begin{subequations}\label{eq:PoissonGlobalMatVec}
\begin{align}
  \widehat{\mat{K}}_e 
  & := 
      \bigl[\begin{array}{@{}c@{\,}c@{}} \mat{A}_{u \hu}^T & \mat{A}_{q \hu}^T \end{array}\bigr]_{\! e}
  \hspace{-0.5ex}
      \begin{bmatrix}
        \mat{Z}_{u\hu} \\
        \mat{Z}_{q\hu} 
      \end{bmatrix}_{\! e}
  \hspace{-0.5ex}
      {+}
      [\mat{A}_{\hu\hu}]_e ,\\
  \hat{\vect{f}}_{e}
  & := [\vecF[\hu]]_e
  {-}\bigl[\begin{array}{@{}c@{\,}c@{}} \mat{A}_{u \hu}^T & \mat{A}_{q \hu}^T \end{array}\bigr]_{\! e}
  \hspace{-0.5ex}
      \begin{Bmatrix}
        \vect{z}_u^f  \\
        \vect{z}_q^f
      \end{Bmatrix}_{\!\! e} .
\end{align}
\end{subequations}

The expressions of the matrices and vectors introduced in this section are detailed in appendix~\ref{app:Poisson}.

\subsection{HDG local postprocess} \label{sc:PoissonPostprocess}

Introduce the discrete functional space
\begin{equation}\label{eq:HDGspacePostProcess}
\VhS(\Omega) {:=}
\lbrace 
v \in \eltwo(\Omega) : \, v\vert_{\Omega_e} \in \Poly[\deg+1](\Omega_e) \forall \Omega_e, \, e {=} 1,\ldots,\numel
\rbrace,
\end{equation}
where $\Poly[\deg+1](\Omega_e)$ denotes the space of polynomial functions of complete degree at most $\deg {+} 1$ in each element $\Omega_e$. 

The HDG postprocess procedure allows to compute a superconvergent approximation $\uS$ of the primal variable by solving an independent local problem in each element, namely
\begin{equation}\label{eq:HDGpostprocess}
\left\lbrace
\begin{aligned}
- \grad {\cdot} (\kappa \grad \uS ) &= \grad {\cdot} (\sqrt{\kappa} \bq_e) &&\text{in $\Omega_e$,}  \\
\bn {\cdot} \kappa \grad \uS &= - \bn {\cdot} \sqrt{\kappa} \bq_e    &&\text{on $\partial\Omega_e$,}
\end{aligned}
\right.
\end{equation}
with the constraint 
\begin{equation}\label{eq:constraintMean}
(\uS, 1)_{\Omega_e} = (u_e, 1)_{\Omega_e} 
\end{equation}
on the mean value of the solution in the element.

For each element $\Omega_e, \ e=1,\ldots,\numel$, the weak form of the postprocess procedure is: find $\uS^h \in \VhS(\Omega_e)$ that satisfies
\begin{subequations}\label{eq:HDGweakPostProcess}
\begin{align}
( \grad \vS, \kappa \grad \uS^h )_{\Omega_e} &= - ( \grad \vS, \sqrt{\kappa} \bq_e^h )_{\Omega_e} , \\
(\uS^h, 1)_{\Omega_e} &= (u_e^h, 1)_{\Omega_e} , \label{eq:constraintDiscrete}
\end{align}
\end{subequations}
for all $\vS \in \VhS(\Omega_e)$.

Using an isoparametric approximation for the functions in the space $\VhS(\Omega)$, the HDG local postprocess gives rise to the linear system
\begin{equation}\label{eq:PoissonHDGdiscretePostProcess}
  \begin{bmatrix}
   \mat{A}_{\star\star} & \mat{a}_{\star\lambda}  \\
   \mat{a}_{\star\lambda}^T & 0
  \end{bmatrix}_{\! e}
  \hspace{-0.5ex}
\begin{Bmatrix}
   \vect{\uS} \\
   \node{\lambda} 
  \end{Bmatrix}_{\!\! e} 
  {=} 
    \begin{bmatrix}
   \mat{0} & \mat{A}_{\star q}  \\
   \mat{a}_{\star \lambda}^T & \mat{0}
  \end{bmatrix}_{\! e}
  \begin{Bmatrix}
   \mathcal{I}^{\star} \vect{u} \\
   \mathcal{I}_{\nsd}^{\star} \vect{q} 
  \end{Bmatrix}_{\!\! e}  ,
\end{equation}
where the saddle-point structure of the problem follows from the imposition of the constraint~\eqref{eq:constraintDiscrete} via the Lagrange multiplier $\lambda$ and $\mathcal{I}^{\star}: \Vh \rightarrow \VhS$ and $\mathcal{I}_{\nsd}^{\star}: [\Vh]^{\nsd} \rightarrow [\VhS]^{\nsd}$ denote the interpolation operators from the spaces of polynomial functions of degree $\deg$ to the ones of degree $\deg {+} 1$ for scalar and vector-valued functions.

The expressions of the matrices and vectors introduced in this section are detailed in appendix~\ref{app:Poisson}.

\section{HDG formulation of the Stokes equations} \label{sc:StokesHDG}

This section presents the formulation of the HDG method for the Stokes equations, extending the framework previously introduced for the Poisson equation. For the sake of simplicity, the present work focuses on the velocity pressure formulation of the Stokes equations. For the Cauchy stress tensor formulation, a tutorial to devise an HDG method based on equal order approximation for all the variables and pointwise symmetric mixed variable is presented in~\cite{MG-GSH-20}.

The open bounded domain $\Omega \subset \RR^{\nsd}$ is characterised now by a boundary partitioned in three portions disjoint by pairs such that $\partial\Omega {=} \Ga[D] \cup \Ga[N] \cup \Ga[S]$, where Dirichlet, Neumann and slip conditions are imposed. The strong form of the Stokes equations is
\begin{equation}\label{eq:StokesStrongForm}
 \left\{\begin{aligned}
 -\grad {\cdot} (\nu \grad \bu - p \Insd) &= \bm{s}       &&\text{in $\Omega$,}\\
   \grad {\cdot} \bu &= 0  &&\text{in $\Omega$,}\\
   \bu &= \buD  &&\text{on $\Ga[D]$,}\\
    \bn {\cdot} (\nu \grad \bu - p \Insd ) &= \bm{g}        &&\text{on $\Ga[N]$,}\\
    \bu {\cdot} \bD + \bn {\cdot} \bigl(\nu \grad \bu -p\Insd \bigr)\bE &= \bm{0} &&\text{on $\Ga[S]$,}\\
 \end{aligned}\right.
\end{equation}
where the pair $(\bu,p)$ denotes the unknown velocity and pressure fields, $\nu>0$ is the kinematic viscosity, $\bn$ is the outward unit normal to the boundary, $\bm{s}$ is the vector of the body forces and $\buD$ and $\bm{g}$ represent the imposed velocity and pseudo-traction on the Dirichlet and Neumann boundaries, respectively. On the slip boundary, matrices $\bD$ and $\bE$ are defined as $\bD {:=} [ \bn,\beta \bt_1 , \ldots , \beta \bt_{\nsd-1} ]$ and $\bE {:=} [ \alpha \bn,\bt_1 , \ldots ,\bt_{\nsd-1} ]$, the tangential vectors $\bt_j$, \ $j {=} 1, \ldots , \nsd {-} 1$ being such that $\{ \bn,\bt_1, \ldots , \bt_{\nsd-1} \}$ form an orthonormal system of vectors. Two scalars, $\alpha$ and $\beta$, represent the penetration and friction coefficient, respectively. For $\alpha,\beta \rightarrow 0$, the case of a perfectly slip condition is retrieved~\cite{MG-GSH-20}.
\begin{remark}\label{rmrk:incompressibility}
The divergence-free condition in equation~\eqref{eq:StokesStrongForm} induces the following compatibility condition on the velocity field
\begin{equation}\label{eq:incompressibleConst}
\langle \buD {\cdot} \bn, 1 \rangle_{\Ga[D]} + \langle \bu {\cdot} \bn, 1 \rangle_{\partial\Omega \setminus \Ga[D]}  = 0 .
\end{equation}
\end{remark}
\begin{remark}\label{rmrk:pConstraint}
In case of a purely Dirichlet boundary value problem, that is $\partial\Omega {=} \Ga[D]$, an additional constraint needs to be introduced to retrieve uniqueness of the pressure field. A common approach relies on imposing the mean value of the pressure in the domain~\cite{donea2003finite,Quarteroni-book}
\begin{equation}\label{eq:pressureConstraintDomain}
  \frac{1}{|\Omega|} ( p,1 )_{\Omega} = 0 ,
\end{equation}
or on the boundary of the domain~\cite{Jay-CG:09,Cockburn-CS:14,Nguyen-NPC:10,MG-GKSH-18}, namely
\begin{equation}\label{eq:pressureConstraint}
  \frac{1}{|\partial \Omega|}  \langle p,1 \rangle_{\partial \Omega} = 0 .
\end{equation}
\end{remark}


\subsection{HDG local and global problems: strong form}

Following the rationale presented in section~\ref{sc:PoissonHDG}, equation~\eqref{eq:StokesStrongForm} is rewritten element-by-element as a system of first-order equations by introducing a mixed variable $\bL {=} -\sqrt{\nu} \grad \bu$, namely
\begin{equation} \label{eq:StokesHDGstrong}
\left\{\begin{aligned}
\bL + \sqrt{\nu} \grad \bu & = \bm{0} 																	&& \text{in $\Omega_e$,} \\
\grad {\cdot} \bigl( \sqrt{\nu} \bL + p\Insd \bigr) &= \bm{s}      														&& \text{in $\Omega_e$,} \\
\grad {\cdot} \bu &= 0  																					&& \text{in $\Omega_e$,} \\
\bu &= \buD     																					&& \text{on $\Ga[D]$,} \\
\bn {\cdot} \bigl(\sqrt{\nu} \bL + p \Insd \bigr) &= -\bm{g}    											&& \text{on $\Ga[N]$,} \\
\bu {\cdot} \bm{D} - \bn {\cdot} \bigl( \sqrt{\nu} \bL + p \Insd \bigr)\bm{E} &= \bm{0}  	&& \text{on $\Ga[S]$,} \\
\jump{\bu \otimes \bn} &=\bm{0}  															&& \text{on $\Gamma$,} \\
\jump{ \bn {\cdot} \bigl( \sqrt{\nu}\bL + p \Insd \bigr) } &= \bm{0}  									&& \text{on $\Gamma$,} \\
\end{aligned} \right.
\end{equation}
for $e {=} 1,\ldots,\numel$.

First, the HDG algorithm performes the hybridisation step by expressing $(\bu_e,p_e,\bL_e)$ in each element $\Omega_e$ as functions of the unknown trace of the velocity $\bhu$ on the element faces via the HDG local problem
\begin{subequations}\label{eq:StokesHDGstrongLocal}
\begin{equation} \label{eq:StokesHDGstrongLocalPb}
\left\{\begin{aligned}
\bL_e + \sqrt{\nu} \grad \bu_e & = \bm{0} 																	&& \text{in $\Omega_e$,} \\
\grad {\cdot} \bigl( \sqrt{\nu} \bL_e + p_e \Insd \bigr) &= \bm{s}      														&& \text{in $\Omega_e$,} \\
\grad {\cdot} \bu_e &= 0  																					&& \text{in $\Omega_e$,} \\
\bu_e &= \buD     																					&& \text{on $\partial\Omega_e \cap \Ga[D]$,} \\
\bu_e &= \bhu     																					&& \text{on $\partial\Omega_e \setminus \Ga[D]$,} 
\end{aligned} \right.
\end{equation}
for $e {=} 1,\ldots,\numel$. Note that equation~\eqref{eq:StokesHDGstrongLocalPb} is a purely Dirichlet boundary value problem. Hence, following remark~\ref{rmrk:pConstraint}, the constraint 
\begin{equation} \label{eq:StokesHDGstrongLocalConstraint}
\frac{1}{|\partial \Omega_e|}  \langle p_e, 1 \rangle_{\partial \Omega} = \rho_e , \quad \text{for $e {=} 1,\ldots,\numel$} ,
\end{equation}
\end{subequations}
is introduced, where $\rho_e$ is an independent variable representing the mean value of the pressure on the boundary $\partial\Omega_e$. It is worth noting that the variable $\rho$ was not present in the HDG approximation of the Poisson equation and its treatment in the \texttt{HDGlab} code will be detailed in section~\ref{sc:solverStokes}.

The HDG global problem thus accounts for the transmission and non-Dirichlet boundary conditions, that is
\begin{subequations}\label{eq:StokesHDGstrongGlobal}
\begin{equation}
\left\{\begin{aligned}
\jump{\bu \otimes \bn} &=\bm{0}  															&& \text{on $\Gamma$,} \\
\jump{ \bn {\cdot} \bigl( \sqrt{\nu}\bL + p \Insd \bigr) } &= \bm{0}  									&& \text{on $\Gamma$,} \\
\bn {\cdot} \bigl(\sqrt{\nu} \bL + p \Insd \bigr) &= -\bm{g}    											&& \text{on $\Ga[N]$,} \\
\bu {\cdot} \bm{D} - \bn {\cdot} \bigl( \sqrt{\nu} \bL + p \Insd \bigr)\bm{E} &= \bm{0}  	&& \text{on $\Ga[S]$,} \\
\end{aligned} \right.
\end{equation}
where, following remark~\ref{rmrk:continuityU}, the first condition is automatically fulfilled.
In addition, the constraint in remark~\ref{rmrk:incompressibility} is rewritten element-by-element in terms of the hybrid variable $\bhu$ leading to
\begin{equation}
\langle \buD {\cdot} \bn, 1 \rangle_{\partial\Omega_e \cap \Ga[D]} {+} \langle \bhu {\cdot} \bn, 1 \rangle_{\partial\Omega_e \setminus \Ga[D]}  = 0 ,
\end{equation}
\end{subequations}
for $e {=} 1,\ldots,\numel$.

\subsection{HDG local and global problems: weak form}

The corresponding weak form of the HDG local problem~\eqref{eq:StokesHDGstrongLocal} is: given $\buD$ on $\Ga[D]$ and $\bhu^h$ on $\Gamma\cup\Ga[N]\cup\Ga[S]$, find $(\bL_e^h,\bu_e^h,p_e^h) \in [\Vh(\Omega_e)]^{\nsd \times \nsd} {\times} [\Vh(\Omega_e)]^{\nsd} {\times} \Vh(\Omega_e)$ that satisfy
\begin{equation}\label{eq:StokesHDGweakLocal}
\left\{\begin{aligned} 
 {-} ( \bW, \bL_e^h )_{\Omega_e}  
 {+} ( \grad {\cdot}(\sqrt{\nu}\bW), \bu_e^h )_{\Omega_e}
&
 {=}   \langle \bn_e {\cdot} \sqrt{\nu} \bW , \buD \rangle_{\partial\Omega_e\cap\Ga[D]} 
 {+} \langle \bn_e {\cdot} \sqrt{\nu} \bW , \bhu^h \rangle_{\partial \Omega_e \setminus \Ga[D] }  ,
\\
  ( \bw, \grad{\cdot}( \sqrt{\nu} \bL_e^h) )_{\Omega_e}  
 {+} ( \bw, \grad p_e^h )_{\Omega_e} 
 {+} \langle \bw , \tau \bu_e^h \rangle_{\partial\Omega_e} 
&
 {=} ( \bw, \bm{s} )_{\Omega_e} 
  + \langle \bw, \tau \buD \rangle_{\partial\Omega_e \cap \Ga[D]}
  + \langle \bw, \tau \bhu^h \rangle_{\partial\Omega_e \setminus \Ga[D]} ,    
\\
   ( \grad q, \bu_e )_{\Omega_e}  
&
 {=} \langle q, \buD {\cdot} \bn_e \rangle_{\partial \Omega_e \cap \Ga[D]}
 {+} \langle q, \bhu^h {\cdot} \bn_e \rangle_{\partial \Omega_e \setminus \Ga[D] } ,
\\
   \langle p_e^h , 1 \rangle_{\partial\Omega_e}  &  {=} |\partial\Omega_e| \rho_e^h ,
\end{aligned}\right.
\end{equation}
for all $(\bW,\bv,q) \in [\Vh(\Omega_e)]^{\nsd \times \nsd} {\times} [\Vh(\Omega_e)]^{\nsd} {\times} \Vh(\Omega_e)$. It is worth noting that, differently from the Poisson case, equation~\eqref{eq:StokesHDGweakLocal} provides $(\bL_e^h,\bu_e^h,p_e^h)$ in terms of two global unknowns, $\bhu^h$ and $\rho^h {:=} \{ \rho_1^h, \ldots , \rho_{\numel}^h \}^T$.

Similarly, the weak form of the HDG global problem~\eqref{eq:StokesHDGstrongGlobal} is: find $\bhu^h \in [\HVh]^{\nsd}$ and $\bm{\rho}^h \in \RR^{\numel}$ such that
\begin{equation} \label{eq:StokesHDGweakGlobal}
\left\{\begin{aligned}
  \sum_{e=1}^{\numel} \Big\{ &
     \langle\bhv , \bn_e {\cdot} \sqrt{\nu} \bL_e^h \rangle_{\partial \Omega_e\setminus (\Ga[D] \cup \Ga[S])}    
      {-} \langle\bhv , \bn_e {\cdot} \sqrt{\nu} \bL_e^h \bE \rangle_{\partial \Omega_e\cap\Ga[S]}    
      \\[-0.5ex] & \hspace{5pt}
      {+} \langle\bhv , p_e^h \bn_e \rangle_{\partial \Omega_e\setminus (\Ga[D] \cup \Ga[S])}    
      {-} \langle\bhv , p_e^h \bn_e {\cdot} \bE \rangle_{\partial \Omega_e\cap\Ga[S]}    
      \\[-0.5ex] & \hspace{25pt}
      {+} \langle\bhv , \tau \bu_e^h \rangle_{\partial \Omega_e\setminus (\Ga[D] \cup \Ga[S])}    
      {-} \langle\bhv , \tau \bu_e^h {\cdot} \bE \rangle_{\partial \Omega_e\cap \Ga[S]}    
      \\[-0.5ex]  & \hspace{43pt}
  	  {-}  \langle \bhv , \tau \bhu^h \rangle_{\partial \Omega_e\setminus (\Ga[D] \cup \Ga[S])}    
      {+}  \langle \bhv , \bhu^h {\cdot} (\bD {+} \tau \bE) \rangle_{\partial \Omega_e \cap \Ga[S]} \! \Big\}  
      = {-}\sum_{e=1}^{\numel} \langle \bhv , \bm{g} \rangle_{\partial \Omega_e \cap \Ga[N]} , 
  \\ & \hspace{195pt}
      \langle \bhu^h {\cdot} \bn_e , 1  \rangle_{\partial \Omega_e \setminus \Ga[D] }  
      = {-} \langle \buD  {\cdot} \bn_e , 1 \rangle_{\partial \Omega_e \cap \Ga[D]} , 
      \\[-0.5ex] & \hspace{300pt} 
      \text{for $e=1,\dotsc,\numel$}   
\end{aligned}\right.
\end{equation}
for all $\bhv \in [\HVh]^{\nsd}$.

\subsection{HDG local and global problems: discrete form} 

The discretisation of the local problem~\eqref{eq:StokesHDGweakLocal} leads to the following linear system of equations
\begin{equation} \label{eq:StokesHDGdiscreteLocal}
  \begin{bmatrix}
    \mat{A}_{LL}    & \mat{A}_{Lu} & \mat{0}              & \mat{0}                   \\
    \mat{A}_{Lu}^T & \mat{A}_{uu} & \mat{A}_{pu}^T & \mat{0}                   \\
    \mat{0}             & \mat{A}_{pu} & \mat{0}              & \mat{a}_{\rho p} \\
    \mat{0}             & \mat{0}          & \mat{a}_{\rho p}^T & 0                   \\
  \end{bmatrix}_{\! e}
  \hspace{-0.5ex}
\begin{Bmatrix}
   \vect{L} \\
   \vect{u} \\
   \vect{p} \\
   \zeta 
  \end{Bmatrix}_{\!\! e} 
  {=}
  \begin{Bmatrix}
    \vecF[L]  \\
    \vecF[u]  \\
    \vecF[p] \\
    0 
  \end{Bmatrix}_{\!\! e} 
  \hspace{-0.5ex}
  {+} 
  \begin{bmatrix}
    \mat{A}_{L\hu} \\
    \mat{A}_{u\hu} \\
    \mat{A}_{p\hu} \\
    \mat{0}
  \end{bmatrix}_{\! e}
  \hspace{-1.5ex}
  \vect{\hu}_e
  {+}
  \begin{Bmatrix}
    \vect{0} \\
    \vect{0} \\
    \vect{0} \\
    1
  \end{Bmatrix}_{\!\! e}
  \hspace{-0.5ex}
  \rho_e,
\end{equation}
where the constraint~\eqref{eq:StokesHDGstrongLocalConstraint} is imposed using the Lagrange multiplier $\zeta$. It is worth noting that the Lagrange multiplier is required to guarantee that equation~\eqref{eq:StokesHDGdiscreteLocal} is well-posed and the computed pressure field is unique but is not utilised in the solution of the global HDG problem. The resulting local elemental solution is given by
\begin{subequations} \label{eq:StokesHDGdiscreteLocalSolution}
\begin{equation} \label{eq:StokesHDGdiscreteLocalPb}
\begin{Bmatrix}
   \vect{L} \\
   \vect{u} \\
   \vect{p} \\
   \zeta 
  \end{Bmatrix}_{\!\! e}
  {=} 
  \begin{Bmatrix}
    \vect{z}_L^f \\
    \vect{z}_u^f \\
    \vect{z}_p^f \\
    \vect{z}_\zeta^f
  \end{Bmatrix}_{\!\! e}
  \hspace{-0.5ex}
  {+}
  \begin{bmatrix}
    \mat{Z}_{L\hu} \\
    \mat{Z}_{u\hu} \\
    \mat{Z}_{p\hu} \\
    \mat{Z}_{\zeta\hu}
  \end{bmatrix}_{\! e}
  \hspace{-1.5ex}
  \vect{\hu}_e
  {+}
  \begin{Bmatrix}
    \vect{z}_L^{\rho} \\
    \vect{z}_u^{\rho} \\
    \vect{z}_p^{\rho} \\
    \vect{z}_\zeta^{\rho}
  \end{Bmatrix}_{\!\! e}
  \hspace{-0.5ex}
  \rho_e,
\end{equation}
with the matrix and vectors defined as
\begin{align} \label{eq:StokesHDGdiscreteLocalMatVec}
  \begin{bmatrix}
    \mat{Z}_{L\hu} \\
    \mat{Z}_{u\hu} \\
    \mat{Z}_{p\hu} \\
    \mat{Z}_{\zeta\hu}
  \end{bmatrix}_{\! e}
  & {:=}
    \begin{bmatrix}
    \mat{A}_{LL}    & \mat{A}_{Lu} & \mat{0}              & \mat{0}                   \\
    \mat{A}_{Lu}^T & \mat{A}_{uu} & \mat{A}_{pu}^T & \mat{0}                   \\
    \mat{0}             & \mat{A}_{pu} & \mat{0}              & \mat{a}_{\rho p} \\
    \mat{0}             & \mat{0}          & \mat{a}_{\rho p}^T & 0                   \\
  \end{bmatrix}_{\! e}^{-1}
  \hspace{-0.5ex}
  \begin{bmatrix}
    \mat{A}_{L\hu} \\
    \mat{A}_{u\hu} \\
    \mat{A}_{p\hu} \\
    \mat{0}
  \end{bmatrix}_{\! e} , \\
  \begin{Bmatrix}
    \vect{z}_L^f \\
    \vect{z}_u^f \\
    \vect{z}_p^f \\
    \vect{z}_\zeta^f
  \end{Bmatrix}_{\!\! e}
   & {:=}
   \begin{bmatrix}
    \mat{A}_{LL}    & \mat{A}_{Lu} & \mat{0}              & \mat{0}                   \\
    \mat{A}_{Lu}^T & \mat{A}_{uu} & \mat{A}_{pu}^T & \mat{0}                   \\
    \mat{0}             & \mat{A}_{pu} & \mat{0}              & \mat{a}_{\rho p} \\
    \mat{0}             & \mat{0}          & \mat{a}_{\rho p}^T & 0                   \\
  \end{bmatrix}_{\! e}^{-1}
  \hspace{-0.5ex}
  \begin{Bmatrix}
    \vecF[L]  \\
    \vecF[u]  \\
    \vecF[p] \\
    0 
  \end{Bmatrix}_{\!\! e} , \\
  \begin{Bmatrix}
    \vect{z}_L^{\rho} \\
    \vect{z}_u^{\rho} \\
    \vect{z}_p^{\rho} \\
    \vect{z}_\zeta^{\rho}
  \end{Bmatrix}_{\!\! e}
   & {:=} 
  \begin{bmatrix}
    \mat{A}_{LL}    & \mat{A}_{Lu} & \mat{0}              & \mat{0}                   \\
    \mat{A}_{Lu}^T & \mat{A}_{uu} & \mat{A}_{pu}^T & \mat{0}                   \\
    \mat{0}             & \mat{A}_{pu} & \mat{0}              & \mat{a}_{\rho p} \\
    \mat{0}             & \mat{0}          & \mat{a}_{\rho p}^T & 0                   \\
  \end{bmatrix}_{\! e}^{-1}
  \hspace{-0.5ex}
  \begin{Bmatrix}
    \vect{0} \\
    \vect{0} \\
    \vect{0} \\
    1
  \end{Bmatrix}_{\!\! e} .
\end{align}
\end{subequations}

The discrete form of the global problem~\eqref{eq:StokesHDGweakGlobal} reads as
\begin{equation}\label{eq:StokesHDGdiscreteGlobal}
  \begin{aligned}
  \sum_{e=1}^{\numel}\Big\{
    \begin{bmatrix} \mat{A}_{\hu L} \mat{A}_{\hu u} & \mat{A}_{\hu p} \end{bmatrix}_{\! e}
    \begin{Bmatrix} \vect{L} \\ \vect{u} \\ \vect{p} \end{Bmatrix}_{\!\! e}
   +
    [\mat{A}_{\hu\hu}]_e \, \vect{\hu}_e \Big\}
  &= 
  \sum_{e=1}^{\numel} [\vecF[\hu]]_e, \\
  \vect{1}^T \, [\mat{A}_{p\hu}]_e \vect{\hu}_e &= - \vect{1}^T \, [\vecF[p]]_e .
  \end{aligned}
\end{equation}

By inserting the solution~\eqref{eq:StokesHDGdiscreteLocalPb} into~\eqref{eq:StokesHDGdiscreteGlobal}, the following system involving the globally-coupled unknowns $\vect{\hu}$ and $\bm{\rho}$ is obtained
\begin{equation}\label{eq:StokesHDGdiscreteGlobalFinal}
  \begin{bmatrix}\widehat{\mat{K}} & \mat{G} \\
                   \mat{G}^T           & \mat{0}  \end{bmatrix}
  \begin{Bmatrix}\vect{\hu} \\ \bm{\rho} \end{Bmatrix} 
  =
  \begin{Bmatrix} \vecHF[\hu] \\ \vecHF[\rho] \end{Bmatrix} ,
\end{equation}
where the matrices and vectors have the form
\begin{subequations}\label{eq:StokesHDGdiscreteGlobalFinalMatVec}
\begin{align}
  \widehat{\mat{K}} 
  & := \Assem_{e=1}^{\numel}
      \bigl[\begin{array}{@{}c@{\,}c@{\,}c@{\,}c@{}} \mat{A}_{\hu L} & \mat{A}_{\hu u} & \mat{A}_{\hu p} & \mat{0} \end{array}\bigr]_{\! e}
  \hspace{-0.5ex}
      \begin{bmatrix}
        \mat{Z}_{L\hu} \\
        \mat{Z}_{u\hu} \\
        \mat{Z}_{p\hu} \\
        \mat{Z}_{\zeta\hu} 
      \end{bmatrix}_{\! e}
  \hspace{-0.5ex}
      {+}
      [\mat{A}_{\hu\hu}]_e ,\\[-2ex]
  \mat{G}^T 
  & := \begin{bmatrix} \vect{1}^T  \, [\mat{A}_{p\hu}]_1 \\ \vect{1}^T  \, [\mat{A}_{p\hu}]_2 \\ \dotsb \\ \vect{1}^T  \, [\mat{A}_{p\hu}]_{\numel}\end{bmatrix}, \\[-2ex]
  \vecHF[\hu]
  & := \Assem_{e=1}^{\numel} [\vecF[\hu]]_e
  {-}\bigl[\begin{array}{@{}c@{\,}c@{\,}c@{\,}c@{}} \mat{A}_{\hu L} & \mat{A}_{\hu u} & \mat{A}_{\hu p} & \mat{0}\end{array}\bigr]_{\! e}
  \hspace{-0.5ex}
      \begin{Bmatrix}
        \vect{z}_L^f  \\
        \vect{z}_u^f  \\
        \vect{z}_p^f \\
        \vect{z}_{\zeta}^f
      \end{Bmatrix}_{\!\! e} , \\[-2ex]
  \vecHF[\rho]  
  & :=   - \begin{bmatrix} \vect{1}^T  \, [\vecF[p]]_1 \\ \vect{1}^T  \, [\vecF[p]]_2 \\ \dotsb \\ \vect{1}^T  \, [\vecF[p]]_{\numel}\end{bmatrix} .
\end{align}
\end{subequations}

The expressions of the matrices and vectors introduced above are detailed in appendix~\ref{app:Stokes}.

\begin{remark}
Differently from the Poisson case, the global problem~\eqref{eq:StokesHDGdiscreteGlobalFinal} features a saddle-point structure, as classical in the context of incompressible flows~\cite{donea2003finite}. The proof of the symmetry of the HDG matrix in equation~\eqref{eq:StokesHDGdiscreteGlobalFinal} can be devised following the rationale described in~\cite{MG-GSH-20}.
\end{remark}

\subsection{HDG local postprocess}  \label{sc:StokesPostprocess}

The HDG postprocess procedure presented in section~\ref{sc:PoissonPostprocess} for the Poisson equation can be extended straightforwardly to the case of the vectorial variable $\bu$.
\begin{remark}
The postprocessing procedure utilised here is inspired by the work of Stenberg~\cite{Stenberg-90-Stokes} and was exploited in the framework of HDG to obtain an improved approximation of the velocity field via the solution of an additional inexpensive element-by-element problem~\cite{Nguyen-NPC:10,RS-SH-18}. Nonetheless, in the context of incompressible flows, it is often of interest retrieving an $\Hdiv$-conforming and exactly divergence-free approximation of the velocity field. For this purpose, alternative postprocessing strategies inspired by the Brezzi-Douglas-Marini (BDM) projection operator~\cite{brezzi1991mixed} were proposed~\cite{Nguyen-CNP:10,Jay-CGNPS-11}. It is worth noting that the above mentioned procedures are suitable only for the velocity-pressure formulation of the incompressible flow equations. In case a formulation based on the Cauchy stress tensor is considered, an additional constraint is required to handle the rigid rotational modes as discussed in~\cite{RS-SGKH-18,MG-GKSH-18,MG-GSH-20}.
\end{remark}

A superconvergent velocity field $\buS$ is thus obtained by solving an independent local problem in each element, namely
\begin{equation}\label{eq:StokesHDGpostprocess}
\left\lbrace
\begin{aligned}
- \grad {\cdot} (\nu \grad \buS ) &= \grad {\cdot} (\sqrt{\nu} \bL_e) &&\text{in $\Omega_e$,}  \\
\bn {\cdot} \nu \grad \buS &= - \bn {\cdot} \sqrt{\nu} \bL_e    &&\text{on $\partial\Omega_e$,}
\end{aligned}
\right.
\end{equation}
with the constraint 
\begin{equation}\label{eq:constraintMeanStokes}
(\buS, 1)_{\Omega_e} = (\bu_e, 1)_{\Omega_e} 
\end{equation}
on the mean value of the velocity in the element.

Hence, for each element $\Omega_e, \ e {=} 1,\ldots,\numel$, the weak form of the postprocess procedure is: find $\buS^h \in [\VhS(\Omega_e)]^{\nsd}$ such that
\begin{subequations}\label{eq:StokesHDGweakPostProcess}
\begin{align}
( \grad \bvS, \nu \grad \buS^h )_{\Omega_e} &= - ( \grad \bvS, \sqrt{\nu} \bL_e^h )_{\Omega_e} , \\
(\buS^h, 1)_{\Omega_e} &= (\bu_e^h, 1)_{\Omega_e} , \label{eq:constraintDiscreteStokes}
\end{align}
\end{subequations}
for all $\bvS \in [\VhS(\Omega_e)]^{\nsd}$.

Using an isoparametric approximation for the functions in the space $[\VhS(\Omega)]^{\nsd}$, the HDG local postprocess for the Stokes equations leads to
\begin{equation}\label{eq:StokesHDGdiscretePostProcess}
  \begin{bmatrix}
   \mat{A}_{\star\star} & \mat{A}_{\star\lambda}  \\
   \mat{A}_{\star\lambda}^T & \mat{0}
  \end{bmatrix}_{\! e}
  \hspace{-0.5ex}
\begin{Bmatrix}
   \vect{\buS} \\
   \bm{\lambda} 
  \end{Bmatrix}_{\!\! e} 
  {=} 
    \begin{bmatrix}
   \mat{0} & \mat{A}_{\star L}  \\
   \mat{A}_{\star \lambda}^T & \mat{0}
  \end{bmatrix}_{\! e}
  \begin{Bmatrix}
   \mathcal{I}_{\nsd}^{\star} \vect{u} \\
   \mathcal{I}_{\nsd \times \nsd}^{\star} \vect{L} 
  \end{Bmatrix}_{\!\! e}  ,
\end{equation}
where $\bm{\lambda}$ is the vector of Lagrange multipliers imposing the constraint~\eqref{eq:constraintDiscreteStokes} on the elemental mean value and $\mathcal{I}_{\nsd \times \nsd}^{\star} {:} [\Vh]^{\nsd \times \nsd} {\rightarrow} [\VhS]^{\nsd \times \nsd}$ denotes the interpolation operator for tensor-valued functions from the space of polynomials of degree $\deg$ to the one of degree $\deg {+} 1$.

The expressions of the matrices and vectors introduced above are detailed in appendix~\ref{app:Stokes}.

\section{The \texttt{HDGlab} repository} \label{sc:repo}

The implementation of the HDG solver for the Poisson and Stokes equations has been made available as an open-source software, released under the terms of the GNU General Public License version 3.0 or any later version (\url{https://www.gnu.org/licenses}) and is freely available from the repository: \url{https://git.lacan.upc.edu/hybridLab/HDGlab}.

\begin{figure}[!tb]
	\centering
	\includegraphics[width=0.25\textwidth]{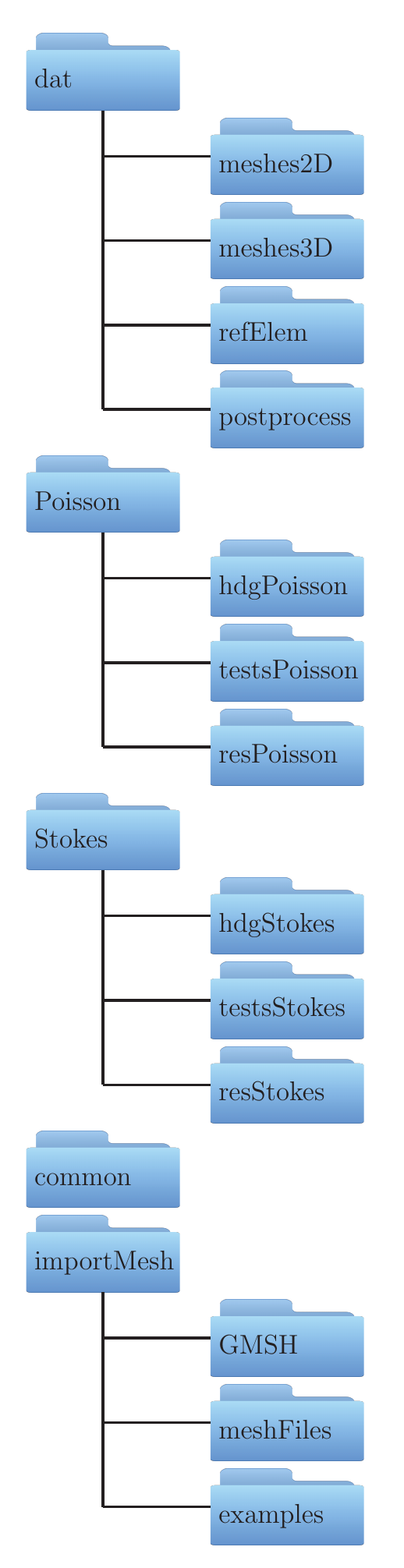}
	\caption{Structure of the \texttt{HDGlab} repository.}
	\label{fig:repository}
\end{figure}

The structure of the repository, illustrated in figure~\ref{fig:repository}, is described in this section.

The directory \texttt{dat} contains the data files. This includes two and three dimensional meshes in the directories \texttt{meshes2D} and \texttt{meshes3D} respectively, the pre-computed reference elements in two and three dimensions in the directory \texttt{refElem} and the data structure required to postprocess high-order HDG solutions in the directory \texttt{postprocess}.

The directory \texttt{Poisson} contains the HDG solver for the Poisson problem as described in section~\ref{sc:PoissonHDG}. The directory \texttt{hdgPoisson} contains the core HDG library for the Poisson equation. The directory \texttt{testsPoisson} is used to organise the functions that describe the setup of the problems to be solved, including the definition of the boundary conditions, source term and, if known, the analytical solution. The directory \texttt{resPoisson} is where the results are saved after an execution of the Poisson solver. 

The directory \texttt{Stokes} contains the HDG solver for the Stokes problem as described in section~\ref{sc:StokesHDG}. The structure of this directory follows the same rationale as the one corresponding to the Poisson solver.

The directory \texttt{common} contains a set of functions that are common to both the Poisson and the Stokes solvers. 

Finally, the directory \texttt{importMesh} contains a library that is provided to import a mesh generated with the open source software \texttt{Gmsh}~\cite{GMSH-09} in \texttt{HDGlab}. The core routines to convert a mesh from \textit{.msh} to \textit{.mat} format are located in the directory \texttt{GMSH}. The directory \texttt{examples} contains some test cases including \texttt{.geo} and \textit{.msh} files, whereas the output of the imported mesh is stored in the directory \texttt{meshFiles}. This library is described in detail in Appendix~\ref{app:Gmsh}.

\section{Data structures} \label{sc:data}

Three data structures are used to manage the mesh, the reference element and the face information required to compute the elemental contributions of the global HDG problem. These three variables are assumed to be an input of the HDG library and a detailed description is provided in this section. To make the developed software more accessible variables of type \textit{struct} are used in this work rather than \textit{class} types.

\subsection{Mesh} \label{sc:mesh}

The variable \texttt{mesh} contains the following information:

\begin{itemize}
	\item \texttt{nsd}: Number of spatial dimensions.
	\item \texttt{optionNodes}: Type of high-order nodal distribution, being an equally-spaced (0) or a Fekete (1) nodal set.
	\item \texttt{nOfNodes}: Number of nodes.
	\item \texttt{X}: Array of dimension \texttt{nOfNodes}$\times$\texttt{nsd} containing the nodal coordinates of the mesh.
	\item \texttt{nOfElements}: Number of elements.	
	\item \texttt{indexT}: Array of dimension \texttt{nOfElements}$\times 2$ containing the connectivity indices of the mesh. The first column contains the first node of the element and the second column contains the last node of the element.
	\item \texttt{pElem}: Array of dimension $1 \times $\texttt{nOfElements} containing the degree of approximation of each element.
	\item \texttt{matElem}: Array of dimension $1 \times $\texttt{nOfElements} containing the material flag for each element.
	\item \texttt{nOfIntFaces}: Number of interior faces (i.e. faces not on the boundary).
	\item \texttt{intFaces}: Array of dimension \texttt{nOfIntFaces}$\times 5$. The first two columns contain the first element sharing this face and its local face number. The next two columns contain the second element sharing this face and its local face number. The last column contains the local node number of the second face that matches with the first local node of the first face.
	\item \texttt{nOfExtFaces}: Number of exterior faces (i.e. faces on the boundary).
	\item \texttt{extFaces}: Array of dimension \texttt{nOfIntFaces}$\times 4$. The first two columns contains the element sharing this face and its local face number. The third column contains the boundary condition flag and the last column contains the flag of the boundary curve or surface.
\end{itemize}

\begin{figure}[!tb]
	\centering
	\includegraphics[width=0.35\textwidth]{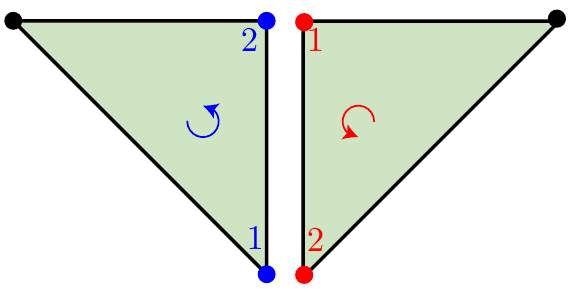}
	\caption{Interface between two triangular elements, showing the orientation of each element and the local node number of each node on the face as seen from the element on the left and on the right of the interface.}
	\label{fig:intFaces2D}
\end{figure}

\begin{figure}[!tb]
	\centering
	\includegraphics[width=0.35\textwidth]{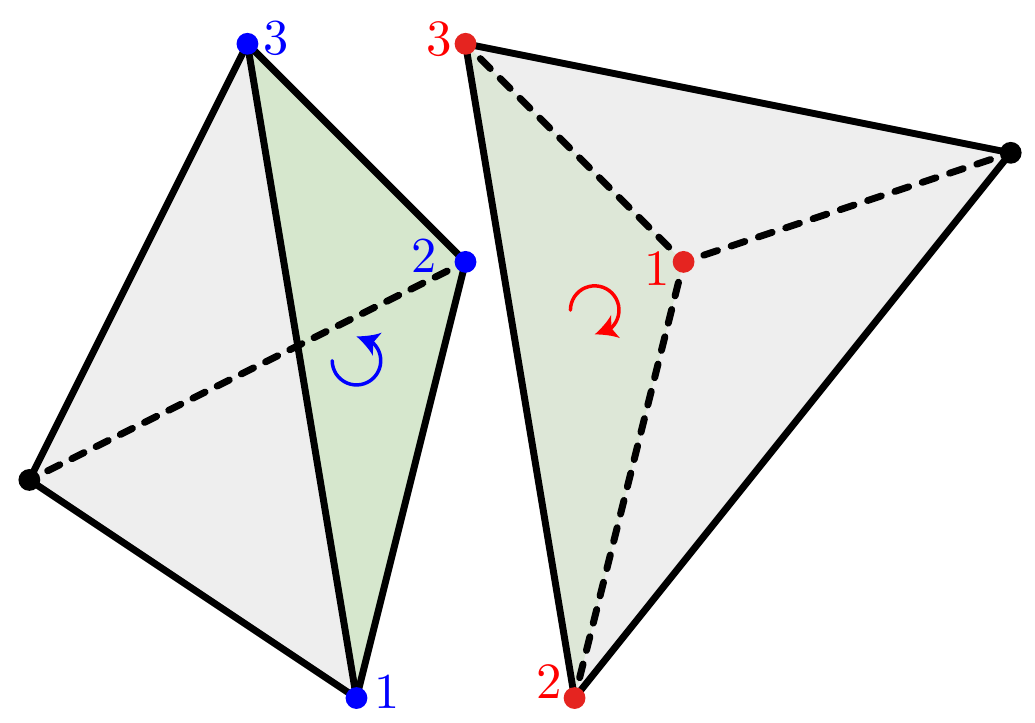}
	\caption{Interface between two tetrahedral elements, showing the orientation of each face and the local node number of each node on the face as seen from the element on the left and on the right of the interface.}
	\label{fig:intFaces3D}
\end{figure}

The information stored in the field \texttt{intFaces} is characteristic of a DG formulation, where integrals on interior faces need to be computed. This is in contrast with a standard CG formulation, where only the field \texttt{extFaces} is needed to impose the boundary conditions. The last column of \texttt{intFaces} is needed to account for the different orientation of an interior face as seen from the element on the left and on the right of a face. It is worth noting that in two dimensions the information could be omitted because the local node number of the second face that matches with the first local node of the first face is always equal to two, as illustrated in figure~\ref{fig:intFaces2D}.
However, in three dimensions this information is required as there are three possible rotations of the local face nodes that do not alter the orientation of the element/face, as depicted in figure~\ref{fig:intFaces3D}. 

\begin{figure}[!tb]
	\centering
	\includegraphics[width=0.35\textwidth]{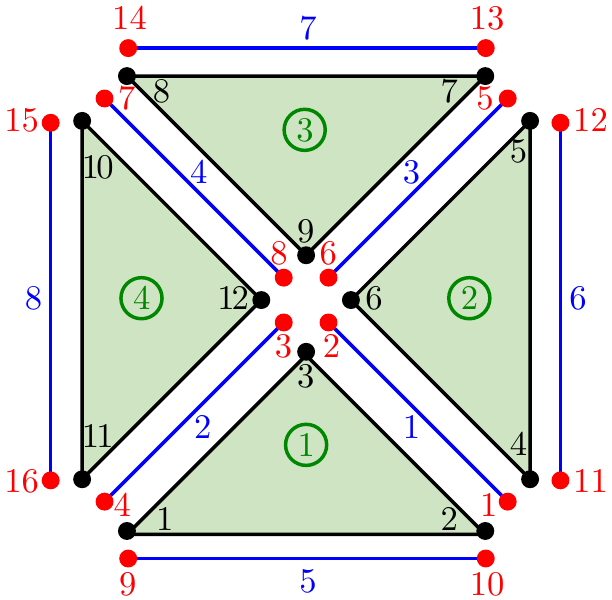}
	\caption{Mesh of $\Omega = [0,1]^2$ with four triangular elements with the global numbering of the mesh elements (in green), the element nodes (in black), the mesh faces (in blue) and the face nodes (in red).}
	\label{fig:mesh4elemHDG}
\end{figure}

To illustrate the \texttt{mesh} data structure, a coarse mesh of the domain $\Omega = [0,1]^2$, with four triangular elements is considered. Figure~\ref{fig:mesh4elemHDG} represents the mesh with the global  numbering of the mesh elements, the element nodes, the mesh faces and the face nodes.

\begin{figure}[!tb]
	\centering
	\includegraphics[width=0.35\textwidth]{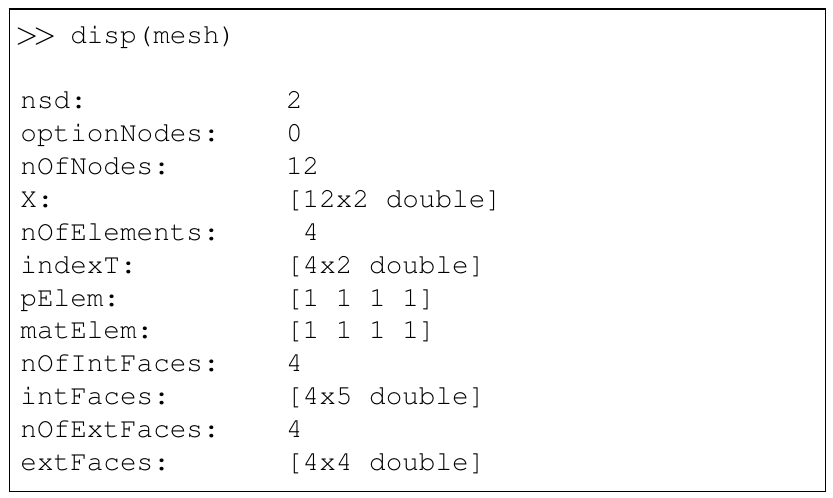}
	\caption{Overview of the data contained in the \texttt{mesh} data structure for the example of figure~\ref{fig:mesh4elemHDG}.}
	\label{fig:mesh4elemHDGA}
\end{figure}

\begin{figure}[!tb]
	\centering
	\includegraphics[width=0.35\textwidth]{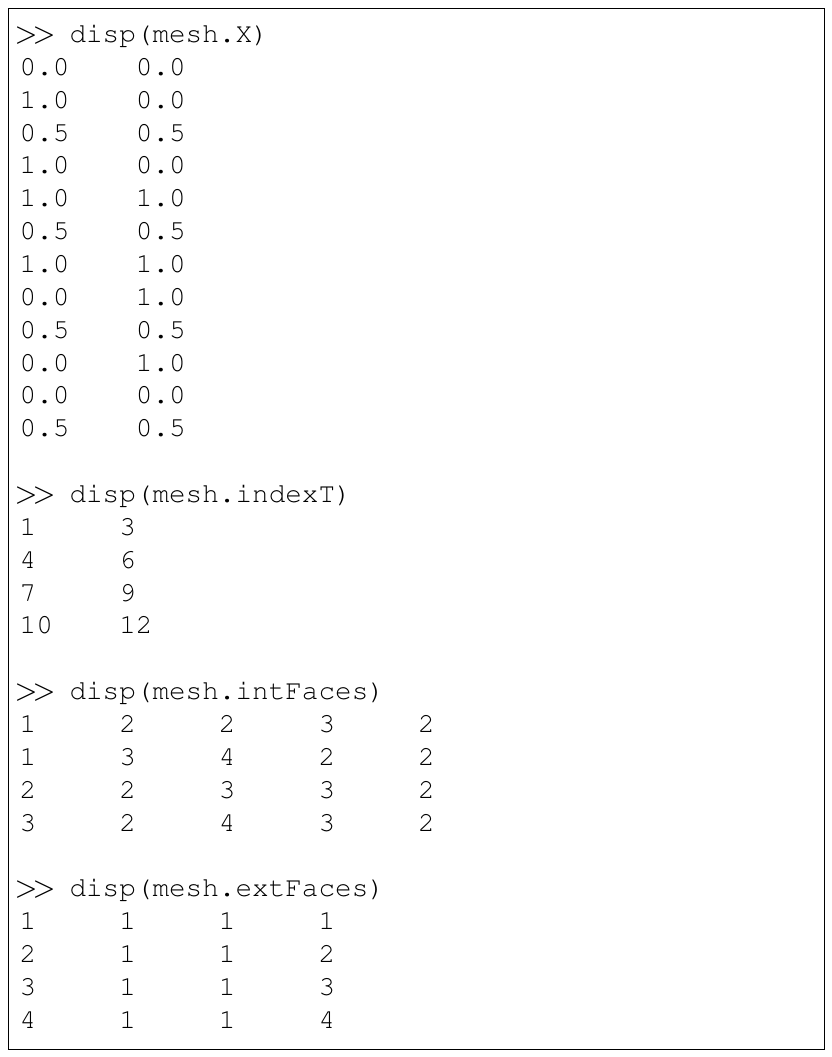}
	\caption{Detail of the fields \texttt{X}, \texttt{indexT}, \texttt{intFaces} and \texttt{extFaces}, corresponding to the data structure \texttt{mesh} of figure~\ref{fig:mesh4elemHDGA}.}
	\label{fig:mesh4elemHDGB}
\end{figure}

An overview of the data contained in the \texttt{mesh} data structure for the example of figure~\ref{fig:mesh4elemHDG} is shown in figure~\ref{fig:mesh4elemHDGA}, with the details of some of the arrays depicted in figure~\ref{fig:mesh4elemHDGB}.

\begin{figure}[!tb]
	\centering
	\includegraphics[width=0.45\textwidth]{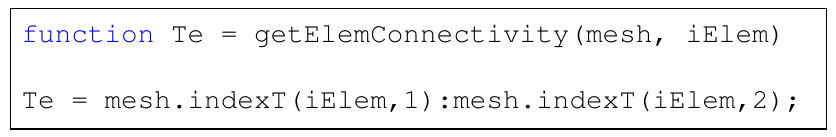}
	\caption{Function to retrieve the connectivity of an element.}
	\label{fig:getElemConnectivity}
\end{figure}

It is worth noting that the connectivity of an element is obtained by using the function shown in figure~\ref{fig:getElemConnectivity}.
This means that the mesh nodes in the array \texttt{X} are duplicated. This functionality is meant to help the handling of meshes with a non-uniform degree of approximation and to provide a seamless way to partition the mesh in case future users would like to parallelise the code. If desired, a different array can be introduced for the connectivity information and the user only needs to redefine the function \texttt{getElemConnectivity}.

\subsection{Reference element} \label{sc:refElem}

As usual in an isoparametric finite element context, the information related to the approximation and numerical integration in an element is stored by means of a reference element, with local coordinates, $\bXi = (\xi_1,\ldots,\xi_{\nsd})$. To easily handle meshes with a non-uniform degree of approximation, the variable \texttt{refElem} is considered an array of dimension $1 \times \texttt{p}_\texttt{max}$, where $\texttt{p}_\texttt{max}$ is the maximum degree of approximation used in all the elements. For each component of the \texttt{refElem} array, the following information is stored:

\begin{itemize}
	\item \texttt{nsd}: Number of spatial dimensions.
	\item \texttt{optionNodes}: Type of high-order nodal distribution, being an equally-spaced (0) or a Fekete (1) nodal set.
	\item \texttt{p}: Degree of approximation.
	\item \texttt{nOfVertices}: Number of element vertices.
	\item \texttt{nOfNodes}: Number of element nodes.	
	\item \texttt{coordinates}: Array of dimension $\texttt{nOfNodes} \times \texttt{nsd}$ containing the local nodal coordinates.	
	\item \texttt{nOfFaces}: Number of element faces.
	\item \texttt{face}: Array of dimension $1 \times \texttt{nOfFaces}$. Each position is a structure containing the following information about an element face:
	\begin{itemize}
		\item \texttt{nodes}: Array containing the element local number of the nodes in the current face.
		\item \texttt{nodesPerm}: Array containing $\texttt{nsd} + 1$ permutations of the face nodes in the field \texttt{nodes}, using the element local number. Each row provides the permutation as required by the field \texttt{intFaces} in the \texttt{mesh} data structure.
		\item \texttt{nodesPermHDG}: Array containing $\texttt{nsd} + 1$ permutations of the face nodes in the field \texttt{nodes}, using the face local number. Each row provides the permutation as required by the field \texttt{intFaces} in the \texttt{mesh} data structure.
	\end{itemize}
	\item \texttt{nOfGauss}: Number of integration points.	
	\item \texttt{gaussWeights}: Array of dimension \texttt{nOfGauss} $\times$ 1, containing the integration weights.	
	\item \texttt{shapeFunctions}: Array of dimension $\texttt{nOfGauss} \times \texttt{nOfNodes} \times (\texttt{nsd}+1)$. When the third index is equal to 1, the array contains the value of the shape functions, for all the nodes at all integration points. When the third index is equal to $r>1$, the array contains the value of the derivative of the shape functions in the $\xi_k$ direction, for all the nodes at all integration points.
	\item \texttt{shapeFunctionsNodesPPp1}: Array of dimension $\texttt{nOfNodes}^\star$ $\times$ $\texttt{nOfNodes}$, where $\texttt{nOfNodes}^\star$ denotes the number of nodes of an element with degree of approximation $\texttt{p} {+} 1$. It contains the value of the shape functions of the current element at the nodes of an element with one extra degree of approximation. This array is only used to perform the HDG postprocess described in sections~\ref{sc:PoissonPostprocess} and~\ref{sc:StokesPostprocess}.
	\item \texttt{shapeFunctionsGaussPPp1}: \ Array of dimension $\texttt{nOfGauss}^\star \times \texttt{nOfNodes}$, where $\texttt{nOfGauss}^\star$ denotes the number of integration points of an element with degree of approximation $\texttt{p} {+} 1$. It contains the value of the shape functions of the current element and its derivatives at the integration points of an element with one extra degree of approximation. This array is only used to perform the HDG postprocess described in sections~\ref{sc:PoissonPostprocess}  and~\ref{sc:StokesPostprocess}.
\end{itemize}

\begin{figure}[!tb]
	\centering
	\includegraphics[width=0.2\textwidth]{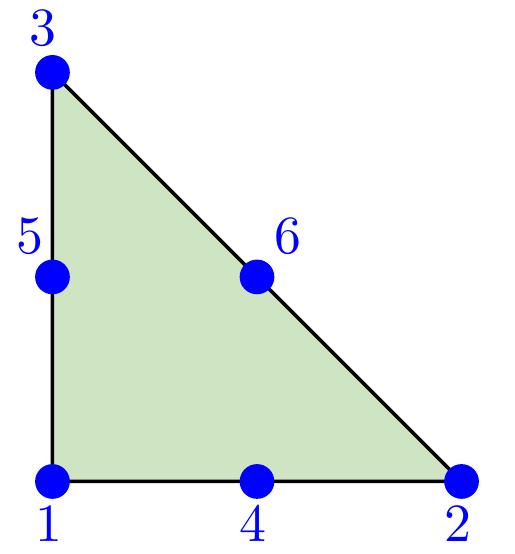}
	\caption{Reference triangular element for $\deg=2$.}
	\label{fig:refElemTriP2}
\end{figure}

\begin{figure}[!tb]
	\centering
	\includegraphics[width=0.45\textwidth]{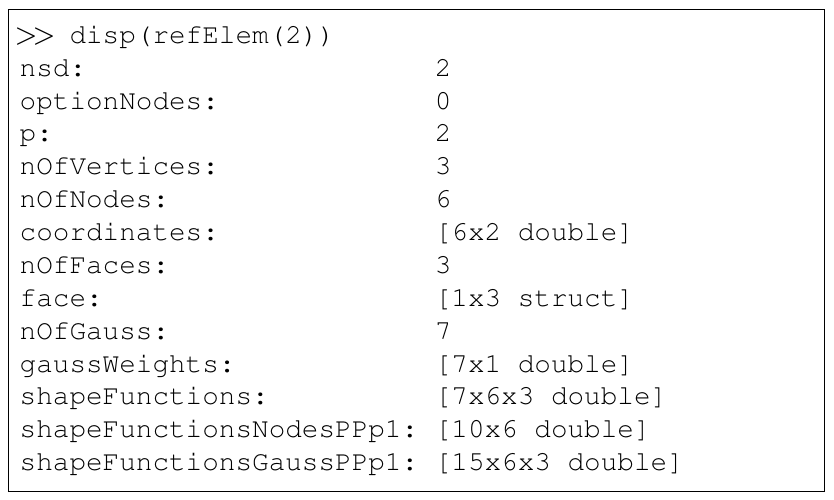}
	\caption{Overview of the data contained in the \texttt{refElem} data structure for the reference triangular quadratic element of figure~\ref{fig:refElemTriP2}.}
	\label{fig:refElemTriP2A}
\end{figure}

To illustrate the \texttt{refElem} data structure, figure~\ref{fig:refElemTriP2} depicts the reference quadratic triangular element.
An overview of the data contained in the \texttt{refElem} data structure for the example of figure~\ref{fig:refElemTriP2} is shown in figure~\ref{fig:refElemTriP2A}.

\begin{figure}[!tb]
	\centering
	\includegraphics[width=0.2\textwidth]{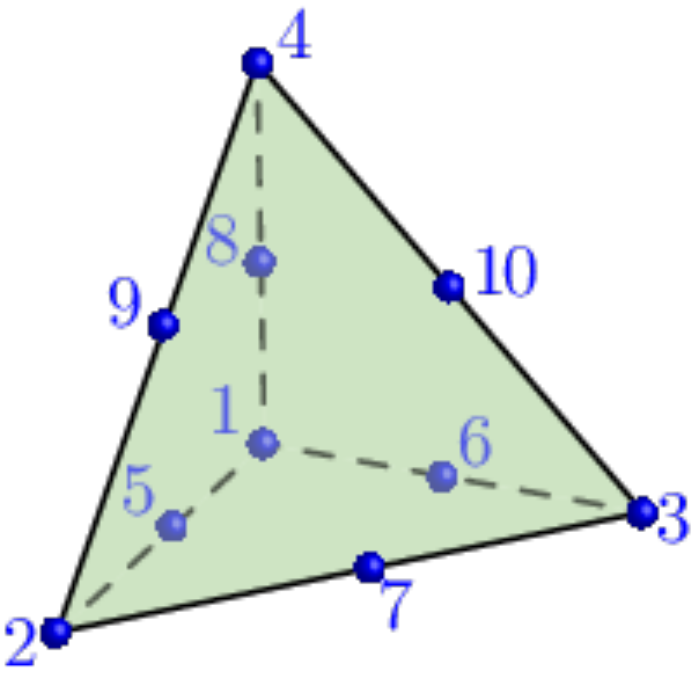}
	\caption{Reference tetrahedral element for $\deg=2$.}
	\label{fig:refElemTetP2}
\end{figure}

\begin{figure}[!tb]
	\centering
	\includegraphics[width=0.45\textwidth]{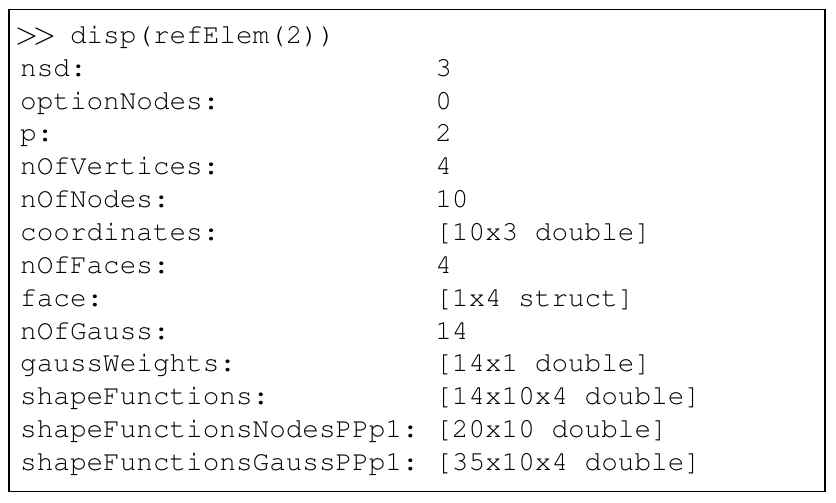}
	\caption{Overview of the data contained in the \texttt{refElem} data structure for the reference tetrahedral quadratic element of figure~\ref{fig:refElemTetP2}.}
	\label{fig:refElemTetP2A}
\end{figure}

Similarly, figure~\ref{fig:refElemTetP2} shows the reference quadratic tetrahedral element and figure~\ref{fig:refElemTetP2A} depicts and overview of the data contained in the \texttt{refElem} data structure.

It is worth noting that the code provided utilises nodal basis functions for the polynomial approximation. However, it is straightforward for future users to change the basis used for the approximation by simply changing the reference element information. With minimum effort it is also possible to incorporate other element types.

\subsection{Reference face} \label{sc:refFace}

\begin{figure}[!tb]
	\centering
	\includegraphics[width=0.45\textwidth]{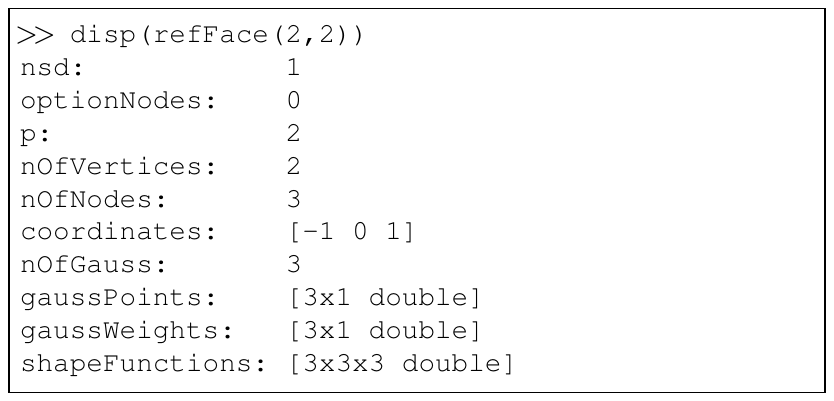}
	\caption{Overview of the data contained in the \texttt{refFace} data structure for the reference face of a quadratic element in two dimensions.}
	\label{fig:refFace2D_Data}
\end{figure}

The information related to the approximation and numerical integration on a face is stored by means of a reference face, with local coordinates, $\bEta = (\eta_1,\ldots,\eta_{\nsd-1})$. To easily handle meshes with a non-uniform degree of approximation, the variable \texttt{refFace} is considered an array of dimension $\deg_\texttt{max} \times \deg_\texttt{max}$, where $\deg_\texttt{max}$ is the maximum degree of approximation used in all the elements. For each diagonal component of the \texttt{refFace} array, the information stored is a subset of the information stored in the \texttt{refElem} data structure. As an example, figure~\ref{fig:refFace2D_Data} offers an overview of the data contained in the diagonal term of \texttt{refFace} corresponding to a quadratic face of a triangular element.

The upper triangular portion of the \texttt{refFace} data structure contains the information associated with the field \texttt{shapeFunctions}. This is only required when a mesh with a non-uniform degree of approximation is employed. The position $(r,s)$ of the array \texttt{refFace} contains the shape functions of degree $r$ evaluated at the integration points of a face with degree of approximation $s$. This is required when computing the integrals in an interior face where the degree of approximation of the elements sharing this face is different. It is worth noting that only the entries in the upper triangle of the array \texttt{refFace} contain relevant information because the degree of approximation used for the hybrid variable in a given face is defined as the maximum between the degree of approximation of the two elements sharing the face.

\section{Preprocess} \label{sc:preprocess}

This section describes the preprocessing stage of the HDG solver. The implementation can be found in the function \texttt{hdgPreprocess}, which produces as an output an updated version of the data structures \texttt{mesh} and \texttt{hdg}. 

\begin{figure}[!tb]
	\centering
	\includegraphics[width=0.45\textwidth]{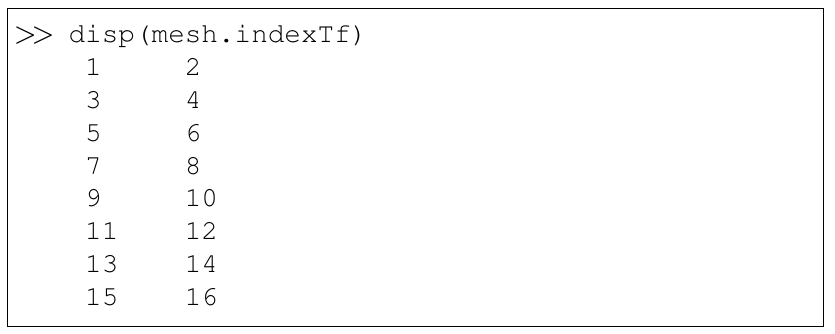}
	\caption{Data contained in \texttt{mesh.indexTf} data structure for the solution of the Poisson equation in the mesh of figure~\ref{fig:mesh4elemHDG}.}
	\label{fig:indexTf_DataPoisson}
\end{figure}

\begin{figure}[!tb]
	\centering
	\includegraphics[width=0.45\textwidth]{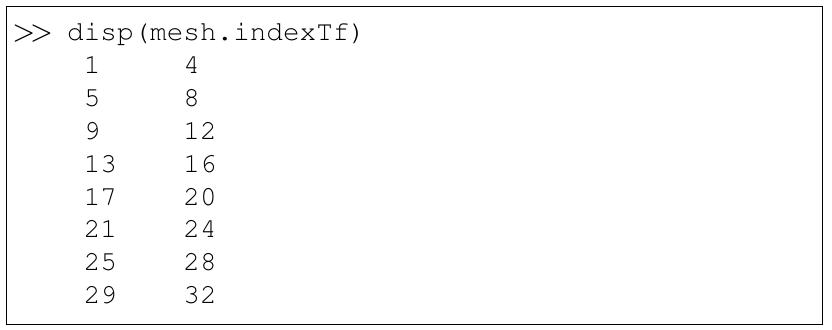}
	\caption{Data contained in \texttt{mesh.indexTf} data structure for the solution of the Stokes equation in the mesh of figure~\ref{fig:mesh4elemHDG}.}
	\label{fig:indexTf_DataStokes}
\end{figure}

The data structure \texttt{mesh} is taken as an input, containing the fields described in section~\ref{sc:mesh}, and a new field, called \texttt{indexTf} is added. This field contains an array of dimension  $\texttt{nOfFaces} \times 2$ featuring the connectivity indices of the mesh skeleton, where $\texttt{nOfFaces}$ is the total number of mesh faces (i.e. interior faces plus exterior faces). The first column contains the first degree of freedom of a face and the second column contains the last degree of freedom of the face. Figure~\ref{fig:indexTf_DataPoisson} shows the data contained in \texttt{indexTf}, after the preprocessing is performed, for the mesh of figure~\ref{fig:mesh4elemHDG} and for a Poisson problem (i.e. scalar unknown).
The same information for a Stokes problem is shown in figure~\ref{fig:indexTf_DataStokes}.

For the Poisson problem, each node has one degree of freedom associated as the hybrid variable contains an approximation of the trace of the solution, which is a scalar quantity. In contrast, for the Stokes problem the global vector of unknowns contains an approximation of the trace of the velocity and an approximation of the mean pressure. Therefore, in two dimensions each face contains $2(\deg+1)$ degrees of freedom for the velocity and one extra degree of freedom per element is introduced for the mean pressure. 

The \texttt{hdg} data structure is also an input of the function \texttt{hdgPreprocess}, containing two user defined parameters, namely:
\begin{itemize}
	\item \texttt{tau}: Value of the constant stabilisation parameter.
	\item \texttt{problem}: String containing the name of the problem to be solved. The code provided contains two model problems, namely \texttt{\textquotesingle Poisson\textquotesingle} and \texttt{\textquotesingle Stokes\textquotesingle}.
\end{itemize}
The structure is updated in the preprocess stage with the following information:
\begin{itemize}
	\item \texttt{faceInfo}: Structure of dimension $1 \times \texttt{nOfElements}$. For each element of the array, the following information provides a link between the element and face information of the mesh:
	\begin{itemize}
		\item \texttt{local2global}: Array of dimension $1 \times \texttt{nOfElementFaces}$, containing the global face number of the faces of the current element.
		\item \texttt{localNumFlux}: Array of dimension $1 \times \texttt{nOfElementFaces}$, containing a flag for the numerical flux function associated with the faces of the current element. For an interior face, a value of zero is set. For a boundary face, the number corresponds to the boundary condition to be imposed and it is linked to the third column of the array \texttt{extFaces} of the \texttt{mesh} data structure described in section~\ref{sc:mesh}.
		\item \texttt{permutations}: Array of dimension $1 \times \texttt{nOfElementFaces}$, containing a flag for the permutation required to ensure that the ordering of the nodes in the global face matches the ordering of the face nodes in the current element.
		\item \texttt{pHat}: Array of dimension $1 \times \texttt{nOfElementFaces}$, containing the degree of approximation used for the hybrid variable in the corresponding faces of the current element.
	\end{itemize} 	
	\item \texttt{nDOFglobal}: Number of global degrees of freedom.
	\item \texttt{vDOFtoSolve}: Number of global degrees of freedom associated with nodes not on a Dirichlet boundary.
\end{itemize} 

In the case of the Stokes equations, three additional fields are introduced in the \texttt{hdg} data structure during the preprocess routine:
\begin{itemize} 	
	\item \texttt{pureDirichlet}: Boolean variable identifying whether the problem has purely Dirichlet boundary conditions.
	\item \texttt{columnsGlobalSystem}: Number of columns in the global system, corresponding to the number of unknowns given by the hybrid velocity and the mean pressure.
	\item \texttt{rowsGlobalSystem}: Number of rows in the global system, including the constraint for the uniqueness of pressure. Note that the value of this variable will differ from \texttt{columnsGlobalSystem} only in the case of purely Dirichlet boundary value problems.
\end{itemize} 

\begin{figure}[!tb]
	\centering
	\includegraphics[width=0.45\textwidth]{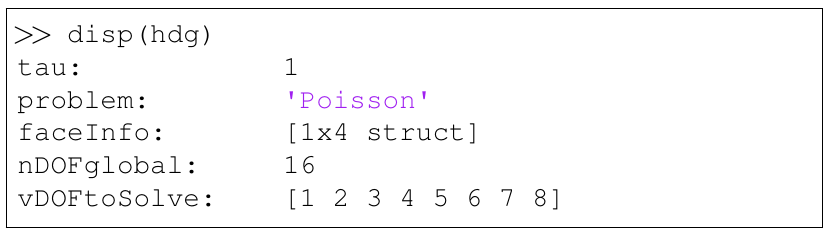}
	\caption{Data contained in the \texttt{hdg} data structure for the mesh of figure~\ref{fig:mesh4elemHDG}.}
	\label{fig:hdg_Data}
\end{figure}

\begin{figure}[!tb]
	\centering
	\includegraphics[width=0.45\textwidth]{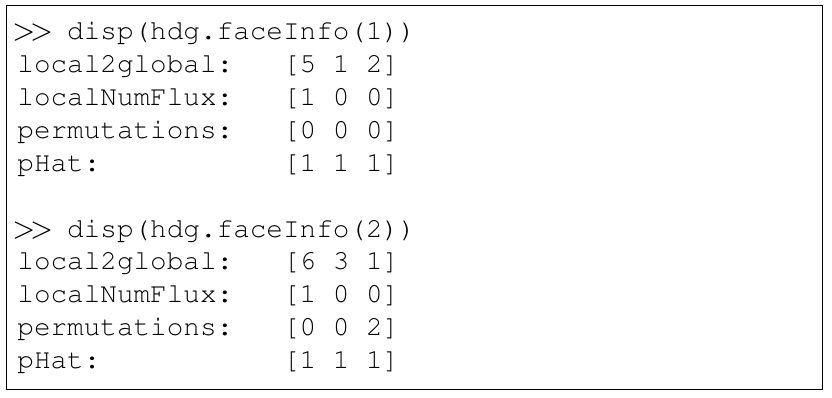}
	\caption{Data contained in the \texttt{hdg.faceInfo} for the first two elements of the mesh of figure~\ref{fig:mesh4elemHDG}.}
	\label{fig:faceInfo_Data}
\end{figure}

The data contained in the \texttt{hdg} data structure, after the preprocess stage, is shown in figure~\ref{fig:hdg_Data} for the Poisson problem on the mesh of figure~\ref{fig:mesh4elemHDG}. 
The data contained in the field \texttt{faceInfo} for the first two elements is also depicted in figure~\ref{fig:faceInfo_Data}.

Two additional simple data structures are defined at the preprocess stage, namely \texttt{problemParams} and \texttt{ctt}. The structure \texttt{problemParams} contains problem-specific information. The current implementation uses this data structure to carry the following information:
\begin{itemize}
	\item \texttt{nOfMat}: Number of materials in the domain. 
	\item \texttt{charLength}: A characteristic length, used to define the value of the stabilisation parameter of the HDG formulation. 
	\item \texttt{example}: An integer that points to the number of a user-defined example. 	
\end{itemize}
In addition, \texttt{problemParams} stores the information on the material parameters. For the Poisson problem:
\begin{itemize}
	\item \texttt{conductivity}: Array of dimension $1 \times \texttt{nOfMat}$ that contains the conductivity of each material present in the domain.
\end{itemize}
For the Stokes problem:
\begin{itemize}
	\item \texttt{viscosity}: A scalar value representing the viscosity coefficient of the fluid.
	\item \texttt{alphaSlip}: A scalar value describing the penetration coefficient for the slip boundary condition.
	\item \texttt{betaSlip}: A scalar value describing the friction coefficient for the slip boundary condition.
\end{itemize}

Finally, the structure \texttt{ctt} contains the following information:
\begin{itemize}
	\item \texttt{iBC\_Interior}: A flag for the numerical flux function to be used on an interior face.
	\item \texttt{iBC\_Dirichlet}: A flag for the numerical flux function to be used on an exterior face where a Dirichlet boundary condition is imposed. 
	\item \texttt{iBC\_Neumann}: A flag for the numerical flux function to be used on an exterior face where a Neumann boundary condition is imposed. 
	\item \texttt{iBC\_Slip}: A flag for the numerical flux function to be used on an exterior face where a slip boundary condition is imposed (only supported for the Stokes problem). 
	\item \texttt{nOfComponents}: Number of components of the primal variable. 
\end{itemize}
The flags used to distinguish the type of face and numerical flux to be considered can be specified by the user and they are linked to the third column of the array \texttt{extFaces} of the \texttt{mesh} data structure described in section~\ref{sc:mesh}.

\section{The \texttt{HDGlab} Poisson solver} \label{sc:solver}

\begin{figure*}[!tb]
	\centering
	\includegraphics[width=0.9\textwidth]{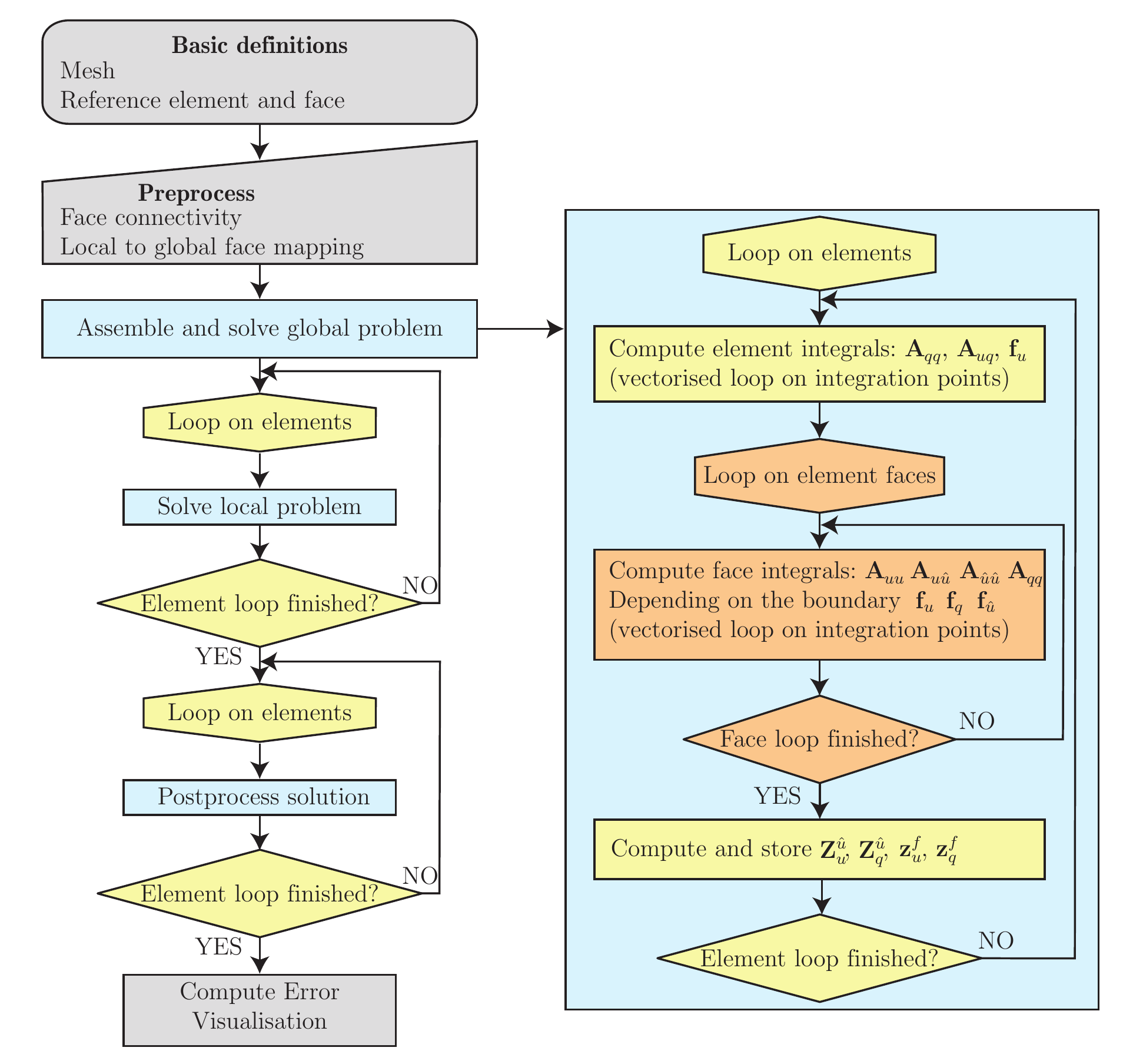}
	\caption{Code workflow diagram.}
	\label{fig:flowChart}
\end{figure*}

A code workflow diagram of the \texttt{HDGlab} Poisson code is shown in figure~\ref{fig:flowChart}.
This section focuses on the core part of the code that involves the assembly and solution of the global system of equations, the element-by-element solution of the local problems and the local postprocess to obtain a superconvergent solution.

\subsection{Global problem} \label{sc:global}

\begin{figure*}[!tb]
	\centering
	\includegraphics[width=0.9\textwidth]{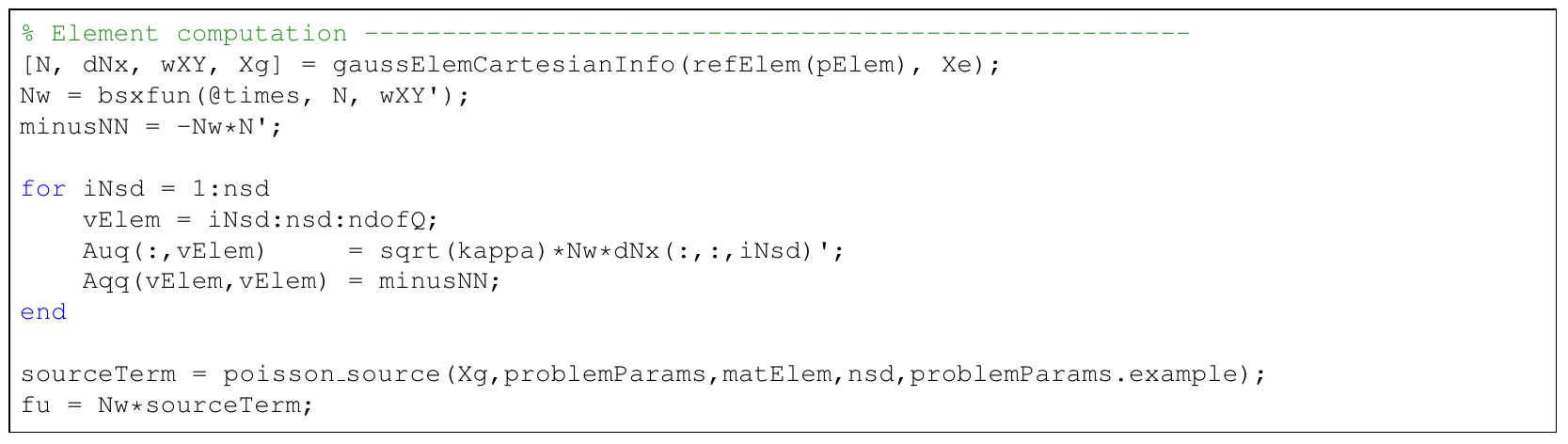}
	\caption{Extract of the function \texttt{hdg\_Poisson\_ElementalMatrices} that computes the element integrals of the HDG formulation for the Poisson problem.}
	\label{fig:elementComputation}
\end{figure*}

The HDG global system of equations is assembled and solved in the \texttt{hdg\_Poisson\_GlobalSystem} function. For each element, \texttt{hdg\_Poisson\_ElementalMatrices} contains two  parts corresponding to the computation of the element integrals and the face integrals respectively. An extract of this function, showing the computation of the elemental matrices $\mat{A}_{qq}$ and $\mat{A}_{uq}$ and the elemental vector $\vect{f}_u$, is displayed in figure~\ref{fig:elementComputation}.
It is worth noting that the loop on integration points is vectorised by using the binary singleton expansion function \texttt{bsxfun}.

In a similar fashion, the second part of the function \texttt{hdg\_Poisson\_ElementalMatrices} performs a loop on the faces of the current element and computes the face integrals that lead to the matrices $\mat{A}_{uu}$, $\mat{A}_{u\hu}$, $\mat{A}_{q\hu}$ and $\mat{A}_{\hu\hu}$. This computation distinguishes between interior and exterior faces and, for the exterior faces, utilises the flag in \texttt{hdg.faceInfo.localNumFlux} to enforce the correct boundary condition. For a Dirichlet face, the vector $\vect{f}_u$ is updated and the vector $\vect{f}_q$ is computed. For a Neumann face, the vector $\vect{f}_{\hu}$ is computed.

One of the distinctive parts of the HDG formulation is found in the loop on faces, where a vector called \texttt{indexFlip} is used to ensure that the ordering of the degrees of freedom corresponding to the hybrid variable, as seen from the current element, matches the global ordering of the degrees of freedom of the vector of unknowns of the global HDG system.

\begin{figure}[!tb]
	\centering
	\includegraphics[width=0.45\textwidth]{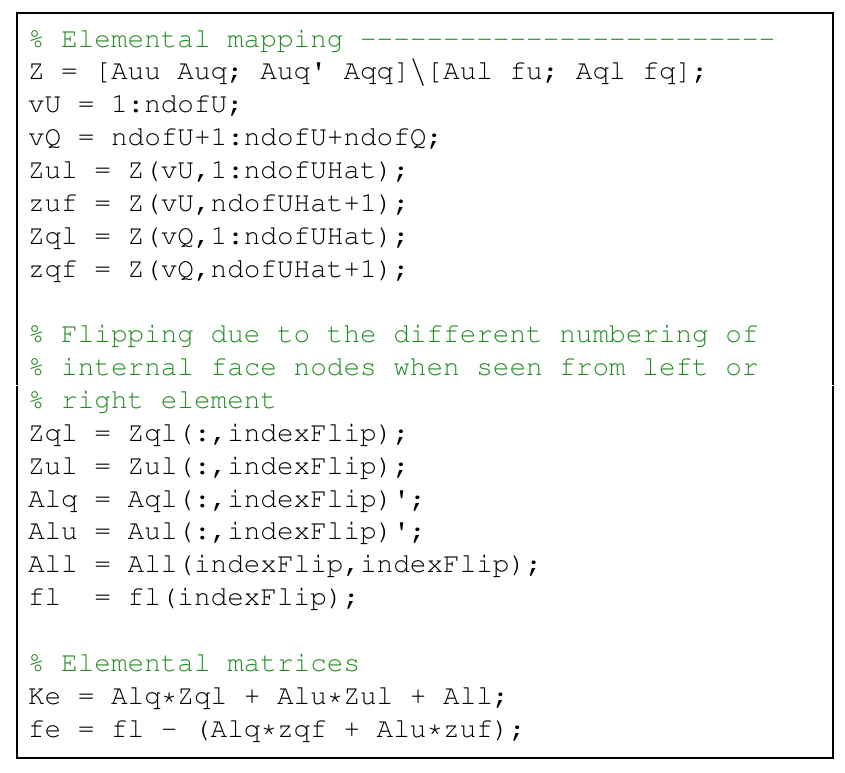}
	\caption{Extract of the \texttt{hdg\_Poisson\_ElementalMatrices} function that computes the elemental matrix $\widehat{\mat{K}}^e$ and vector $\vecHF[]^e$.}
	\label{fig:elemContributions}
\end{figure}

After all the elemental matrices and vectors are computed, the elemental contributions to the global system are prepared to be assembled, namely $\widehat{\mat{K}}^e$ and $\vecHF[]^e$, as shown in the extract of figure~\ref{fig:elemContributions}.
It is worth noting that at this stage the matrices $\mat{Z}_{u\hu}$ and $\mat{Z}_{q\hu}$ and the vectors $\vect{z}_u^f$ and $\vect{z}_q^f$, defined in equation~\eqref{eq:PoissonHDGdiscreteLocalSolution}, are stored in the data structure \texttt{local} in order to be used during the solution of the element-by-element local problems. For large problems, the user might choose to save these matrices to disk before solving the global system of equations.

\subsection{Local problem} \label{sc:local}

\begin{figure}[!tb]
	\centering
	\includegraphics[width=0.45\textwidth]{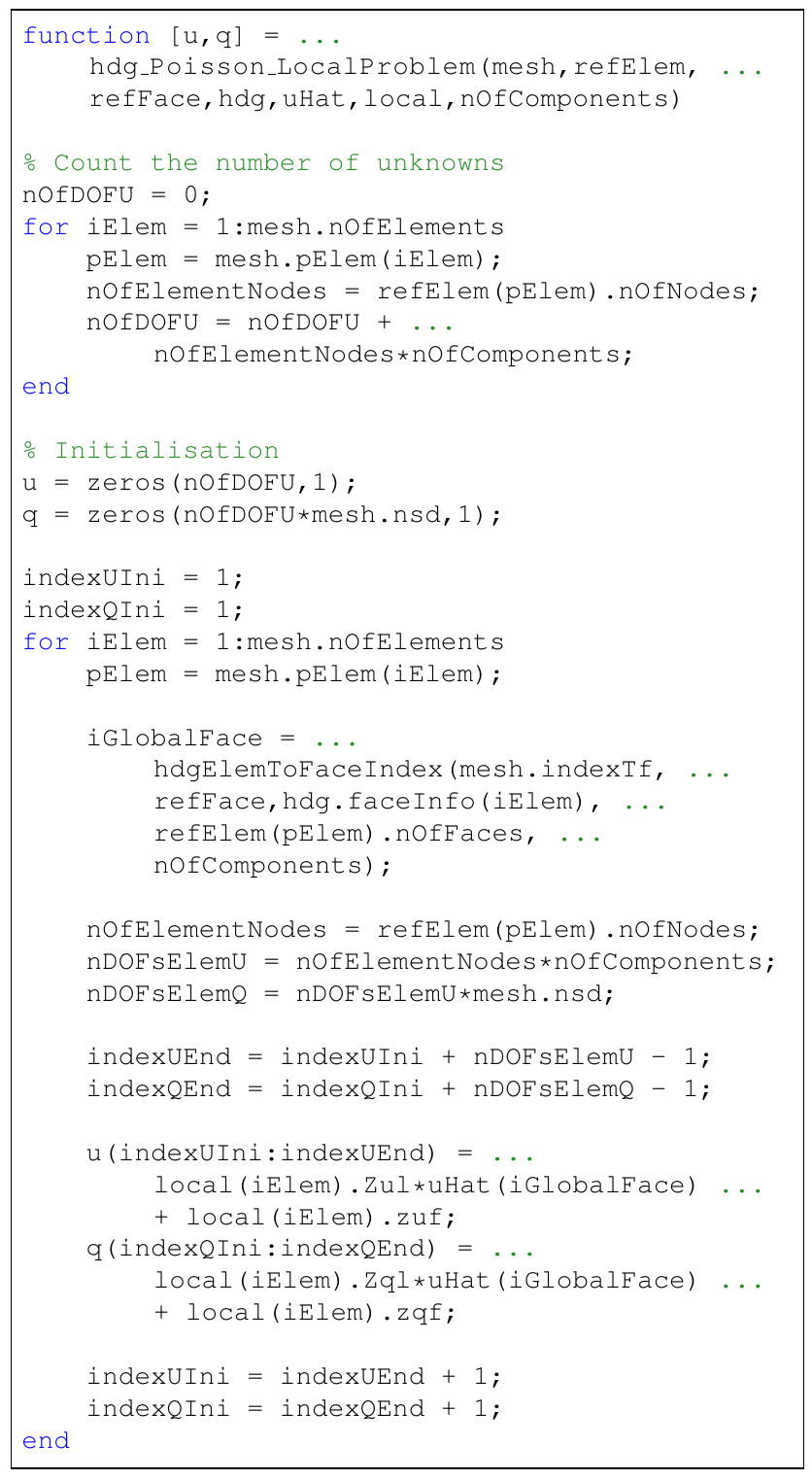}
	\caption{Function \texttt{hdg\_Poisson\_LocalProblem} to solve the element-by-element local problems.}
	\label{fig:localProblem}
\end{figure}

After the global system of equations is solved, the function \texttt{hdg\_Poisson\_LocalProblem}, shown in figure~\ref{fig:localProblem}, is called to solve the local problems.
This function is just the implementation of equation~\eqref{eq:PoissonHDGdiscreteLocalPb}. The only aspect that requires attention is the indexing of the global vectors for the primal and mixed variables. This is managed by the function \texttt{hdgElemToFaceIndex}, which is designed to work for variables with any number of components. It is also worth noting that this function accounts for the possibility to have a non-uniform degree of approximation.

\subsection{Local postprocess} \label{sc:postprocess}

\begin{figure*}[!tb]
	\centering
	\includegraphics[width=0.9\textwidth]{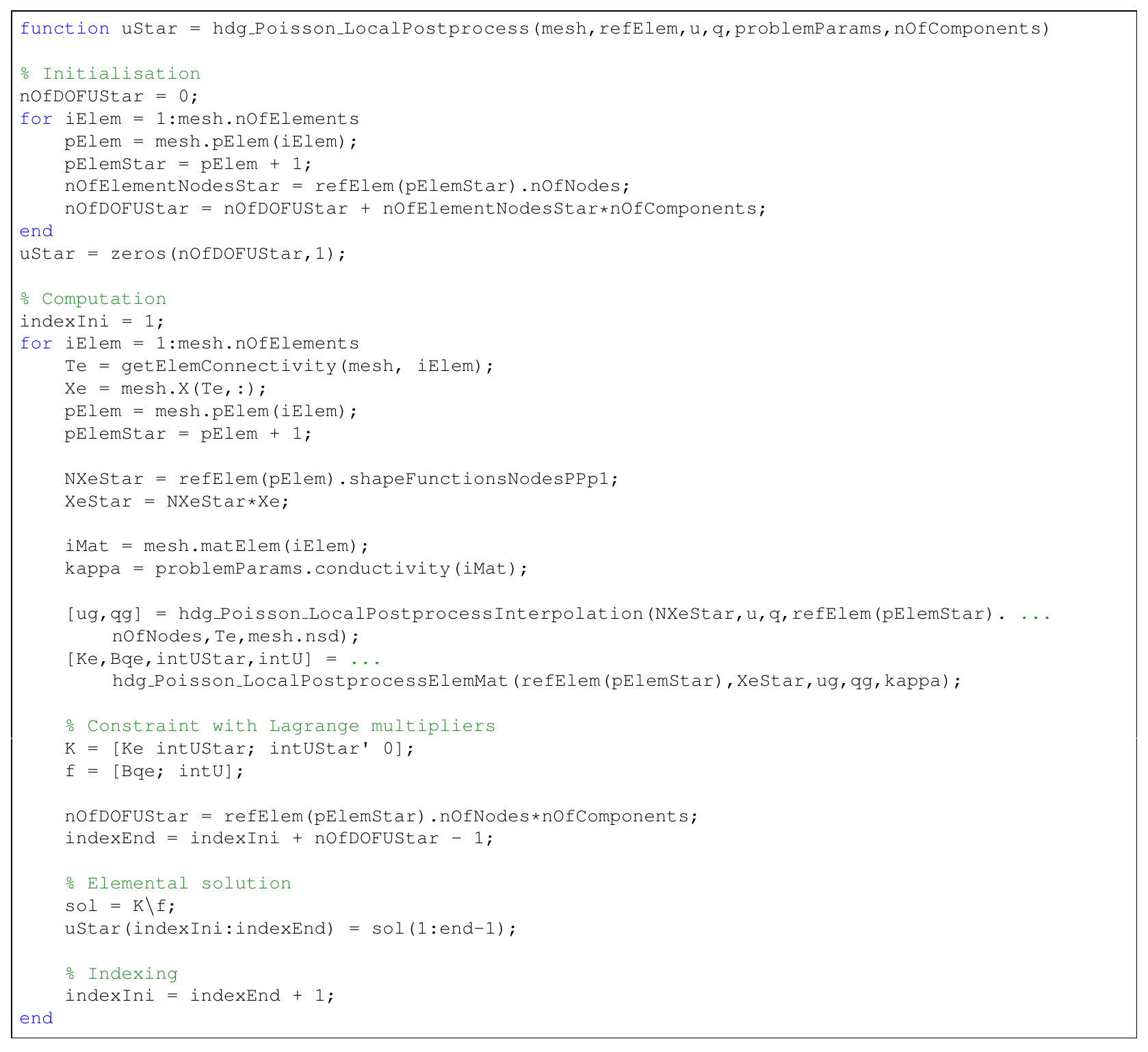}
	\caption{Function \texttt{hdg\_Poisson\_LocalPostprocess} to perform a local, element-by-element, postprocess of the solution.}
	\label{fig:localPostprocess}
\end{figure*}

As discussed in section~\ref{sc:PoissonPostprocess}, once the primal and mixed variables are computed, it is possible to perform a local, element-by-element, postprocess procedure to obtain a more accurate, superconvergent, approximation of the solution. This is implemented in the function \texttt{hdg\_Poisson\_LocalPostprocess}, shown in figure~\ref{fig:localPostprocess}.

A key aspect in the function \texttt{hdg\_Poisson\_LocalPostprocess} is the computation of the elemental  matrices and vectors in equation~\eqref{eq:PoissonHDGdiscretePostProcess}, which is implemented in the subroutine \texttt{hdg\_Poisson\_LocalPostprocessElemMat}.

It is worth emphasising that the implementation assumes that no extra geometric information is known at this stage. Therefore, to compute the high-order nodal distribution of degree $\deg+1$, the nodal distribution in the reference element is mapped to the physical space by using the isoparametric mapping of degree $\deg$. This implies that, for curved elements, a subparametric formulation is considered. This formulation can lead to a suboptimal rate of convergence for the postprocessed solution as demonstrated in~\cite{RS-SH-18,RS-19}, where a NURBS-enhanced implementation was proposed.

\section{The \texttt{HDGlab} Stokes solver} \label{sc:solverStokes}

In this section, the \texttt{HDGlab} solver for the Stokes equations is presented. It is worth noting that the code features the same structure introduced in the previous section for the Poisson case. Hence, only the differences with respect to the Poisson solver will be detailed.

\subsection{A vector-valued problem} \label{sc:stokesVectorial}

The HDG global system of equations is assembled and solved in the \texttt{hdg\_Stokes\_GlobalSystem} function. More precisely, the element and face integrals are computed by the function \texttt{hdg\_Stokes\_ElementalMatrices} for each element.

The first difference with respect to the Poisson code is represented by the vectorial nature of the primal and hybrid variables and by the tensor-valued mixed variable. For the sake of computational efficiency, the representation of the second-order tensor $\vect{L}$ is written as a vector by rows, namely
$$
\vect{L} = \begin{cases}
[\node{L}_{11} \ \node{L}_{12} \ \node{L}_{21} \ \node{L}_{22}]^T , & \text{in 2D} , \\
[\node{L}_{11} \ \node{L}_{12} \ \node{L}_{13} \ \node{L}_{21} \ \node{L}_{22} \ \node{L}_{23} \ \node{L}_{31} \ \node{L}_{32} \ \node{L}_{33}]^T , & \text{in 3D} . 
\end{cases}
$$

\begin{figure}[!tb]
	\centering
	\includegraphics[width=0.45\textwidth]{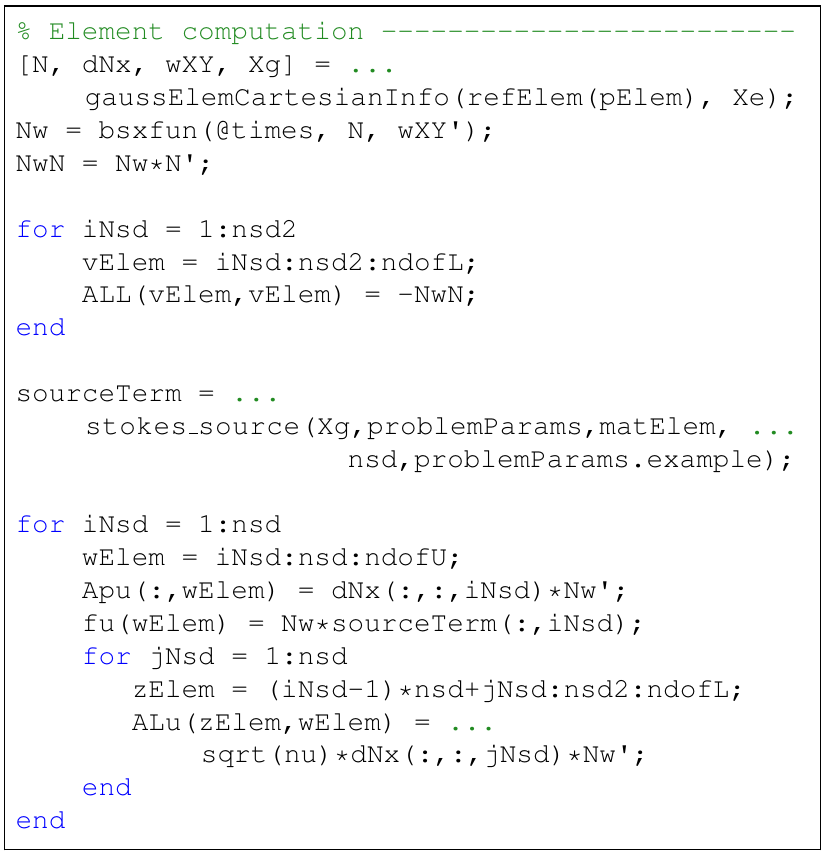}
	\caption{Extract of the \texttt{hdg\_Stokes\_ElementalMatrices} function that computes the element integrals of the HDG formulation for the Stokes problem.}
	\label{fig:elementContribStokes}
\end{figure}

Figure~\ref{fig:elementContribStokes} reports the computation of the elemental matrices $\mat{A}_{LL}$, $\mat{A}_{Lu}$ and $\mat{A}_{pu}$ and the elemental vector $\vecF[u]$. It is worth noting that the assembly of these matrices and this vector account for the appropriate numbering of the vector-valued and tensor-valued unknowns. Similarly to the Poisson case, the loop on integration points is vectorised via the command \texttt{bsxfun}.

\subsection{Slip boundary conditions} \label{sc:stokesSlip}

In the second part of \texttt{hdg\_Stokes\_ElementalMatrices}, the face integrals are computed within a loop on the element faces. More precisely, the matrices $\mat{A}_{L\hu}$, $\mat{A}_{uu}$, $\mat{A}_{u\hu}$ and $\mat{A}_{p\hu}$ are computed for the local problem, whereas the matrix $\mat{A}_{\hu\hu}$ is computed for the global problem. In addition, on the Neumann faces the vector $\vecF[\hu]$ is computed, whereas on the Dirichlet faces the vectors $\vecF[L]$ and $\vecF[p]$ are computed  and the vector $\vecF[u]$ is updated. 
\begin{remark}
In case of Dirichlet and Neumann boundary conditions, the remaining matrices involved in the global problem are such that
\begin{subequations} \label{eq:matGlobSym}
\begin{align}
\mat{A}_{\hu L} &= \mat{A}_{L\hu}^T , \\ 
\mat{A}_{\hu u} &= \mat{A}_{u\hu}^T , \\ 
\mat{A}_{\hu p} &= \mat{A}_{p\hu}^T .
\end{align}
\end{subequations}
\end{remark}

\begin{figure}[!tb]
	\centering
	\includegraphics[width=0.45\textwidth]{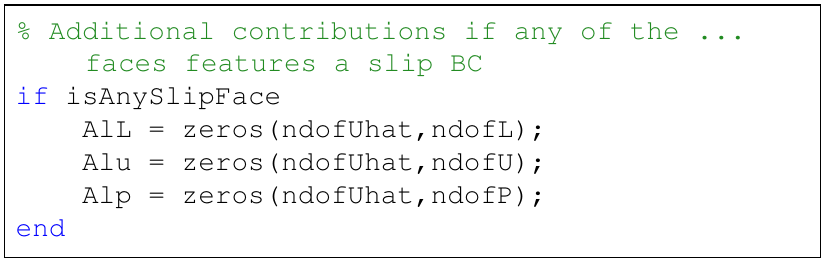}
	\caption{Extract of the \texttt{hdg\_Stokes\_ElementalMatrices} function that initialises the elemental matrices associated with the slip boundaries in the global problem for the Stokes problem.}
	\label{fig:slipBCMatGlob}
\end{figure}

In case slip boundary conditions are also considered, the properties in equation~\eqref{eq:matGlobSym} no longer hold. In this context, the matrices $\mat{A}_{\hu L}$, $\mat{A}_{\hu u}$ and $\mat{A}_{\hu p}$ are computed in the function \texttt{hdg\_Stokes\_ElementalMatrices}. Figure~\ref{fig:slipBCMatGlob} displays the initialisation of the above elemental matrices which are then computed in the loop on faces when the flag in \texttt{hdg.faceInfo.localNumFlux} matches \texttt{ctt.iBC\_Slip}.

\begin{figure}[!tb]
	\centering
	\includegraphics[width=0.45\textwidth]{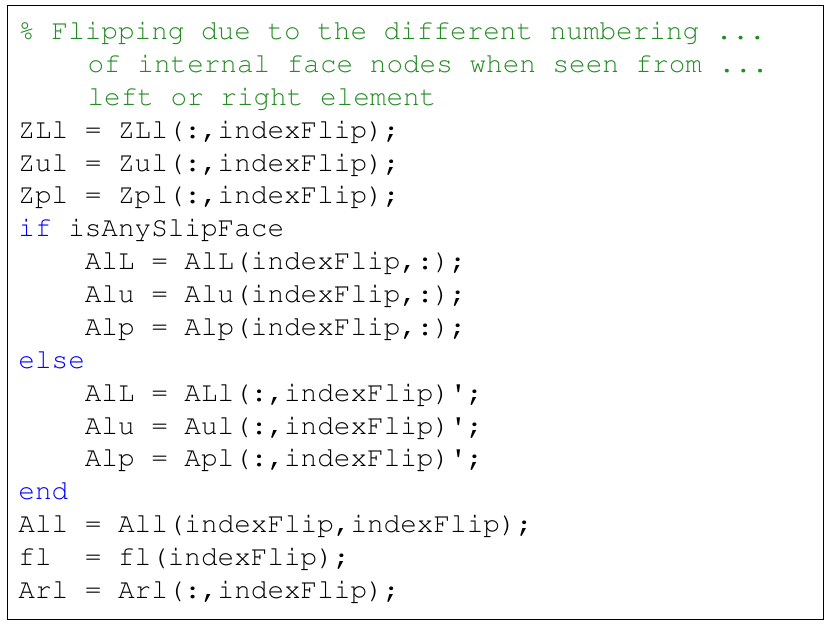}
	\caption{Extract of the \texttt{hdg\_Stokes\_ElementalMatrices} function that defines the elemental matrices of the global problem for the Stokes case, depending on the boundary conditions imposed.}
	\label{fig:slipBC_flip}
\end{figure}

A specific treatment of the case in which slip boundary conditions are considered is also required for the definition of the elemental matrices of the global problem with appropriate ordering of the degrees of freedom of the hybrid variable using the \texttt{indexFlip} vector, see figure~\ref{fig:slipBC_flip}.

\subsection{Additional constraint in the local problem} \label{sc:stokesLocalSaddlePoint}

A major difference between the Poisson and Stokes case is the structure of the local problems~\eqref{eq:PoissonHDGdiscreteLocal} and~\eqref{eq:StokesHDGdiscreteLocal}. Despite both matrices are symmetric, the one arising from the discretisation of the Poisson equation is positive definite, whereas it is indefinite in the Stokes case. More precisely, the matrix in~\eqref{eq:StokesHDGdiscreteLocal} features a saddle-point structure, as classical in the context of finite element approximations of incompressible flow problems~\cite{donea2003finite}. In addition, since the HDG local problem is a purely Dirichlet boundary value problem, the constraint~\eqref{eq:StokesHDGstrongLocalConstraint} is introduced using a Lagrange multiplier $\zeta$.

\begin{figure*}[!tb]
	\centering
	\includegraphics[width=0.9\textwidth]{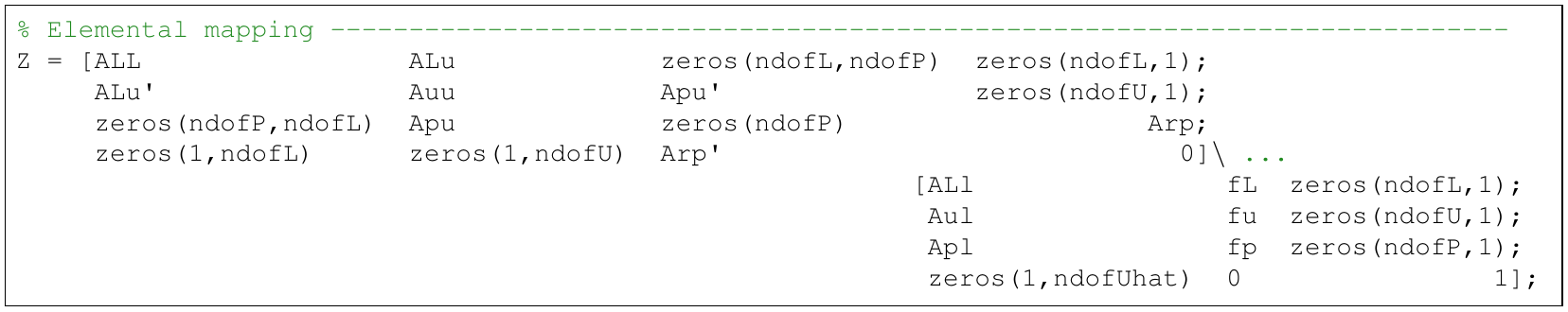}
	\caption{Extract of the \texttt{hdg\_Stokes\_ElementalMatrices} function that computes the matrices and vectors in equation~\eqref{eq:StokesHDGdiscreteLocal}.}
	\label{fig:stokesHybridisation}
\end{figure*}

The structure of the symmetric indefinite matrix involved in the local problem is displayed on the left-hand side of figure~\ref{fig:stokesHybridisation}. On the right-hand side, the blocks of the first and last columns are associated with the contribution of $\vect{\hu}$ and $\bm{\rho}$, respectively, whereas the second column is related to the independent term of the equation. The figure reports the hybridisation stage in which the elemental matrices and vectors defined in~\eqref{eq:StokesHDGdiscreteLocalSolution} are computed for each element. The output of this computation is stored in the data structure \texttt{local} to be successively utilised in the solution of the element-by-element local problems. 

\begin{figure*}[!tb]
	\centering
	\includegraphics[width=0.9\textwidth]{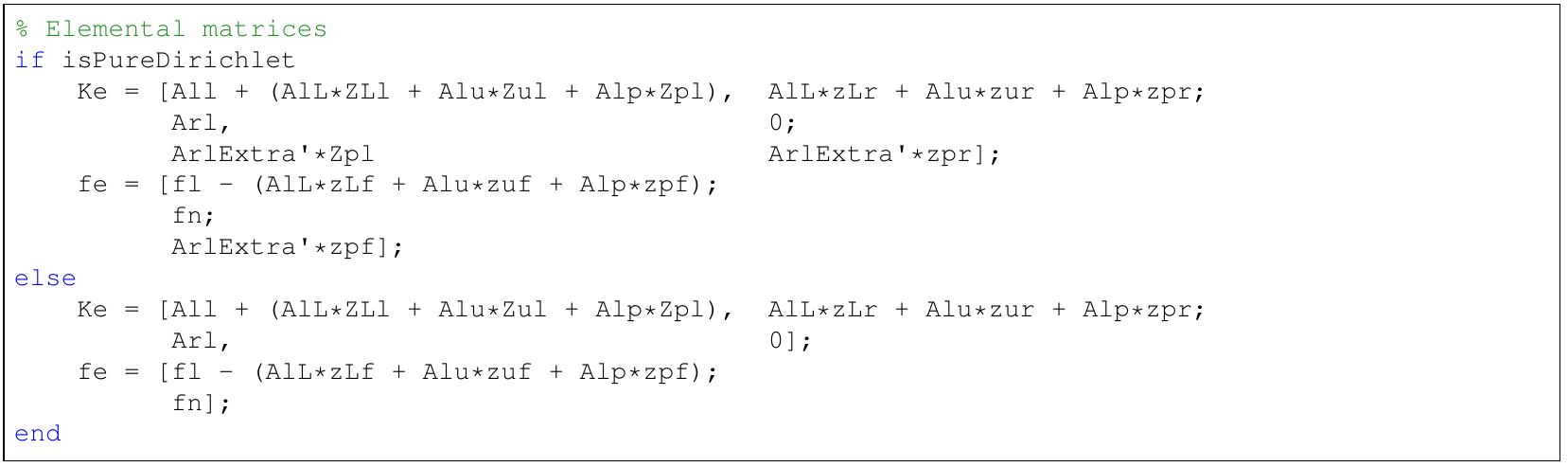}
	\caption{Extract of the \texttt{hdg\_Stokes\_ElementalMatrices} function that computes the block matrix and vector in equation~\eqref{eq:StokesHDGdiscreteGlobalFinal}.}
	\label{fig:stokesElemMatrix}
\end{figure*}

\begin{figure}[!tb]
	\centering
	\includegraphics[width=0.45\textwidth]{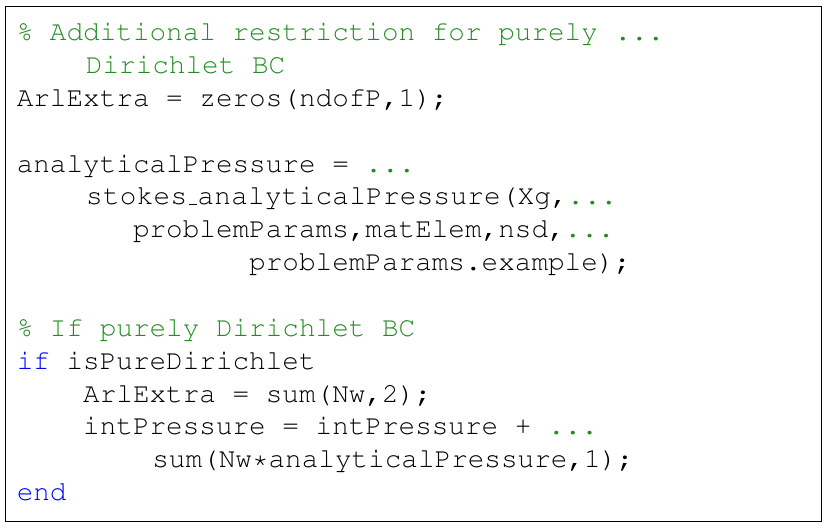}
	\caption{Extract of the \texttt{hdg\_Stokes\_ElementalMatrices} function that computes the vector $\mat{a}_{\overline{\rho} \rho}$ for the pressure constraint.}
	\label{fig:pressureConstraint}
\end{figure}

Finally, \texttt{hdg\_Stokes\_ElementalMatrices} computes the elemental contribution to the matrix in the global problem~\eqref{eq:StokesHDGdiscreteGlobalFinal} as reported in figure~\ref{fig:stokesElemMatrix}. It is worth noting that the variable \texttt{hdg.pureDirichlet} is utilised here to discriminate the construction of the global matrix of a purely Dirichlet boundary value problem. More precisely, besides the blocks $\widehat{\mat{K}}$, $\mat{G}$ and $\mat{G}^T$, also the vector $\mat{a}_{\overline{\rho} \rho}$ (i.e. \texttt{ArlExtra}) arising from the imposition of the global constraint~\eqref{eq:pressureConstraintDomain} is taken into account. An extract of the \texttt{hdg\_Stokes\_ElementalMatrices} function computing such a vector is displayed in figure~\ref{fig:pressureConstraint}.

\begin{remark}
The assembly of the block matrix reported in figure~\ref{fig:stokesElemMatrix} does not exploit the symmetry of the terms in equation~\eqref{eq:StokesHDGdiscreteGlobalFinal} to its full potential.
Indeed, the rationale of \texttt{HDGlab} being educational, the matrix $\mat{G}$ is constructed by inserting the solution~\eqref{eq:StokesHDGdiscreteLocalPb} of the local problem into the global problem~\eqref{eq:StokesHDGdiscreteGlobal}, leading to
\begin{equation} \label{eq:matG}
  \mat{G} 
  := \Assem_{e=1}^{\numel} 
  \bigl[\begin{array}{@{}c@{\,}c@{\,}c@{\,}c@{}} \mat{A}_{\hu L} & \mat{A}_{\hu u} & \mat{A}_{\hu p} & \mat{0}\end{array}\bigr]_{\! e}
  \hspace{-0.5ex}
      \begin{Bmatrix}
        \vect{z}_L^{\rho}  \\
        \vect{z}_u^{\rho}  \\
        \vect{z}_p^{\rho} \\
        \vect{z}_{\zeta}^{\rho}
      \end{Bmatrix}_{\!\! e} .
\end{equation}
The interested reader can thus employ the provided code to numerically verify that the matrix $\mat{G}$ defined in equation~\eqref{eq:matG} is the transpose of the one introduced in~\eqref{eq:StokesHDGdiscreteGlobalFinalMatVec}. The formal proof can be devised following the rationale described in~\cite{MG-GSH-20}.
\end{remark}

\subsection{Assembly of the global system} \label{sc:stokesAssembly}

\begin{figure}[!tb]
	\centering
	\includegraphics[width=0.45\textwidth]{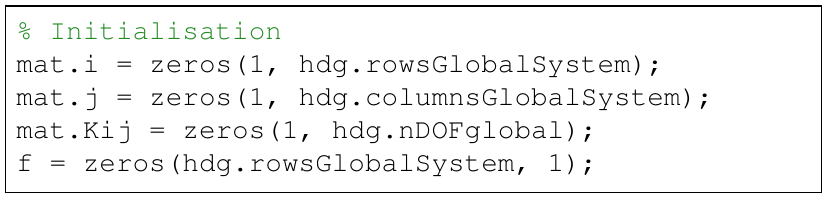}
	\caption{Extract of the \texttt{hdg\_Stokes\_Globalystem} function that initialises the structures required for the assembly of the global system.}
	\label{fig:initialiseStokesGlobal}
\end{figure}

As described in section~\ref{sc:StokesHDG}, the global problem for the Stokes equations features a saddle-point structure. Hence, the assembly of the global system presents major differences with respect to the Poisson case previously discussed. First, figure~\ref{fig:initialiseStokesGlobal} reports the initialisation of the structures required to perform the assembly. It is worth recalling that the value of \texttt{hdg.rowsGlobalSystem} differs from the one of \texttt{hdg.columnsGlobalSystem} only if Dirichlet conditions are imposed on all boundaries. In this case, the additional constraint~\eqref{eq:pressureConstraintDomain} is considered to guarantee the well-posedness of the problem.

\begin{figure}[!tb]
	\centering
	\includegraphics[width=0.45\textwidth]{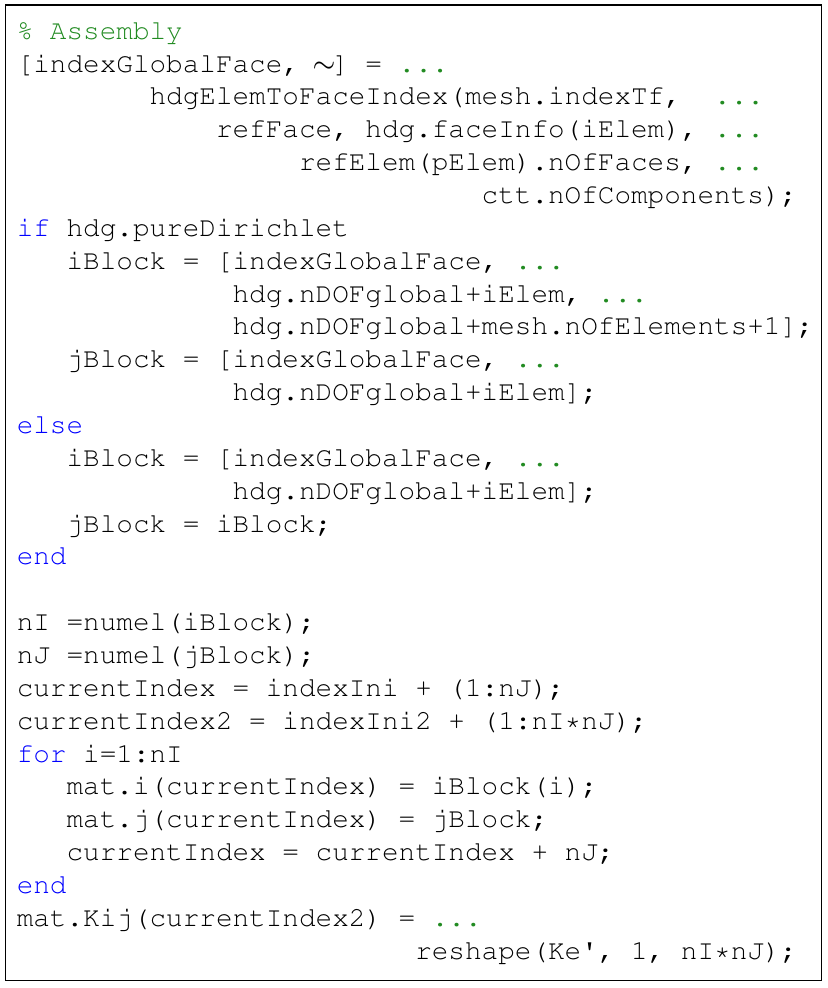}
	\caption{Extract of the \texttt{hdg\_Stokes\_Globalystem} function that constructs the structures required for the assembly of the global system.}
	\label{fig:stokesAssembly}
\end{figure}

The construction of the structures to perform the assembly of the global system is displayed in figure~\ref{fig:stokesAssembly}. According to the variable \texttt{hdg.pureDirichlet}, the above mentioned constraint is introduced as an extra line in the system. 
%
%

\begin{figure}[!tb]
	\centering
	\includegraphics[width=0.45\textwidth]{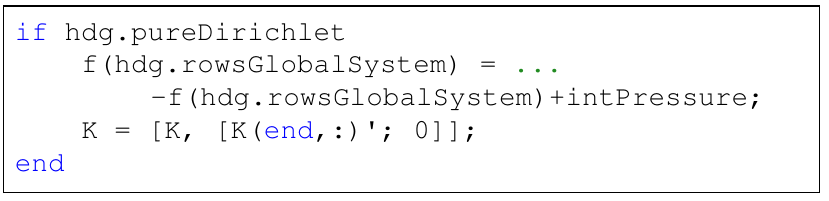}
	\caption{Extract of the \texttt{hdg\_Stokes\_Globalystem} function to impose the constraint in the global system for purely Dirichlet boundary value problems.}
	\label{fig:stokesGlobalConstraint}
\end{figure}

Of course, the matrix arising from the operations displayed in figure~\ref{fig:stokesAssembly} is rectangular. In order to retrieve a square matrix, an extra column is added to the matrix and the constraint is imposed via a Lagrange multiplier (Fig.~\ref{fig:stokesGlobalConstraint}).

\subsection{Three unknowns in the local problem} \label{sc:stokesLocal}

\begin{figure*}[!tb]
\centering
\includegraphics[width=0.9\textwidth]{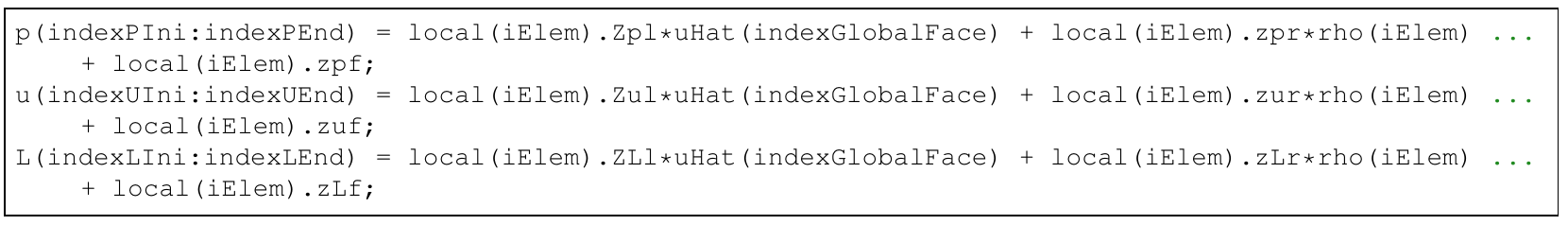}
\caption{Extract of the \texttt{hdg\_Stokes\_LocalProblem} function where the elemental values of the primal and mixed unknowns are computed.}
\label{fig:stokesLocal}
\end{figure*}

The structure of the code for the local problem in the Stokes case replicates the one presented in figure~\ref{fig:localProblem} for the Poisson equation and is thus omitted. The only difference lies in the computation of three variables in each element, namely the pressure $\vect{p}$, the velocity $\vect{u}$ and the gradient of velocity $\vect{L}$. An extract of the function \texttt{hdg\_Stokes\_LocalProblem} is displayed in figure~\ref{fig:stokesLocal}, focusing on the operations in equation~\eqref{eq:StokesHDGdiscreteLocalSolution}.

\section{Visualisation} \label{sc:visualisation}

The use of a high-order functional approximation means that non-standard techniques are required to produce a reliable representation of the solution in each element. With the increased popularity of high-order methods in recent years, different strategies to efficiently display high-order solutions have been proposed~\cite{remacle2007efficient,nelson2012elvis,loseille2018vizir}. The \texttt{HDGlab} provides the capability to accurately display high-order solutions, including curved isoparametric elements. 

Three data structures are employed in the visualisation. The data structure \texttt{plotOpts} is used to collect all the user defined options available for the visualisation. It contains the following information:
\begin{itemize}
	\item \texttt{resolution}: Takes value 1 for a faster but less accurate representation of high-order solutions and value 2 for a slower but more accurate representation of high-order solutions.
	\item \texttt{fieldsWithMesh}: Plots the solution whilst displaying the mesh.
	\item \texttt{fieldsWithNodes}: Plots the solution whilst displaying the nodes.
	\item \texttt{componentsToPlot}: A user-defined vector containing the components of the solution to be represented.
	\item \texttt{alphaFace}: Sets the transparency between 0 and 1.
\end{itemize}
In addition, for the Stokes equation, two problem-specific options are available in the data structure \texttt{plotOpts}:
\begin{itemize}
	\item \texttt{componentsU}: A boolean variable allowing the predefined visualisation of all the components of the velocity vector. This functionality relies on the definition of the vector \texttt{componentsToPlot}.
	\item \texttt{moduleU}: A boolean variable allowing the visualisation of the module of the velocity.
\end{itemize}

The data structure \texttt{postproc} is provided for triangular and tetrahedral elements with equally-spaced and Fekete nodal sets in the directory \texttt{dat/postprocess}. In two dimensions, the data structure \texttt{postproc} contains the following information:
\begin{itemize}
	\item \texttt{nOfNodesPlot}: Number of nodes used to display the high-order solution in each element.
	\item \texttt{nodesPlot}: Array of dimension $\texttt{nOfNodesPlot} \times 2$. Coordinates of the nodes, in the reference element, used to display the high-order solution.
	\item \texttt{nSubElemsOnePlot}: Number of subelements used to display the high-order solution in each element.
	\item \texttt{connecNodesPlot}: Array of dimension $\texttt{nSubElemsOnePlot} \times 3$ accounting for the connectivity of the submesh used to display the high-order solution in each element.
	\item \texttt{nOfEdges}: Number of edges of the element.
	\item \texttt{edgeNodesSplit}: Array of dimension $\texttt{5r} \times \texttt{nOfEdges}$, where \texttt{r} is the resolution selected by the user in the data structure \texttt{plotOpts}. The $i$-th column contains the local number of the list of \texttt{nodesPlot} that belong to the $i$-th edge. 
	\item \texttt{elem}: Array of dimension $1 \times \texttt{p}_\texttt{max}$, where $\texttt{p}_\texttt{max}$ is the maximum degree of approximation used in all the elements. The component $\texttt{p}$ of \texttt{elem} contains a field called $\texttt{N}$, of dimension $\texttt{nOfNodesPlot} \times \texttt{p}(\texttt{p}+1)/2$, that stores the value of the shape functions of order $\texttt{p}$ at the positions given by \texttt{nodesPlot}. This information is used to interpolate the solution at the nodes of the submesh, providing a more accurate representation of the high-order solution.
	\item \texttt{face}: Array of dimension $1 \times \texttt{p}_\texttt{max}$. The component $\texttt{p}$ of \texttt{elem} contains a field called $\texttt{N}$, of dimension $\texttt{nOfNodesPlot} \times (\texttt{p}+1)$, that stores the value of the shape functions of order $\texttt{p}$ at the positions of an edge. This information is used to interpolate the solution at the edges of the submesh.
\end{itemize}

\begin{figure}[!tb]
	\centering
	\includegraphics[width=0.23\textwidth]{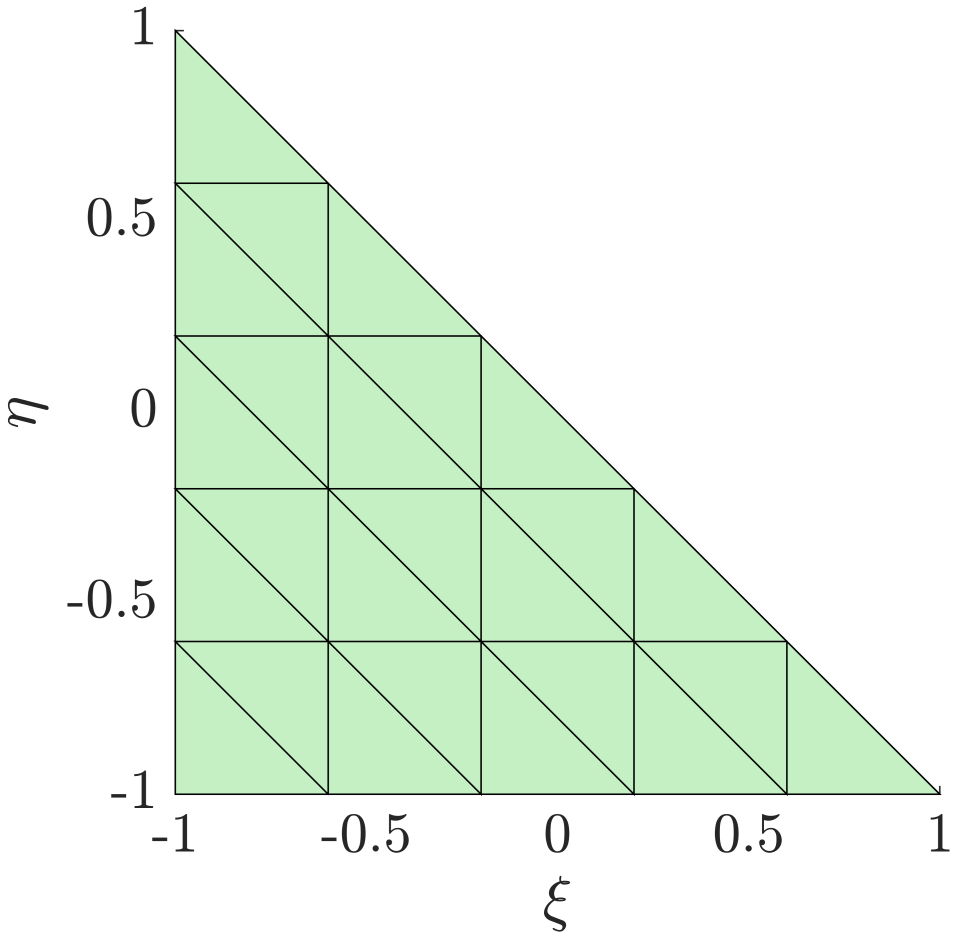}
	\includegraphics[width=0.23\textwidth]{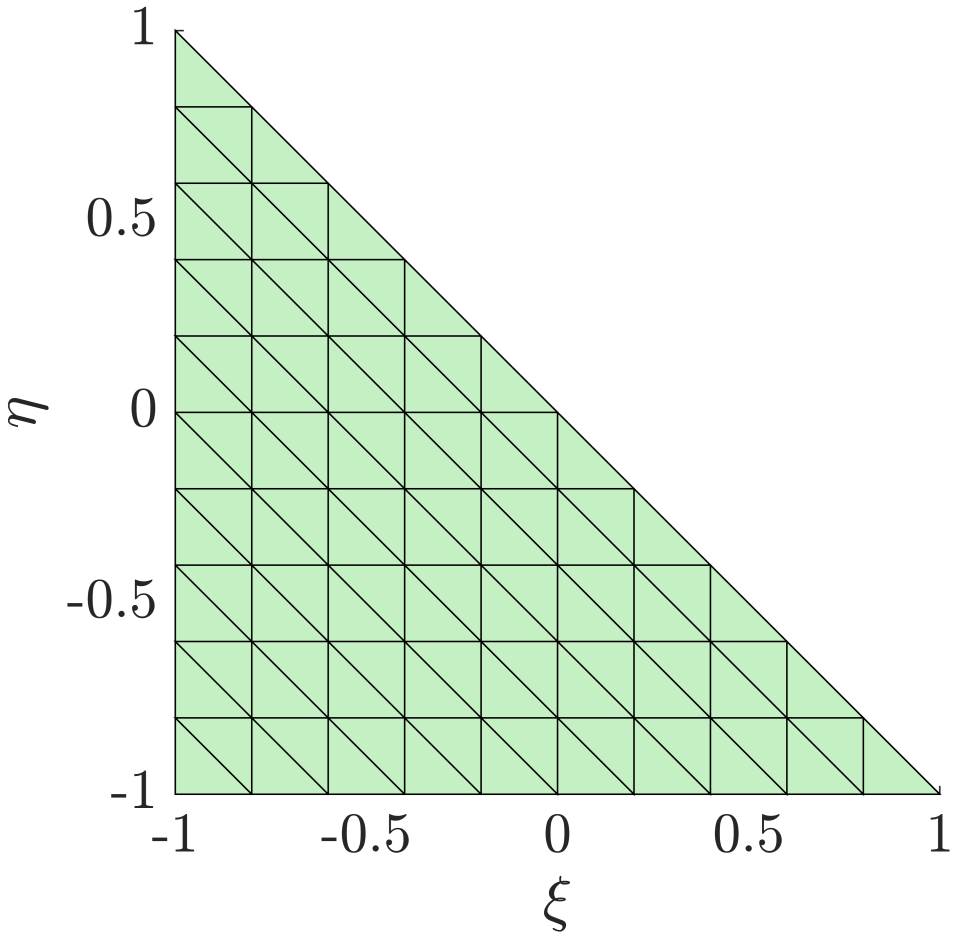}
	\caption{Submesh of the reference element used to provide an accurate representation of high-order solutions in each element. The left picture corresponds to \texttt{resolution}=1 and the right picture to \texttt{resolution}=2.}
	\label{fig:postprocessTri}
\end{figure}

\begin{figure}[!tb]
	\centering
	\includegraphics[width=0.23\textwidth]{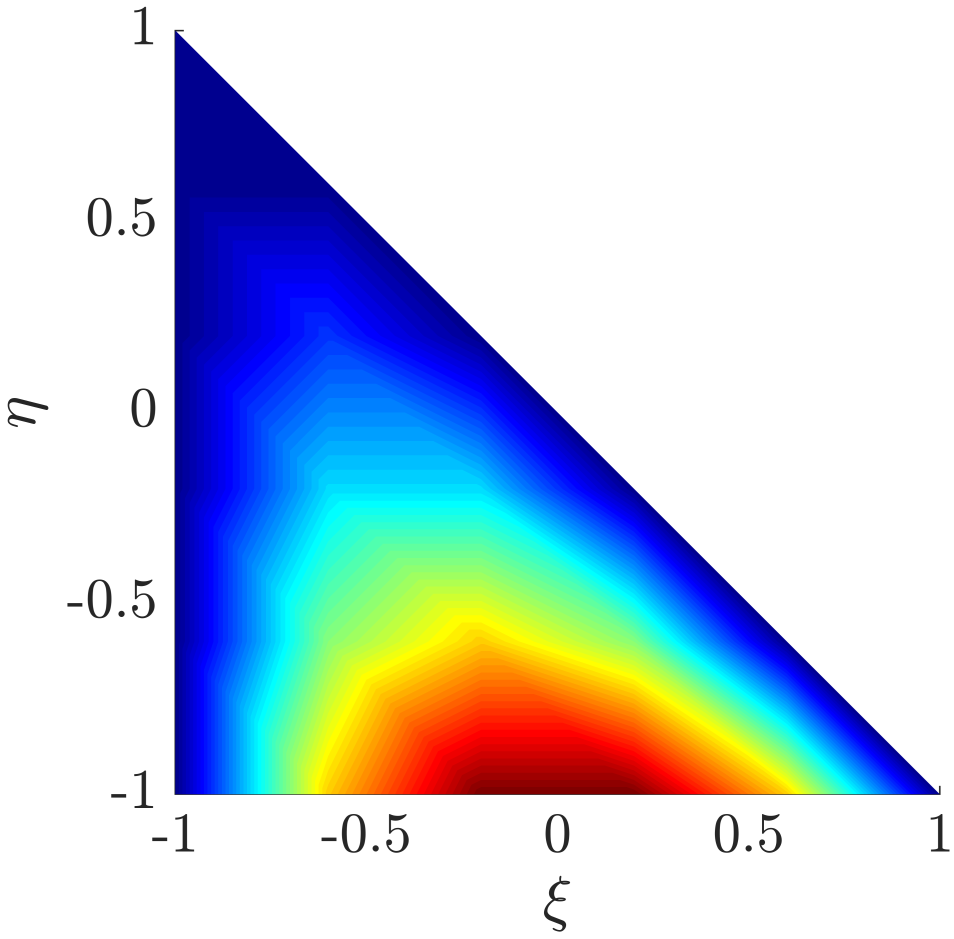}
	\includegraphics[width=0.23\textwidth]{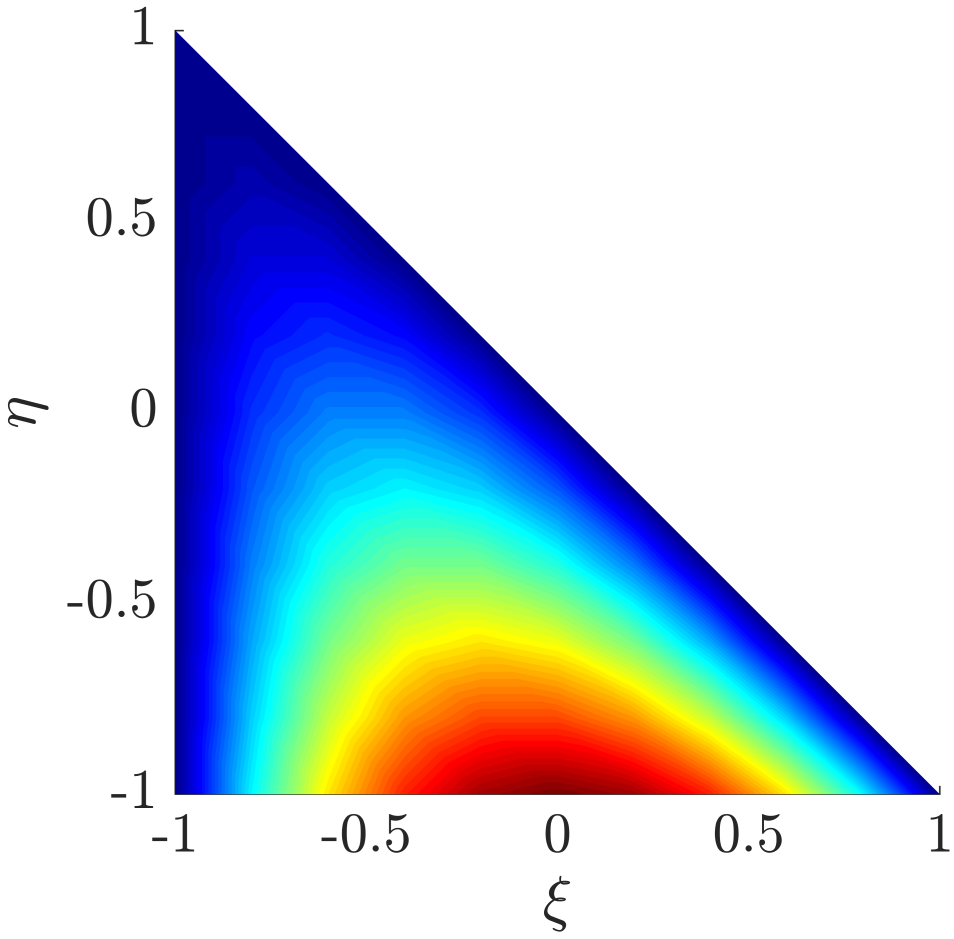}
	\caption{Shape function of the fourth node of a quadratic triangular element using \texttt{resolution}=1 (left) and \texttt{resolution}=2 (right).}
	\label{fig:postprocessTriP2}
\end{figure}

The submeshes used for a triangular element with \texttt{resolution}=1 and \texttt{resolution}=2 are displayed in figure~\ref{fig:postprocessTri}.
To illustrate the effect of the user-defined parameter \texttt{resolution} on the visualisation, figure~\ref{fig:postprocessTriP2} depicts the shape function associated to the fourth node of the reference quadratic triangular element, shown in figure~\ref{fig:refElemTriP2}, using \texttt{resolution}=1 and \texttt{resolution}=2.

In three dimensions the data structure \texttt{postproc} contains the following information: 
\begin{itemize}
	\item \texttt{Face}: Structure that contains:
	\begin{itemize}
		\item  \texttt{nElemPlot}: Number of subelements used to display the high-order solution in each element face.
		\item  \texttt{connecPlot}: Array of dimension $\texttt{nElemPlot} \times 3$ accounting for the connectivity of the submesh used to display the high-order solution in each element face. 		
		\item  \texttt{nNodesPlot}: Number of nodes used to display the high-order solution on each element face. It is given by $(5\texttt{r}+1)(5\texttt{r}+2)/2$, where \texttt{r} is the resolution selected by the user in the data structure \texttt{plotOpts}.
		\item  \texttt{edgeNodesPlot}: Array of dimension $1 \times (15\texttt{r}+1)$, where \texttt{r} is the resolution selected by the user in the data structure \texttt{plotOpts}. It contains the local number of the face nodes that belong to the edges of the face.
	\end{itemize}	
	\item \texttt{Elem}: Array of dimension $1 \times \texttt{p}_\texttt{max}$, where $\texttt{p}_\texttt{max}$ is the maximum degree of approximation used in all the elements. The component $\texttt{p}$ of \texttt{Elem} contains:
	\begin{itemize}
		\item  \texttt{face}: Array of dimension $1 \times 4$ where the $i$-th component contains the local number of the vertices on the $i$-th face and the value of the element shape functions of order $\texttt{p}$ at the nodal distribution used for plotting the solution on the $i$-th face.
		\item  \texttt{coord}: Array of dimension $\texttt{p}(\texttt{p}+1)(\texttt{p}+2)/6$ that contains the nodal distribution on the reference tetrahedral for an approximation of degree $\texttt{p}$.
	\end{itemize}
	\item \texttt{faceVertices}: Array of dimension $4 \times 3$. The $i$-th row contains the local number of the vertices on the $i$-th face.
\end{itemize}

The third data structure utilised during the postprocess stage is called \texttt{visual} and it is built by the function \texttt{buildSubmeshPostprocess2D} in two dimensions and by the routine \texttt{buildSubmeshPostprocess3D} in three dimensions. This data structure contains the following information: 
\begin{itemize}
	\item \texttt{X}: Physical coordinates of all the nodes of the submesh used to display the high-order solution.
	\item \texttt{T}: Connectivities of all the subelements used to display the high-order solution.
	\item \texttt{Xnodes}: Physical coordinates of all the vertices of the mesh. This field is only used if the user sets \texttt{fieldsWithNodes}=1 in the data structure \texttt{plotOpts}.
	\item \texttt{edges}: Structure containing the list of nodes of the submesh that form the high-order representation of the physical edges of the mesh.
\end{itemize}

In a separate function,  the data structure is updated by adding the field \texttt{U} that contains the interpolated values of the solution on the submesh used to display the high-order solution. This action is performed by the function called \texttt{interpolateSolutionPostprocess2D} and \texttt{interpolateSolutionPostprocess3D} in two and three dimensions respectively. 

It is worth noting that the function that creates the data structure \texttt{visual} is independent on the field to be represented and, therefore, it is only called once. Instead, the second function that updates the data structure \texttt{visual} with the field \texttt{U} depends upon the field to be represented. Therefore, several calls can be made to the function updating \texttt{visual}  without the need to build the submesh again. It is also important to note that the function that updates the data structure \texttt{visual} with the field \texttt{U} accepts elemental and nodal fields.

Once all information is available in the data structure \texttt{visual}, the \texttt{postprocessField2D} function is used to plot the high-order solution. 

In three dimensions there is an extra option available that consists of representing the solution only in a region of the computational domain. The user can set the value of a string, called \texttt{conditionPlot}, that specifies a region in the physical space. Before \texttt{visual} is computed, the function \texttt{selectFacesToPlot3D} computes the list of faces in the computational mesh that satisfy the condition given by \texttt{conditionPlot}. Then, the submesh and interpolation of the solution is only performed over the faces that satisfy the condition specified by the user.

Finally, the visualisation function also accounts for the need to represent the superconvergent solution obtained after the local postprocess described in sections \ref{sc:PoissonPostprocess} and \ref{sc:StokesPostprocess}. To simplify the implementation, the visualisation builds a \texttt{mesh} data structure where the degree of approximation in each element is the degree used for the computation plus one. With this information, the same functions used to display the high-order primal solution can be used to postprocess the higher order superconvergent solution.

\section{Numerical examples} \label{sc:examples}

In this section, several numerical examples showing the capabilities of the \texttt{HDGlab} solvers for the Poisson and Stokes equations are presented. As mentioned in sections~\ref{sc:PoissonHDG} and~\ref{sc:StokesHDG}, the choice of the stabilisation parameter $\tau$ is critical for the accuracy of the HDG approximation. For the examples involving the Poisson equation, the definition $\tau {=} c_{\!_P} \kappa/\ell$ is considered, where $\ell$ is a characteristic length of the domain and $c_{\!_P}$ a scaling factor selected equal to $1$~\cite{MG-SpinaGH-20}. Following~\cite{MG-GSH-20}, the stabilisation for the Stokes cases is defined as $\tau {=} c_{\!_S} \nu/\ell$, $\ell$, the scaling factor being $c_{\!_S} {=} 3$.

\subsection{Optimal convergence properties} \label{sc:stokesConvergence}

The optimal convergence properties of the proposed HDG implementation are presented for the Stokes flow, using test cases with analytical solution, in two and three dimensions. Uniform meshes of triangular and tetrahedral elements with Fekete nodal sets are utilised.

First, the two-dimensional Wang flow~\cite{wang1991exact} in the unit square domain $\Omega {=} [0,1]^2$ is considered. The analytical velocity field is
\begin{equation} \label{eq:wangVelocity}
\bu(\bx) =
\begin{Bmatrix} 
2 a x_2 - b \lambda \cos(\lambda x_1) \exp\{-\lambda x_2\} \\[1ex] 
b \lambda \sin(\lambda x_1) \exp\{-\lambda x_2\}
\end{Bmatrix} ,
\end{equation}
whereas the pressure field is uniformly zero in $\Omega$. The coefficients $a$, $b$ and $\lambda$ in~\eqref{eq:wangVelocity} are selected such that $a {=} b {=} 1$ and $\lambda {=} 10$ and the kinematic viscosity $\nu$ is set to 1. The source term $\bm{s}$ and the boundary conditions are computed starting from the analytical solution above. More precisely, a pseudo-traction $\bm{g}$ is applied on the bottom surface $\Ga[N] {:=} \{(x_1,x_2) \in \Omega \ | \ x_2 {=} 0\}$ and Dirichlet data $\buD$ are imposed on the remaining boundaries $\Ga[D] {=} \partial \Omega \setminus \Ga[N]$.

\begin{figure}
\centering
\includegraphics[width=0.45\textwidth]{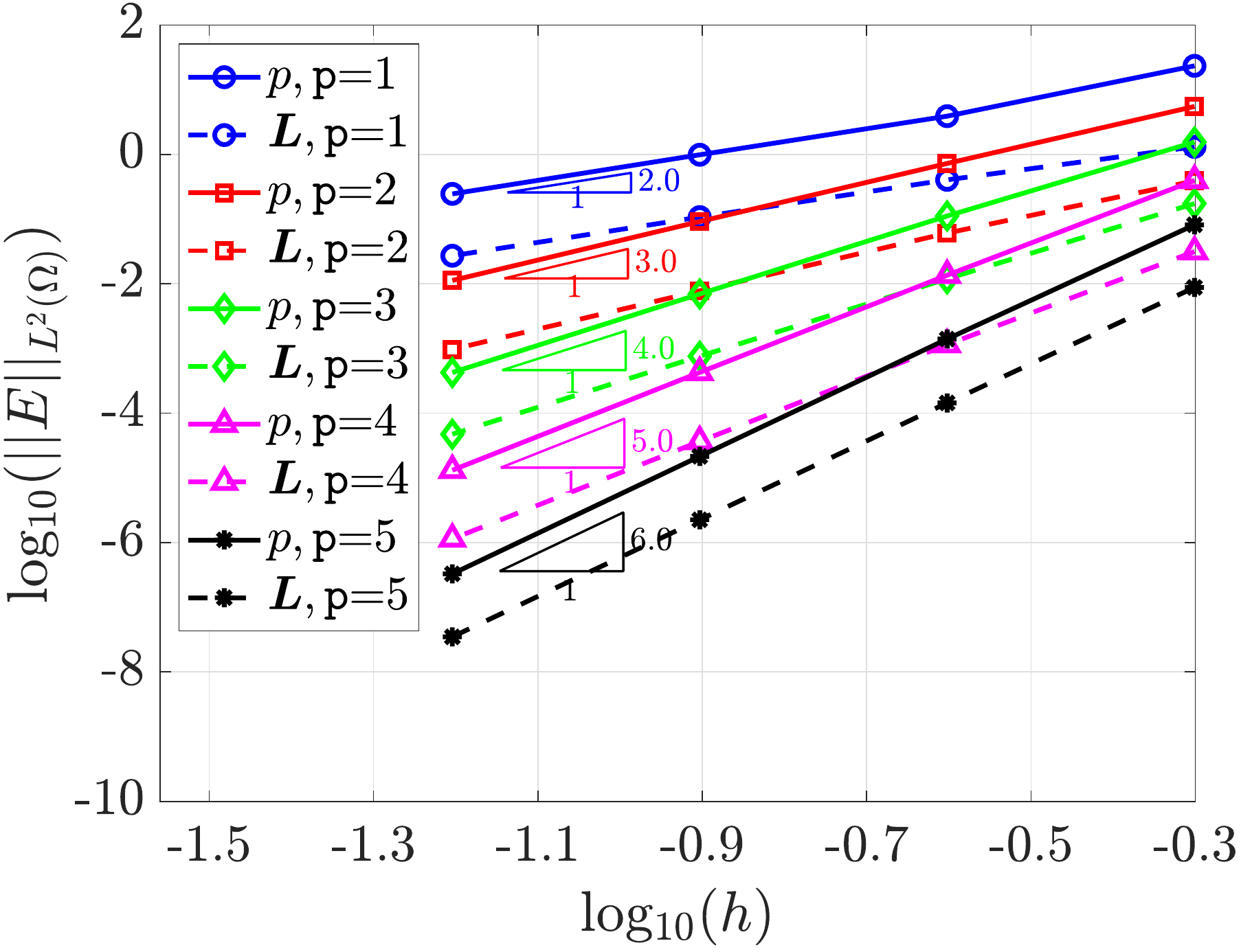}
\includegraphics[width=0.45\textwidth]{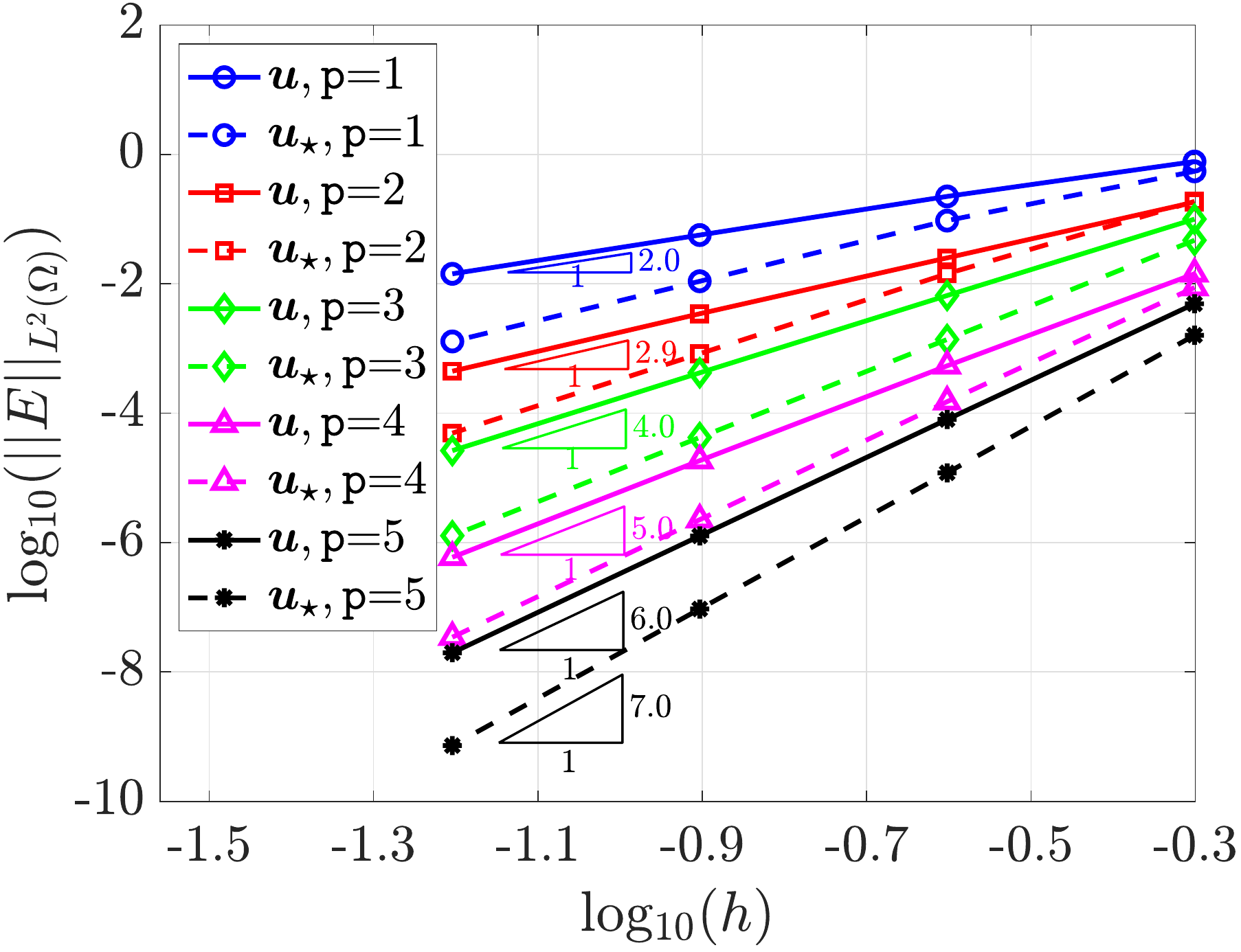}
\caption{Two-dimensional Wang flow. Convergence of the $\eltwo(\Omega)$ error of pressure, $p$, mixed variable, $\bL$ (left), primal, $\bu$, and postprocessed, $\bu_\star$, velocities (right) as a function of the characteristic mesh size $h$ for polynomial degree of approxiomation $\deg {=} 1, \ldots, 5$.}
\label{fig:stokesConvergenceWang}
\end{figure}

Figure~\ref{fig:stokesConvergenceWang} displays the convergence history of the relative error, measured in the $\eltwo(\Omega)$ norm, of the primal, mixed and postprocessed variables as a function of the characteristic mesh size.
Optimal convergence of order $\deg {+} 1$ is observed for velocity, $\bu$, pressure, $p$, and gradient of velocity, $\bL$, whereas superconvergence of order $\deg {+} 2$ is achieved by the postprocessed velocity $\buS$. 

The following example involves a three-dimensional Stokes flow in the unit cube $\Omega {=} [0,1]^3$, with the following manufactured solution
\begin{align} \label{eq:StokesAnalytic3D}
\bu(\bx) &=
\begin{Bmatrix} 
n +(x_3 - x_2) \sin(x_1 - n) \\
m - x_2 \biggl[ \left( x_3 - \frac{1}{2} x_2 \right) \cos(x_1 - n) + \left( x_1 - \frac{1}{2} x_2 \right) \cos(x_3 - n) \biggr] \\[1ex] 
n + (x_1 - x_2) \sin(x_3 - n) 
\end{Bmatrix} , \\
p(\bx) &= x_1 (1 - x_1) + x_2 (1 - x_2) + x_3 (1 - x_3) ,
\end{align}
where $m {=} 1$ and $n {=} \tfrac{1}{2}$. The kinematic viscosity is set to $\nu {=} 1$, a Neumann datum is imposed on the boundary $\Ga[N] {:=} \{(x_1,x_2,x_3) \in \Omega \ | \ x_3 {=} 0\}$, whereas Dirichlet conditions are prescribed on $\Ga[D] {=} \partial\Omega \setminus \Ga[N]$. 

\begin{figure} 
\centering
\includegraphics[width=0.45\textwidth]{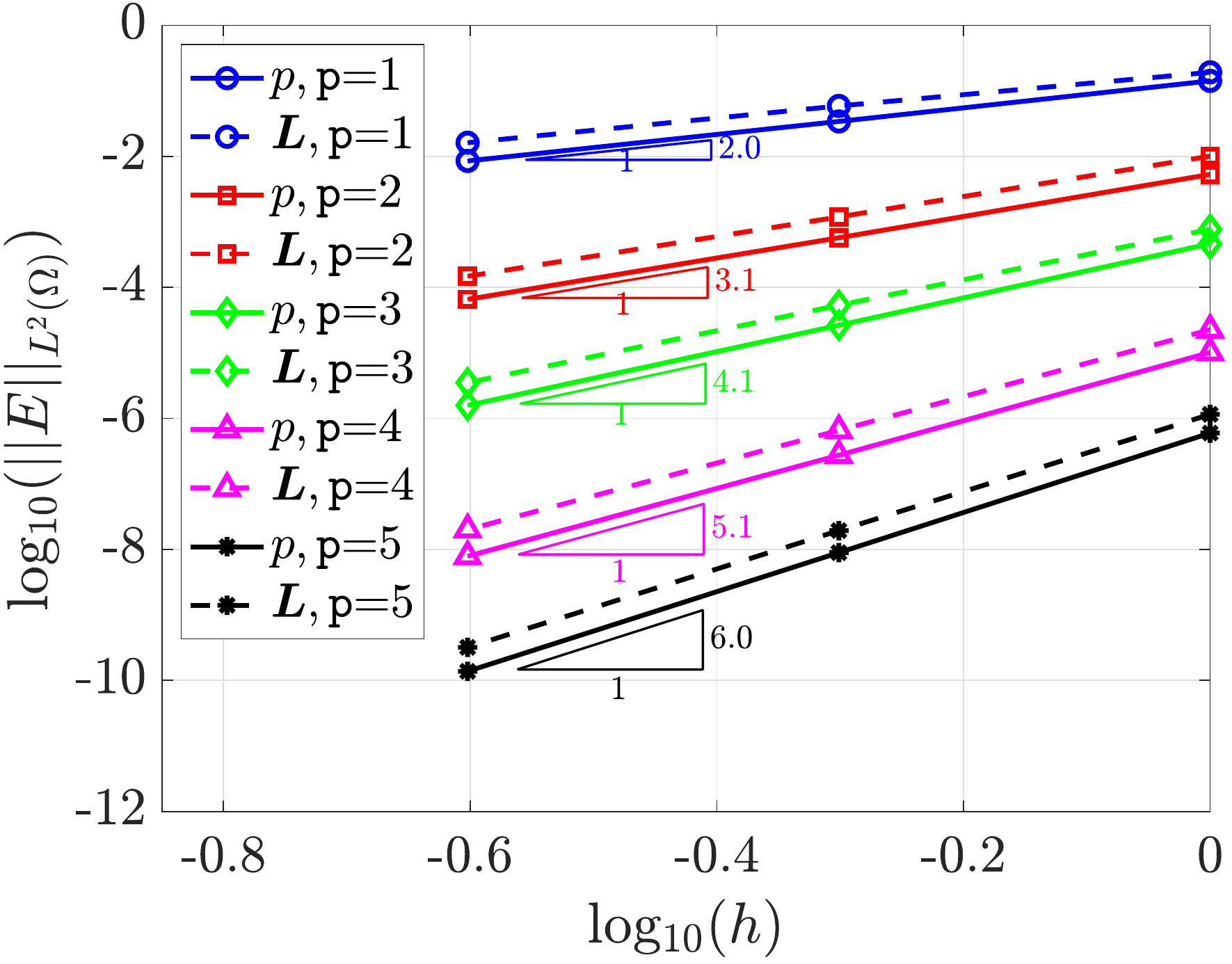}
\includegraphics[width=0.45\textwidth]{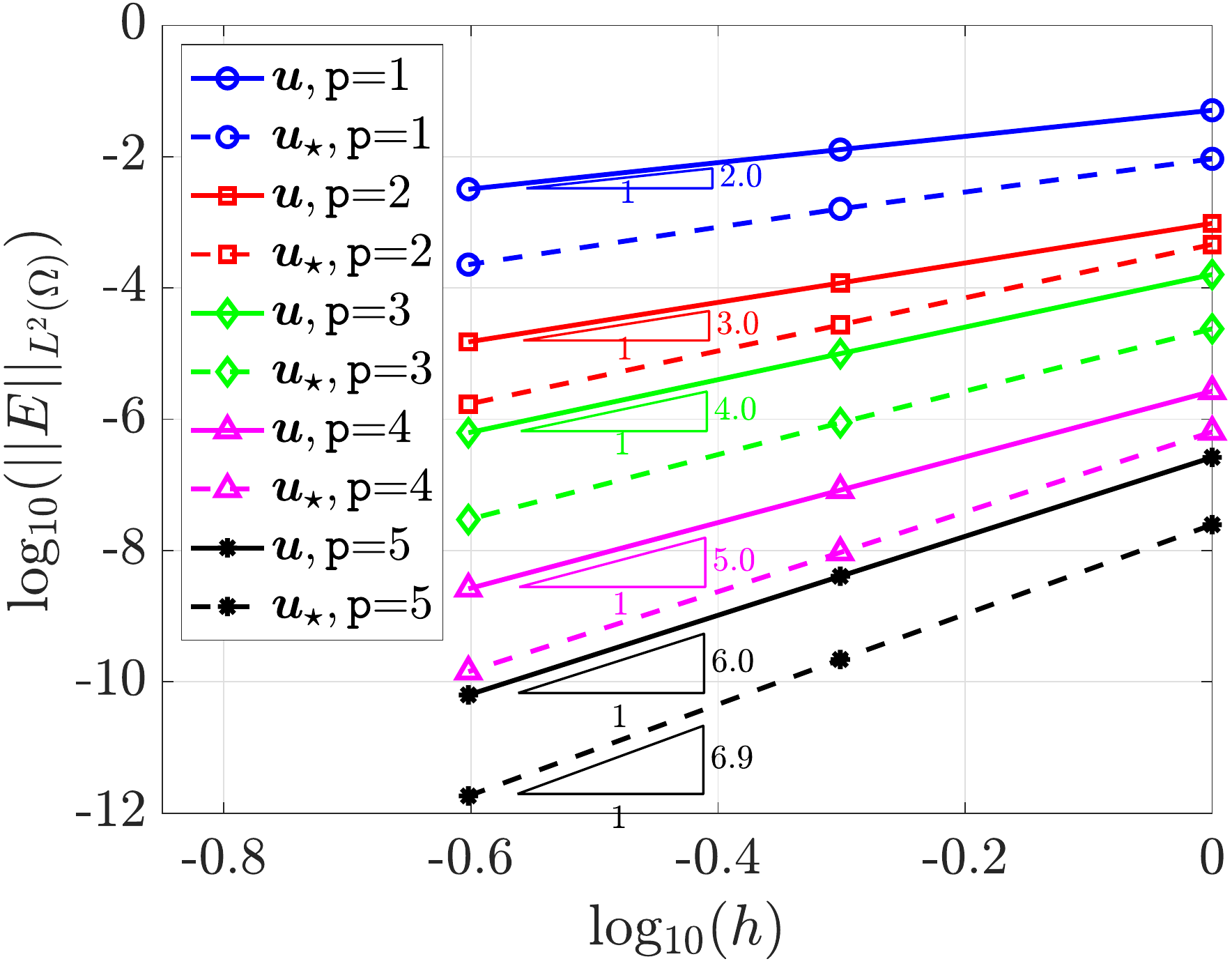}
\caption{Three-dimensional manufactured Stokes flow. Convergence of the $\eltwo(\Omega)$ error of pressure, $p$, mixed variable, $\bL$ (left), primal, $\bu$, and postprocessed, $\bu_\star$, velocities (right) as a function of the characteristic mesh size $h$ for polynomial degree of approxiomation $\deg {=} 1, \ldots, 5$.}
\label{fig:stokesConvergenceCube}
\end{figure}

The optimal convergence of order $\deg {+} 1$ of the relative $\eltwo(\Omega)$ error for velocity, pressure and gradient of velocity and the superconvergence of the postprocessed velocity are confirmed in the 3D case by the results in figure~\ref{fig:stokesConvergenceCube}.

\subsection{High-order curved meshes} \label{sc:stokesCouette}

The coaxial Couette flow~\cite{childs2010rotating} is considered to show the optimal convergence properties of the \texttt{HDGlab} solver using a high-order isoparametric approximation in a domain featuring curved boundaries. 

This test consists of an incompressible viscous flow within two coaxial circular cylinders of infinite length and radius $\Rint {=} 1$ and $\Rext {=} 5$, respectively. The computational domain is defined as a section of the 3D cylinders, that is, $\Omega {=} \{ (x_1,x_2) \in \RR^2 \ | \  \Rint \leq r \leq \Rext \}$, where $r {:=} \sqrt{x_1^2 {+} x_2^2}$ is the distance to the axis of the cylinders.
%
%
Dirichlet boundary conditions enforcing the value of the angular velocities $\wInt {=} 0$ and $\wExt {=} 1/\Rext$ are imposed on the internal and external boundary, respectively. The analytical expression of the azimuthal component of the velocity is
\begin{equation} \label{eq:couetteAnalytical}
u_{\phi} = \frac{ \Rext^2 \wExt -\Rint^2 \wInt }{ \Rext^2 -\Rint^2 } r + \frac{ ( \wInt - \wExt) \Rext^2 \Rint^2 }{ \Rext^2 -\Rint^2 } \frac{1}{r} .
\end{equation}
Of course, being a purely Dirichlet boundary value problem, the constraint on the mean value of pressure is introduced to enforce the field to be uniformly equal to 1 in the domain.

\begin{figure}
\centering
\includegraphics[width=0.35\textwidth]{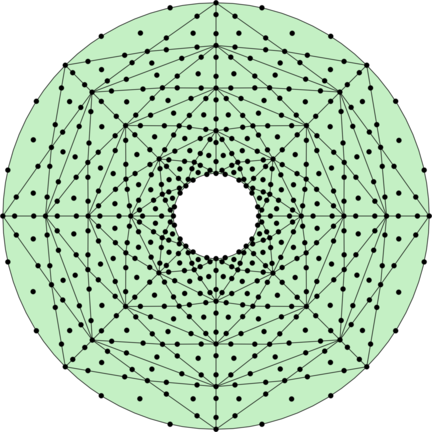}
\includegraphics[width=0.40\textwidth]{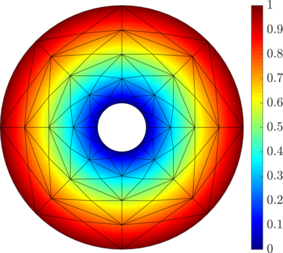}
\caption{First level of refinement of the third-order mesh used for the convergence study of the two-dimensional Couette flow (left) and module of the computed velocity (right).}
\label{fig:stokesCouette}
\end{figure}

A set of high-order uniformly refined meshes with Fekete nodal distribution is constructed using the strategy described in~\cite{poya2016unified}. Figure~\ref{fig:stokesCouette} displays the first level of mesh refinement featuring 128 triangular elements of polynomial degree 3 and the module of the computed velocity.

\begin{figure}
\centering
\includegraphics[width=0.45\textwidth]{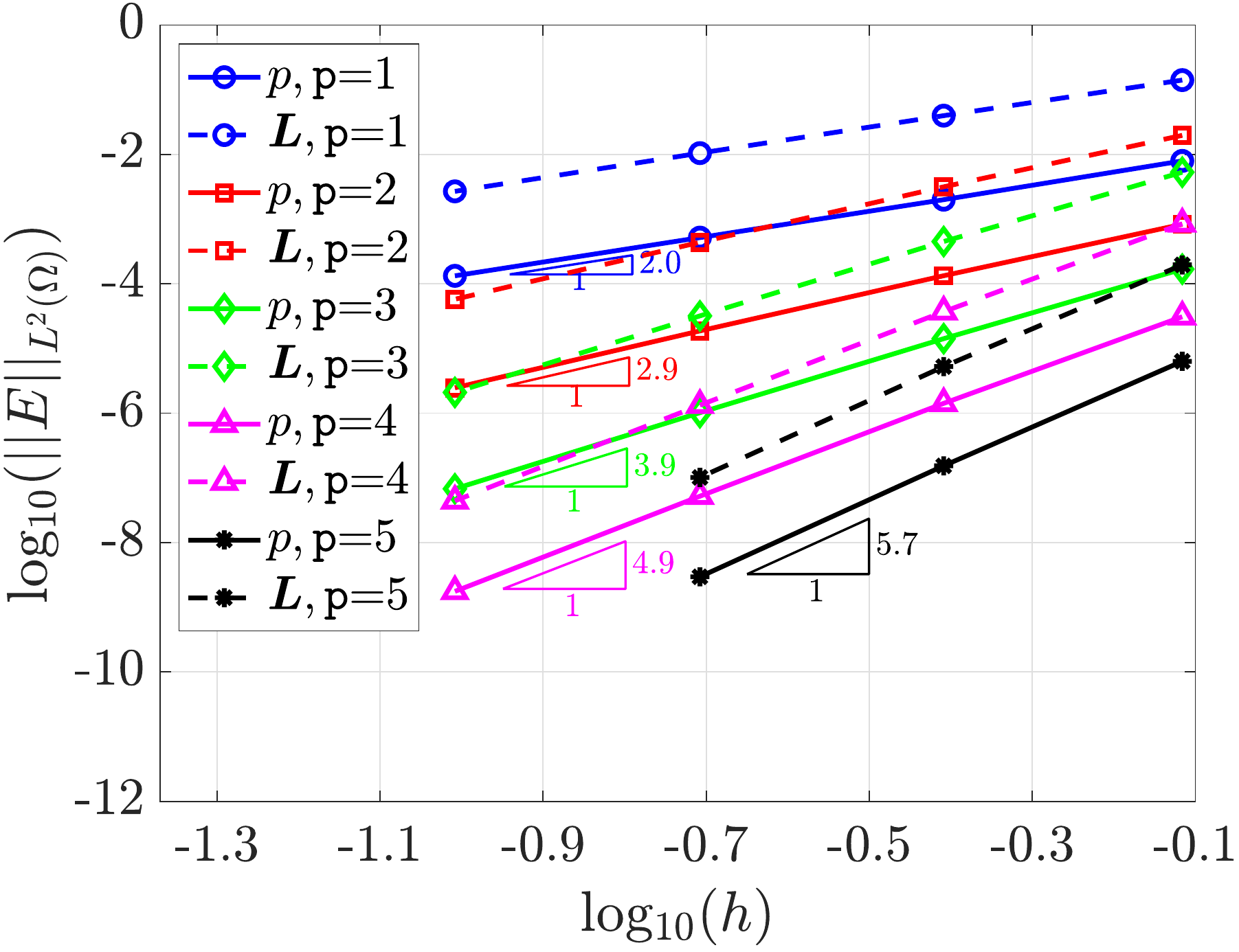}
\includegraphics[width=0.45\textwidth]{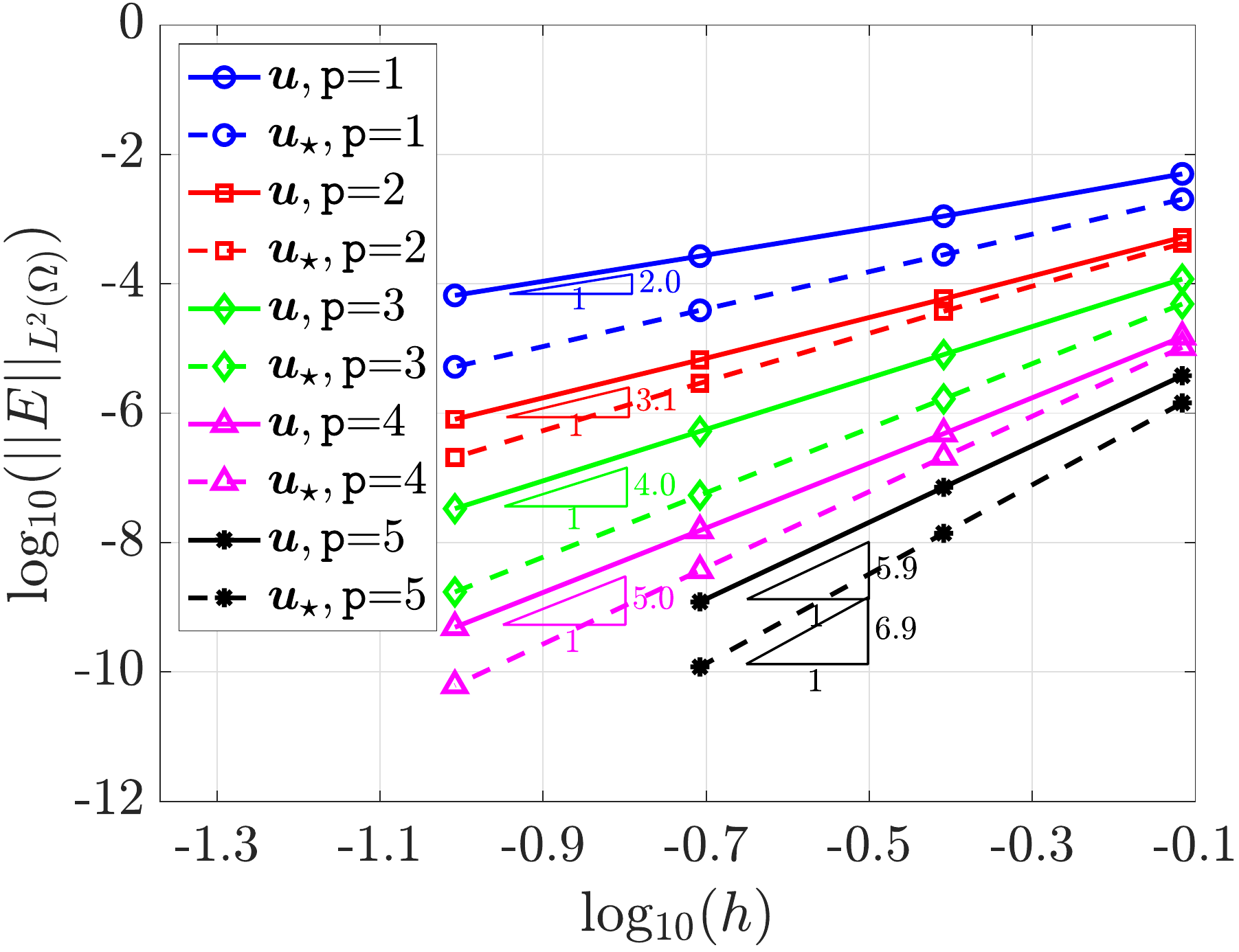}
\caption{Two-dimensional Couette flow. Convergence of the $\eltwo(\Omega)$ error of pressure, $p$, mixed variable, $\bL$ (left), primal, $\bu$, and postprocessed, $\bu_\star$, velocities (right) as a function of the characteristic mesh size $h$ for polynomial degree of approxiomation $\deg {=} 1, \ldots, 5$.}
\label{fig:stokesConvergenceCouette}
\end{figure}

The convergence of the relative error, measured in the $\eltwo(\Omega)$ norm, of the primal, mixed and postprocessed variables is reported in figure~\ref{fig:stokesConvergenceCouette} as a function of the characteristic mesh size. Optimal convergence of the primal and mixed variables and superconvergence of the postprocessed variable is achieved also in presence of high-order curved meshes.

\subsection{Non-uniform degree of approximation} \label{sc:stokesNonUniform}

In this section, the flexibility of \texttt{HDGlab} to devise a non-uniform polynomial degree approximation in the domain is presented. This case, inspired by the study on micromixers in~\cite{Farahinia-FZ-19}, consists of the flow in a microchannel with five obstacles. The problem setup features a parabolic inlet velocity profile and homogeneous Dirichlet and Neumann conditions on the top/bottom walls and on the outlet, respectively.

The channel has dimensions $[0,6.6] \times [-0.5,0.5]$ and the obstacles, attached to the top and bottom walls have thickness $0.2$ and height $0.5$. A mesh with local element size ranging between $0.08$ and $0.19$ is generated without any specific \emph{a priori} refinement. It is worth noting that only two mesh elements are defined along the thickness of the obstacles.

\begin{figure*}
\centering
\subfigure[Degree distribution]{\includegraphics[width=0.9\textwidth]{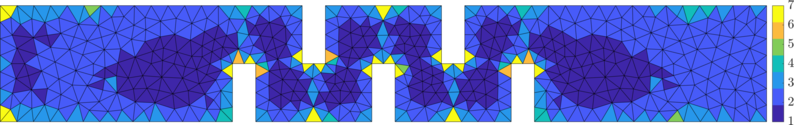}}

\hspace{3pt} \subfigure[Module of the velocity]{\includegraphics[width=0.91\textwidth]{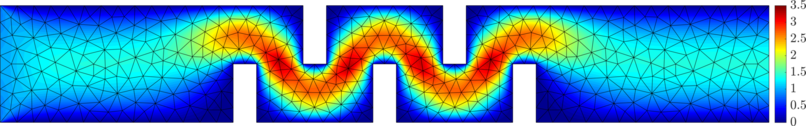}}
\caption{(a) Mesh and degree of approximation and (b) module of the velocity for the flow in a microchannel with obstacles using non-uniform polynomial degree $\deg$ between $1$ and $7$.}
\label{fig:stokesChannel}
\end{figure*}

To capture the complex flow features among the obstacles, a high-order non-uniform polynomial degree distribution is generated following the adaptivity strategy described in~\cite{RS-SH-18}. The resulting degree of approximation in each element is displayed in figure~\ref{fig:stokesChannel}. High-order polynomials are employed in correspondance of the tip of the obstacles where localised flow features appear, whereas low-order approximations are utilised in region further away.

The module of the velocity on the described mesh is also reported in figure~\ref{fig:stokesChannel}. It is worth noting that the mesh structure featuring the non-uniform polynomial approximation is provided as a datum for this test case. The corresponding simulation is performed seamlessly in \texttt{HDGlab} and no specific intervention is required to the user.

\subsection{Stokes flow past a sphere} \label{sc:stokesSphere}

Finally, the Stokes flow past a sphere is considered. The domain $\Omega {=} [-H,L] \times [-H,H] \times [-H,H] \setminus \Ball$ is defined, with $H {=} 5$, $L {=} 10$ and $\Ball$ being the sphere of radius 1 centred in the point $(0,0,0)$. Exploiting the symmetry of the problem, only a quarter of the domain is meshed and slip conditions are imposed on the corresponding symmetry planes. Homogeneous Neumann and no-slip conditions are applied on the outlet and on the surface of the sphere, respectively. On the inlet and on the remaining lateral and top planes, a Dirichlet boundary conditions with the analytical velocity is enforced.

\begin{figure*}
\centering
\subfigure[Module of the velocity]{\includegraphics[width=0.45\textwidth]{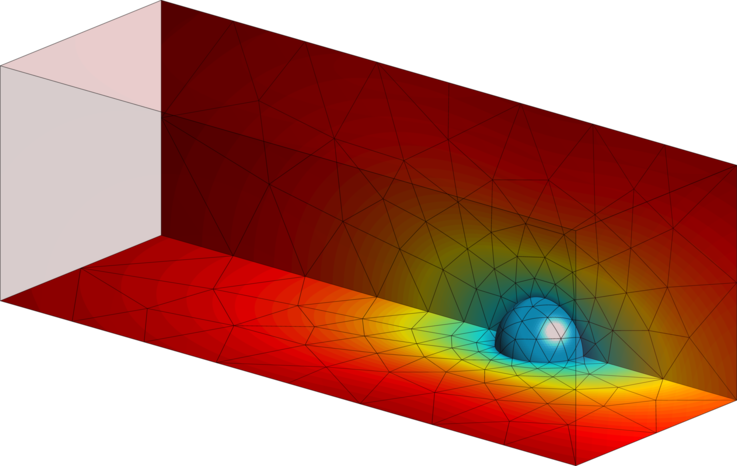}}
\hfill
\subfigure[Pressure]{\includegraphics[width=0.45\textwidth]{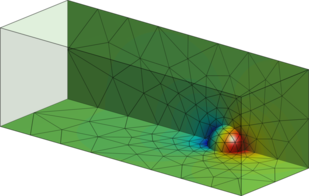}}
\caption{(a) Module of the velocity and (b) pressure field of the external flow past a quarter of a sphere using polynomial degree $\deg {=} 6$.}
\label{fig:sphereVelocityPressure}
\end{figure*}

A high-order mesh featuring 1,036 tetrahedral elements is generated via the solid mechanics analogy described in~\cite{poya2016unified,xie2013generation}. Figure~\ref{fig:sphereVelocityPressure} displays the module of the velocity and the pressure field computed using an isoparametric approximation of degree 6.

It is worth noting that the computations in sections~\ref{sc:stokesConvergence},~\ref{sc:stokesCouette},~\ref{sc:stokesNonUniform} and~\ref{sc:stokesSphere} are performed using a unique \texttt{HDGlab} solver for the Stokes equations, independently on the number of spatial dimensions of the problem under analysis. Indeed, \texttt{HDGlab} provides a seamless implementation of the HDG method, in which all relevant information is extracted from the mesh structure and the user is required to specify only the physical parameters and the boundary conditions to setup a test case.

\subsection{Applications of the Poisson solver} \label{sc:applications}

\begin{figure}
	\centering
	\includegraphics[width=0.45\textwidth]{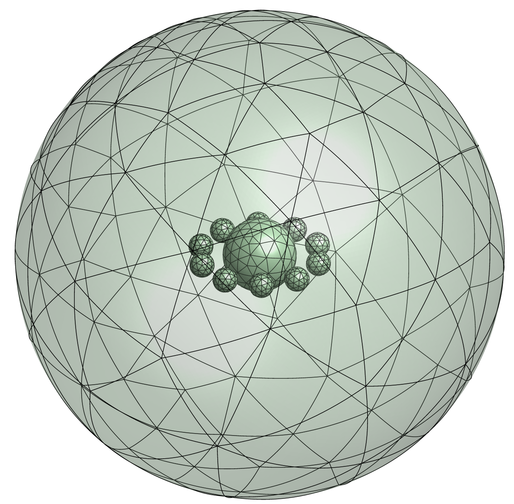}
	\includegraphics[width=0.45\textwidth]{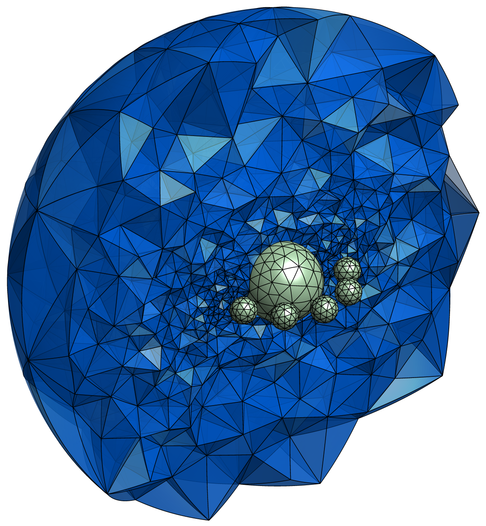}
	\caption{Two views of the quadratic tetrahedral mesh used to solve the electrostatic problem.}
	\label{fig:conductingSpheresMesh}
\end{figure}

The next example, taken from~\cite{liu2009fast}, shows the solution of an electrostatic problem governed by the Poisson equation in three dimensions. The domain of interest corresponds to the exterior of 11 conducting spheres and it is discretised with a mesh of 35,895 quadratic tetrahedral elements, as shown in figure~\ref{fig:conductingSpheresMesh}.
This figure shows one of the implemented capabilities of the postprocessing library to display the faces corresponding to the exterior and interior faces of the mesh separately, with different colour and transparencies in each case. The first plot in figure~\ref{fig:conductingSpheresMesh} is produced by selecting the faces corresponding to the far field boundary and using a transparency. In a second phase, the exterior faces corresponding to the conducting spheres are displayed with no transparency. It is worth noting that the far field boundary and the boundary corresponding to the conducting spheres could be distinguished using either the boundary condition flag or simply imposing a condition on the faces to be displayed. The second plot of figure~\ref{fig:conductingSpheresMesh} is also obtained in two stages. First, the interior faces satisfying a condition corresponding to a positive $x_2$ coordinate are displayed with a transparency. Second, the exterior faces corresponding to the conducting spheres is displayed with no transparency. When representing the interior and exterior faces, a constant element field is used to assign different colours and aid the visualisation. 

\begin{figure}
	\centering
	\subfigure[Conducting  spheres]{\includegraphics[width=0.45\textwidth]{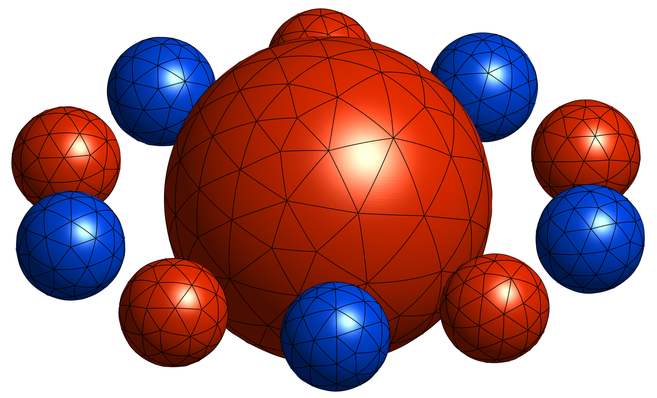}\label{fig:conductingSpheresBC}}
	\subfigure[Electric field]{\includegraphics[width=0.45\textwidth]{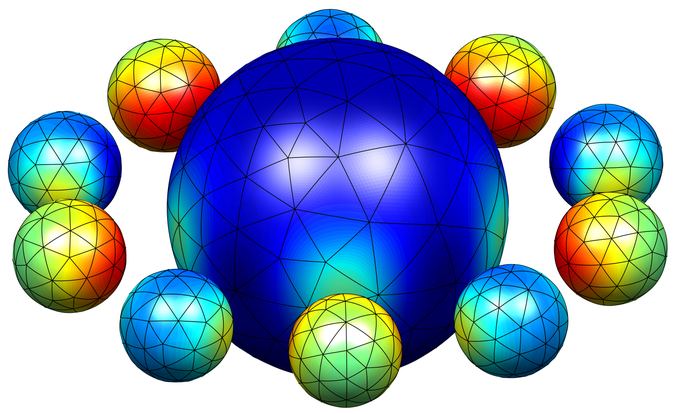}\label{fig:conductingSpheresSol}}
	\caption{(a) The 11 conducting spheres coloured according to the boundary condition imposed. Red colour is used for the spheres where a positive potential is imposed, whereas blue is used for the spheres where a negative potential is imposed. (b) Magnitude of the electric field on the surface of the 11 conducting spheres.}	
	\label{fig:conductingSpheresFig}
\end{figure}

Dirichlet boundary conditions are considered in the whole boundary of the computational domain. A positive electrostatic potential of magnitude $5$ is imposed on the central sphere and five of the surrounding spheres, whereas a negative potential of magnitude $-5$ is imposed on the remaining spheres. On the outer boundary a zero potential is imposed. Figure~\ref{fig:conductingSpheresBC} shows the 11 conducting spheres coloured according to the boundary condition imposed.
This figure is produced by using the boundary condition flag to select the first set and the second set of spheres separately.
After solving the problem with \texttt{HDGlab}, not only the primal variable corresponding to the electrostatic potential is obtained but also its gradient, which corresponds to the electric field in this example. Figure~\ref{fig:conductingSpheresSol} shows the magnitude of the electric field on the surface of the 11 spheres.

\begin{figure}
	\centering
	\includegraphics[width=0.45\textwidth]{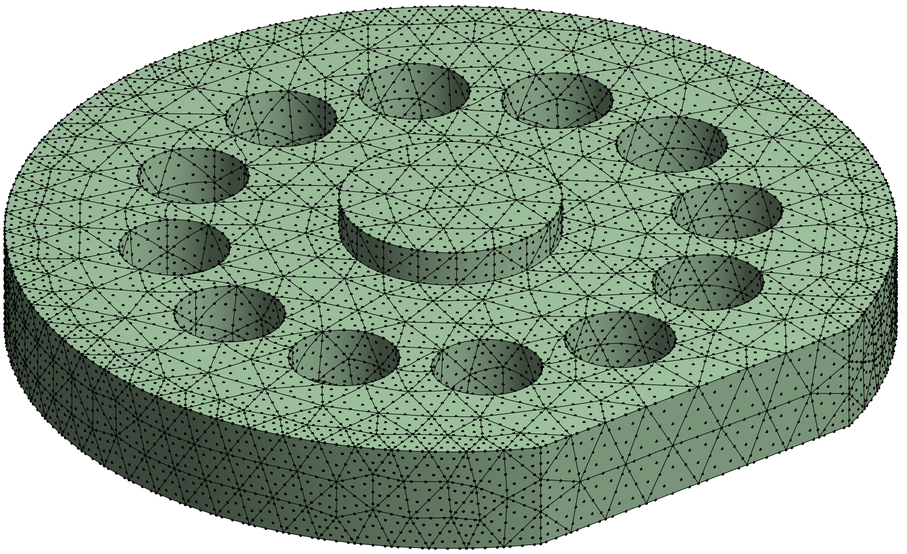}
	\includegraphics[width=0.45\textwidth]{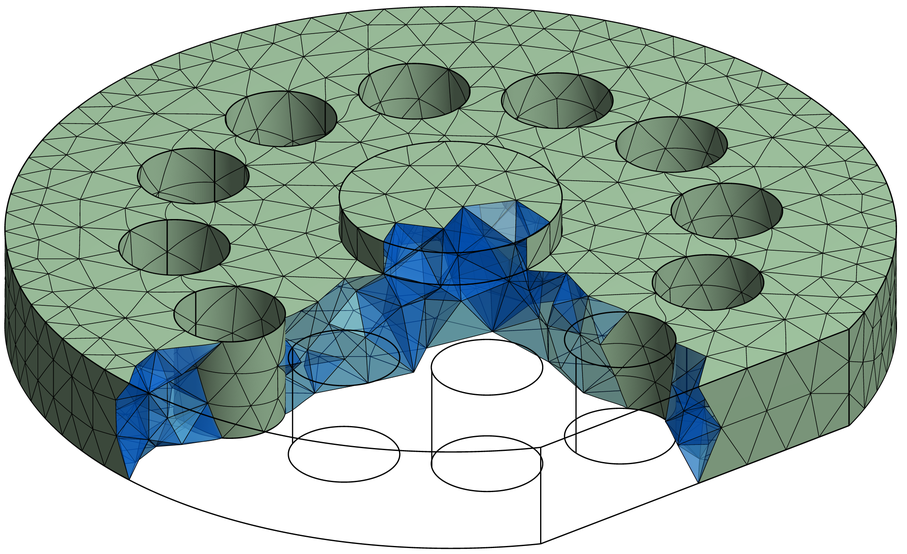}
	\caption{Two views of the fourth order tetrahedral mesh used to solve the heat transfer problem.}
	\label{fig:cylinderWithRingsMesh}
\end{figure}

The following example involves the solution of a heat transfer problem in a three dimensional mechanical component. The domain is discretised with a mesh of 6,465 elements with $\deg=4$, as illustrated in figure~\ref{fig:cylinderWithRingsMesh}.
The first plot in figure~\ref{fig:cylinderWithRingsMesh} includes the high-order nodal distribution and the second plot in figure~\ref{fig:cylinderWithRingsMesh} shows the possibilities offered by the postprocessing library provided within \texttt{HDGlab}. In fact, it shows an extra functionality that can be added to display the intersection curves of the underlying CAD geometry when the user have access to this data.

\begin{figure}
	\centering
	\subfigure[Mechanical component]{\includegraphics[width=0.45\textwidth]{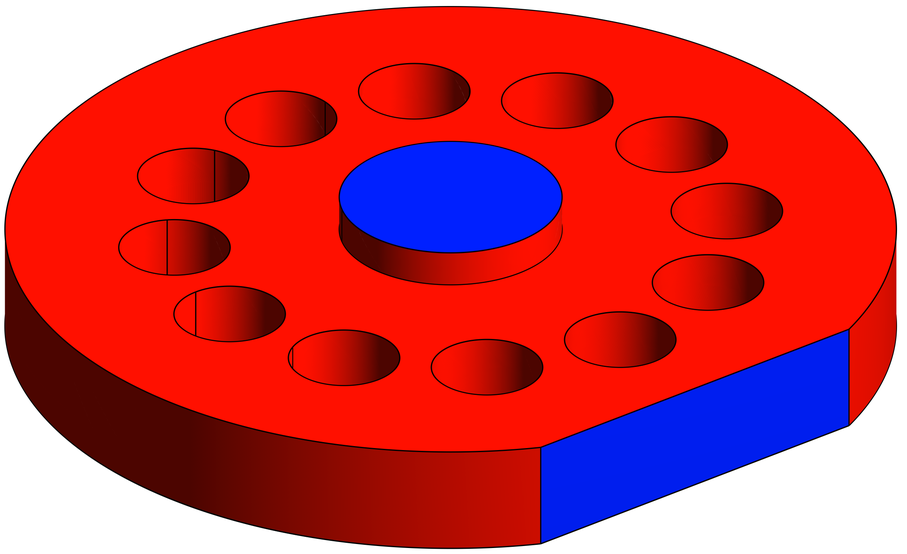}\label{fig:cylinderWithRingsBCs}}
	\subfigure[Temperature distribution]{\includegraphics[width=0.45\textwidth]{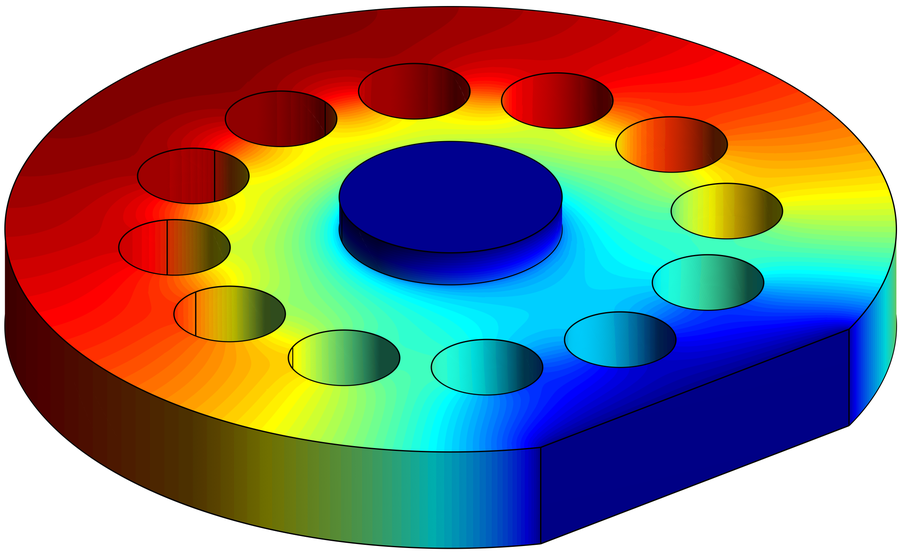}\label{fig:cylinderWithRingsSol}}
	\caption{(a) The mechanical component coloured according to the boundary condition imposed. Red colour denotes a homogeneous Neumann boundary condition and the blue colour a Dirichlet boundary condition. (b) Temperature distribution on the surface of the mechanical component.}
	\label{fig:cylinderWithRingsFig}
\end{figure}

Dirichlet and Neumann boundary conditions are imposed on the blue and red portions of the boundary, respectively, as shown in figure~\ref{fig:cylinderWithRingsBCs}.
In the Dirichlet part of the boundary a temperature equal to 10 is imposed, whereas the Neumann boundary condition is homogeneous, imposing that part of the boundary is perfectly insulated. The temperature field obtained after solving the problem with \texttt{HDGlab} is shown in figure~\ref{fig:cylinderWithRingsSol}.

\begin{figure}
	\centering
	\includegraphics[width=0.45\textwidth]{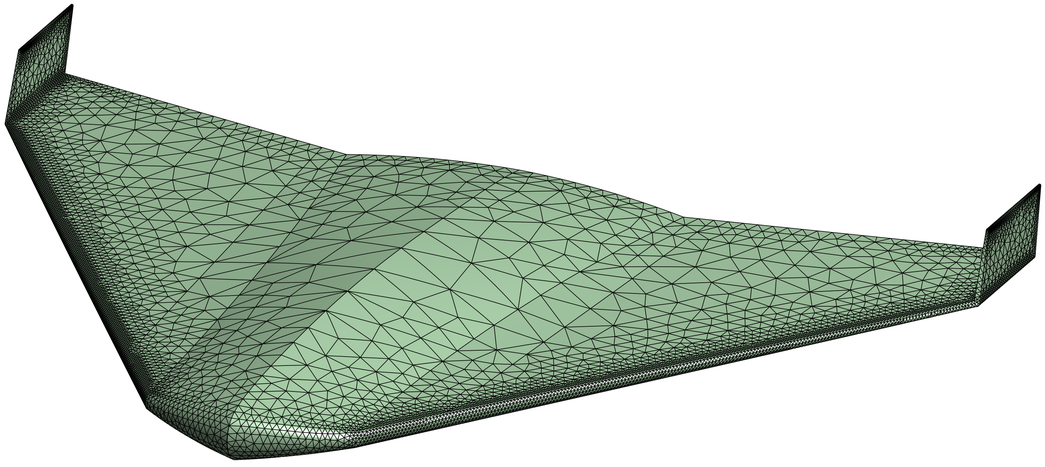}
	\caption{Surface mesh of a generic UAV.}
	\label{fig:uavMesh}
\end{figure}

\begin{figure}
	\centering
	\includegraphics[width=0.45\textwidth]{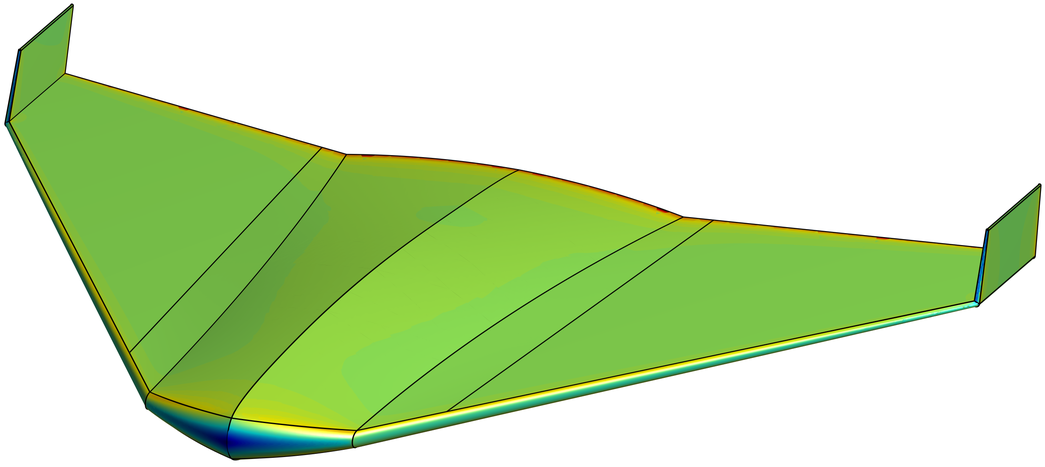}
	\includegraphics[width=0.45\textwidth]{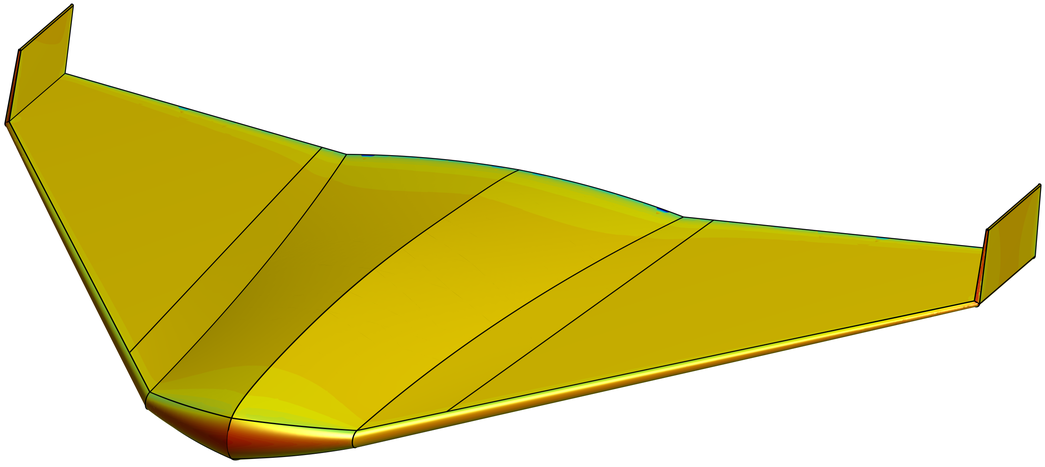}
	\caption{Magnitude of the velocity field and pressure field on the surface of a generic UAV.}
	\label{fig:uavSol}
\end{figure}

The last example considers the computation of the potential flow past a generic unmanned aerial vehicle (UAV). The domain is discretised using 101,923 tetrahedra of degree $\deg {=} 2$, as represented in figure~\ref{fig:uavMesh}.
A Neumann boundary condition, corresponding to a unit velocity, is imposed on the inflow part of the boundary and a Dirichlet boundary condition corresponding to zero potential on the outflow. On the rest of the boundary a homogeneous Neumann boundary condition is enforced to represent a physical wall.
After solving the problem with \texttt{HDGlab}, the flow potential and the velocity field are obtained. Figure~\ref{fig:uavSol} shows the magnitude of the velocity field on the surface of the UAV and the pressure field, computed using the Bernoulli equation.

This last example shows the applicability of the developed \texttt{HDGlab} to problems involving complex geometries, where the resulting global system has more than one million of equations. Despite the code was not developed with computational efficiency in mind, it still enables the interested users to solve relatively large problems and, with some additional improvements in terms of performance, it can be used for larger three dimensional problems.

\section{Concluding remarks} \label{sc:conclusions}

An open-source Matlab implementation of the HDG method for elliptic problems has been presented, the so-called \texttt{HDGlab}. The code is capable of solving problems governed by the Poisson and Stokes equations using high-order simplicial elements, including curved elements, by means of an isoparametric formulation.

\texttt{HDGlab} provides a suitable environment to those interested in the programming of hybrid methods in general and the HDG method in particular. The paper describes the HDG formulation employed for the Poisson and Stokes solvers and provides a very detailed description of the code, with particular emphasis in the data structures utilised. The presentation of the code involves the preprocess, computation and postprocess stages, and includes detailed explanations of the most relevant functions. A set of examples is presented to illustrate the use and the potential of \texttt{HDGlab}. The examples go from simple test cases with a known analytical solution, used to demonstrate the numerical properties of the HDG method, to more complex such as the computation of the potential flow around a UAV using high-order curved meshes.

\texttt{HDGlab} is available as an open-source software, released under the terms of the GNU General Public License version 3.0 or any later version (\url{https://www.gnu.org/licenses}) and is freely available from the repository: \url{https://git.lacan.upc.edu/hybridLab/HDGlab}.

\appendix 

\section{Matrices and vectors for the Poisson solver}
\label{app:Poisson}

In this appendix, the expressions of the matrices and vectors appearing in the discrete form of the HDG local and global problems for the Poisson equation are presented.

An isoparametric formulation is considered and the coordinates $\bXi {=} \{\xi_1,\ldots,\xi_{\nsd} \}^T$ of a reference element $\widetilde{\Omega}(\bXi)$ are mapped to the coordinates $\bx {=} \{x_1,\ldots,x_{\nsd} \}^T$ of the local element $\Omega(\bx)$ by the transformation
$$
\bx(\bXi) = \sum_{i=1}^{\nen} N_i(\bXi) \bx_i ,
$$
where $\{\bx_i\}_{i=1,\ldots,\nen}$ denotes the vector of the nodal coordinates of the element. Hence, the isoparametric approximations introduced in~\eqref{eq:isoParam} are defined in a reference element $\widetilde{\Omega}(\bXi)$ for $u$ and $\bq$ and on a reference face $\widetilde{\Gamma}(\bEta)$, $\bEta {=} \{\eta_1,\ldots,\eta_{\nsd-1} \}^T$ for $\hu$, the corresponding shape functions being $N(\bXi)$ and $\hN(\bEta)$, respectively.

For the sake of readability, introduce the compact forms of the shape functions and their derivatives, namely
\begin{subequations}\label{eq:shapeFunction}
\begin{align}
  \Nmat & := \begin{bmatrix} N_1 & N_2 & \dots & N_{\nen} \end{bmatrix}^T , \\
  \NmatHat &:= \begin{bmatrix} \hN_1 & \hN_2 & \dots & \hN_{\nfn} \end{bmatrix}^T , \\
  \NmatSub[n] & := \begin{bmatrix} N_1\bn & N_2\bn & \dots & N_{\nen}\bn \end{bmatrix}^T , \\
  \NmatSub[\nsd] &:= \begin{bmatrix} N_1\Insd & N_2\Insd & \dots & N_{\nen}\Insd \end{bmatrix}^T , \\
  \Qmat & := \begin{bmatrix} (\bJ^{-1} \grad \! N_1)^T & (\bJ^{-1} \grad \! N_2)^T& \dots &  (\bJ^{-1} \grad \! N_{\nen})^T\end{bmatrix}^T , 
\end{align}
\end{subequations}
where $\bn$ denotes the outward unit normal vector to a face and $\bJ$ is the Jacobian of the isoparametric mapping. 

The solution of the HDG local problem~\eqref{eq:PoissonHDGdiscreteLocal} involves the matrices and vectors
\begin{align*}
  [\mat{A}_{uu}]_e := & \sumfa \tauF 
                                   \sumgf \Nmat(\bXigf) \Nmat^T\!(\bXigf) \abs{\bJ(\bXigf)} \wgf , \\[-0.5ex]
[\mat{A}_{u q}]_e := & \sqrt{\kappa} \sumge \Nmat(\bXige) \Qmat^T\!(\bXige) \abs{\bJ(\bXige)} \wge , \\[-0.5ex]
[\mat{A}_{qq} ]_e := & -\sumge \NmatSub[\nsd](\bXige) \NmatSub[\nsd]^T\!(\bXige) \abs{\bJ(\bXige)} \wge , \\[-0.5ex]
[\mat{A}_{u \hu}]_e := & \sumfa \tauF 
                                      \biggl(  \sumgf \Nmat(\bXigf) \NmatHat^T\!\!(\bXigf) \abs{\bJ(\bXigf)} \wgf \biggr)
                                      \bigl(1 - \chiF[D] \bigr) , \\[-0.5ex]
[\mat{A}_{q \hu}]_e := & \sqrt{\kappa} \sumfa 
                                      \biggl(  \sumgf \NmatSub[n](\bXigf) \NmatHat^T\!\!(\bXigf) \abs{\bJ(\bXigf)} \wgf \biggr)
                                      \bigl(1 - \chiF[D] \bigr) , \\[-0.5ex]
[\vecF[u]]_e := &\sumge \Nmat(\bXige) s\bigl(\bx(\bXige) \bigr) \abs{\bJ(\bXige)} \wge
						   + \sumfa \tauF 
                               \biggl(  \sumgf \Nmat(\bXigf) \uD\!\bigl(\bx(\bXigf) \bigr) \abs{\bJ(\bXigf)} \wgf \biggr)
                               \chiF[D] , \\[-0.5ex]
[\vecF[q]]_e := & \sqrt{\kappa} \sumfa 
                              \biggl(  \sumgf \NmatSub[n](\bXigf) \uD\!\bigl(\bx(\bXigf) \bigr) \abs{\bJ(\bXigf)} \wgf \biggr)
                              \chiF[D] , 
\end{align*}
where $\numfaE$ denotes the number of faces of the element $\Omega_e$, $\bXige$ are the $\nipe$ integration points defined on the reference element and $\bXigf$ are the $\nipf$ integration points of the reference face. The corresponding weights for the integration points are denoted by $\wge$ and $\wgf$, respectively. In addition, the indicator function $\chiF[\square]$ is defined as
\begin{equation} \label{eq:indicatorFunction}
\chiF[\square] := \begin{cases}
  1 & \text{ if face \texttt{f} belongs to $\Ga[\square]$} , \\
  0 & \text{ otherwise} .
  \end{cases}
\end{equation}

Similarly, for the HDG global problem~\eqref{eq:PoissonHDGdiscreteGlobal} the following matrix and vector
\begin{align*}
[\mat{A}_{\hu \hu}]_e := & - \sumfa \tauF 
                                              \biggl(  \sumgf \NmatHat(\bXigf) \NmatHat^T\!\!(\bXigf) \abs{\bJ(\bXigf)} \wgf \biggr)
                                               \bigl(1 - \chiF[D] \bigr) , \\[-0.5ex]
[\vecF[\hu]]_e := & - \sumfa 
                               \biggl(  \sumgf \NmatHat(\bXigf) g\bigl(\bx(\bXigf) \bigr) \abs{\bJ(\bXigf)} \wgf \biggr)
                               \chiF[N] , 
\end{align*}
are defined.

To describe the terms appearing in the HDG postprocess~\eqref{eq:PoissonHDGdiscretePostProcess}, the compact forms
\begin{subequations}\label{eq:shapeFunctionStar}
\begin{align} 
  \NmatStar & := \begin{bmatrix} N_1^{\star} & N_2^{\star} & \dots & N_{\nenS}^{\star} \end{bmatrix}^T , \\
  \NmatStarSub[\nsd] & := \begin{bmatrix} N_1^{\star} \Insd & N_2^{\star} \Insd & \dots & N_{\nenS}^{\star} \Insd \end{bmatrix}^T , \\
  \QmatStar & := \begin{bmatrix} (\bJs^{-1} \grad \! N_1^{\star})^T & \! (\bJs^{-1} \grad \! N_2^{\star})^T& \! \dots \! & \! (\bJs^{-1} \grad \! N_{\nenS}^{\star})^T\end{bmatrix}^T , 
\end{align}
\end{subequations}
are introduced, where $\{ N_i^{\star} \}_{i=1,\ldots,\nenS}$ denote the shape functions for the discretisation of $\uS$, $\bJs$ is the Jacobian of the corresponding  isoparametric transformation and $\nenS$ is the number of elemental nodes of the approximation.

The following matrices and vector are thus defined as
\begin{align*}
[\mat{A}_{\star\star}]_e := & \kappa \sumgS \QmatStar(\bXigS) \QmatStar^T\!(\bXigS) \abs{\bJs\!(\bXigS)} \wgS , \\[-0.5ex]
[\mat{a}_{\star\lambda}]_e := & \sumgS \NmatStar(\bXigS) \abs{\bJs\!(\bXigS)} \wgS , \\[-1.5ex]
[\mat{A}_{\star q}]_e := & - \sqrt{\kappa} \sumgS \QmatStar(\bXigS) \NmatStarSub[\nsd]^T\!(\bXigS) \abs{\bJs\!(\bXigS)} \wgS , 
\end{align*}
where $\bXigS$ and $\wgS$ are the $\nipS$ integration points and the weights associated with the higher order approximation in the space $\VhS$.

\section{Matrices and vectors for the Stokes solver}
\label{app:Stokes}

In this appendix, the expressions of the matrices and vectors appearing in the discrete form of the HDG local and global problems for the Stokes equations are presented. It is worth recalling that in this problem, the primal, $\vect{u}$, and hybrid, $\vect{\hu}$, variables are $\nsd$-dimensional vector-valued unknowns and the mixed variable $\vect{L}$ is a tensor-valued unknown of dimension $\nsd {\times} \nsd$.

In addition to the compact forms of the shape functions presented in equations~\eqref{eq:shapeFunction} and~\eqref{eq:shapeFunctionStar}, the following definitions are introduced for the vectorial problem
\begin{subequations}\label{eq:shapeFunctionVectorial}
\begin{align}
  \NmatSub[\msd] &:= \begin{bmatrix} N_1 \Imsd & N_2 \Imsd & \dots & N_{\nen}  \Imsd \end{bmatrix}^T , \\
  \NmatStarSub[\msd] & := \begin{bmatrix} N_1^{\star} \Imsd & N_2^{\star} \Imsd & \dots & N_{\nenS}^{\star} \Imsd \end{bmatrix}^T , \\
  \NmatHatSub[\nsd] &:= \begin{bmatrix} \hN_1 \Insd & \hN_2 \Insd & \dots & \hN_{\nfn}  \Insd \end{bmatrix}^T , \\
  \NmatHatSub[E] &:= \begin{bmatrix} \hN_1 \bE & \hN_2 \bE & \dots & \hN_{\nfn}  \bE \end{bmatrix}^T , \\
  \NmatHatSub[D] &:= \begin{bmatrix} \hN_1 \bD & \hN_2 \bD & \dots & \hN_{\nfn}  \bD \end{bmatrix}^T , 
\end{align}
where $\msd {:=} \nsd^2$. Moreover, for each component $j {=} 1, \ldots, \nsd$, the compact forms
\begin{align}
  \NmatSub[n_{\! j}] &:= \begin{bmatrix} N_1 n_j  & N_2 n_j  & \dots & N_{\nen} n_j  \end{bmatrix}^T , \\
  \QmatSub[j] & := \begin{bmatrix} [\bJ^{-1} \grad \! N_1]_j & [\bJ^{-1} \grad \! N_2]_j & \dots &  [\bJ^{-1} \grad \! N_{\nen}]_j \end{bmatrix}^T , \\
  \QmatStarSub[j] & := \begin{bmatrix} [\bJs^{-1} \grad \! N_1^{\star}]_j & [\bJs^{-1} \grad \! N_2^{\star}]_j & \dots &  [\bJs^{-1} \grad \! N_{\nenS}^{\star}]_j \end{bmatrix}^T , 
\end{align}
\end{subequations}
are defined, $n_j$ being the $j$-th component of the outward unit normal vector $\bn$ to the face.

The solution of the HDG local problem~\eqref{eq:StokesHDGdiscreteLocal} involves the matrices and vectors
\begin{align*}
  [\mat{A}_{LL}]_e := & -\sumge \NmatSub[\msd](\bXige) \NmatSub[\msd]^T\!(\bXige) \abs{\bJ(\bXige)} \wge , \\[-0.5ex]
  [\mat{A}_{Lu}]_e := & \sqrt{\nu} \sumJ \sumge
  									\QmatSub[j](\bXige) \NmatSub[\nsd]^T\!(\bXige) \abs{\bJ(\bXige)} \wge , \\[-0.5ex]
  [\mat{A}_{uu}]_e := & \sumfa \tauF 
                                   \sumgf \NmatSub[\nsd](\bXigf) \NmatSub[\nsd]^T\!(\bXigf) \abs{\bJ(\bXigf)} \wgf , 
                                   \\[-0.5ex]
  [\mat{A}_{p u}]_e := & \sumge \Qmat(\bXige) \NmatSub[\nsd]^T\!(\bXige) \abs{\bJ(\bXige)} \wge , \\[-0.5ex]
  [\mat{a}_{\rho p} ]_e := & \sumfa 
                                          \sumgf \Nmat(\bXigf) \abs{\bJ(\bXigf)} \wgf , \\[-0.5ex]
  [\mat{A}_{L \hu}]_e := & \sqrt{\nu} \sumfa 
                                      \biggl(  \sumJ \sumgf \NmatSub[n_{\! j}](\bXigf) \NmatHatSub[\nsd]^T\!(\bXigf) \abs{\bJ(\bXigf)} \wgf \biggr)
                                      \bigl(1 - \chiF[D] \bigr) , 
                                      \\[-0.5ex]
  [\mat{A}_{u \hu}]_e := & \sumfa \tauF 
                                      \biggl(  \sumgf \NmatSub[\nsd](\bXigf) \NmatHatSub[\nsd]^T\!(\bXigf) \abs{\bJ(\bXigf)} \wgf \biggr)
                                      \bigl(1 - \chiF[D] \bigr) , \\[-0.5ex]
  [\mat{A}_{p \hu}]_e := & \sumfa 
                                      \biggl(  \sumgf \NmatSub[n](\bXigf) \NmatHatSub[\nsd]^T\!(\bXigf) \abs{\bJ(\bXigf)} \wgf \biggr)
                                      \bigl(1 - \chiF[D] \bigr) , 
\end{align*}
\begin{align*}
  [\vecF[L]]_e := & \sqrt{\nu} \sumfa 
                                      \biggl(  \sumJ \sumgf \NmatSub[n_{\! j}](\bXigf) \buD\!\bigl(\bx(\bXigf) \bigr) \abs{\bJ(\bXigf)} \wgf \biggr)
                                      \chiF[D] , \\[-0.5ex]
  [\vecF[u]]_e := &\sumge \NmatSub[\nsd](\bXige) \bm{s}\bigl(\bx(\bXige) \bigr) \abs{\bJ(\bXige)} \wge 
		 				   + \sumfa \tauF 
                               \biggl(  \sumgf \NmatSub[\nsd](\bXigf) \buD\!\bigl(\bx(\bXigf) \bigr) \abs{\bJ(\bXigf)} \wgf \biggr)
                               \chiF[D] , \\
  [\vecF[p]]_e := & \sumfa 
                              \biggl(  \sumgf \NmatSub[n](\bXigf) \buD\!\bigl(\bx(\bXigf) \bigr) \abs{\bJ(\bXigf)} \wgf \biggr)
                              \chiF[D] .
\end{align*}

Similarly, the matrices and vectors for the global problem~\eqref{eq:StokesHDGdiscreteGlobal} are
\begin{align*}
  [\mat{A}_{\hu L}]_e := & \sqrt{\nu} \sumfa \Bigg\{ {-}
                                        \biggl(  \sumJ \sumgf \NmatHatSub[E](\bXigf) \NmatSub[n_{\! j}]^T\!(\bXigf) \abs{\bJ(\bXigf)} \wgf \biggr)
                                      \chiF[S] \\[-1.5ex]
  										& {+} \biggl(  \sumJ \sumgf \NmatHatSub[\nsd](\bXigf) \NmatSub[n_{\! j}]^T\!(\bXigf) \abs{\bJ(\bXigf)} \wgf \biggr)
                                      \! \bigl(1 - \chiF[D] \bigr) \! \bigl(1 - \chiF[S] \bigr) \! \Bigg\} , \\[-0.5ex]
  [\mat{A}_{\hu u}]_e := & \sumfa \tauF \Bigg\{ {-}
                                      \biggl(  \sumgf \NmatHatSub[E](\bXigf) \NmatSub[\nsd]^T\!(\bXigf) \abs{\bJ(\bXigf)} \wgf \biggr)
                                      \chiF[S] \\[-1.5ex]
                                      & {+} \biggl(  \sumgf \NmatHatSub[\nsd](\bXigf) \NmatSub[\nsd]^T\!(\bXigf) \abs{\bJ(\bXigf)} \wgf \biggr)
                                      \bigl(1 - \chiF[D] \bigr) \bigl(1 - \chiF[S] \bigr) \Bigg\} , \\[-0.5ex]
  [\mat{A}_{\hu p}]_e := & \sumfa \Bigg\{ {-}
                                      \biggl(  \sumgf \NmatHatSub[E](\bXigf) \NmatSub[n]^T\!(\bXigf) \abs{\bJ(\bXigf)} \wgf \biggr)
                                      \chiF[S] \\[-1.5ex]
                                      & {+} \biggl(  \sumgf \NmatHatSub[\nsd](\bXigf) \NmatSub[n]^T\!(\bXigf) \abs{\bJ(\bXigf)} \wgf \biggr)
                                      \bigl(1 - \chiF[D] \bigr) \bigl(1 - \chiF[S] \bigr) \Bigg\} , \\[-0.5ex]
  [\mat{A}_{\hu \hu}]_e := & \sumfa  \Bigg\{
                                      \biggl(  \sumgf \NmatHatSub[D](\bXigf) \NmatHatSub[\nsd]^T\!(\bXigf) \abs{\bJ(\bXigf)} \wgf \biggr)
                                      \chiF[S] \\[-1.5ex]
                                      & {+} \tauF \biggl(  \sumgf \NmatHatSub[E](\bXigf) \NmatHatSub[\nsd]^T\!(\bXigf) \abs{\bJ(\bXigf)} \wgf \biggr)
                                      \chiF[S] \\[-1.5ex]
                                      & {-} \tauF  \biggl(  \sumgf \NmatHatSub[\nsd](\bXigf) \NmatHatSub[\nsd]^T\!(\bXigf) \abs{\bJ(\bXigf)} \wgf \biggr)
                                      \bigl(1 - \chiF[D] \bigr) \bigl(1 - \chiF[S] \bigr) \Bigg\} , \\[-0.5ex]
  [\vecF[\hu]]_e := & - \sumfa 
                               \biggl(  \sumgf \NmatHatSub[\nsd](\bXigf) \bm{g}\bigl(\bx(\bXigf) \bigr) \abs{\bJ(\bXigf)} \wgf \biggr)
                               \chiF[N] .
\end{align*}

In addition, if a problem with purely Dirichlet boundary conditions is solved, the constraint~\eqref{eq:pressureConstraintDomain} gives rise to the vector
\begin{equation}
[\mat{a}_{\overline{\rho} \rho}]_e := \sumge \Nmat(\bXige) \abs{\bJ(\bXige)} \wge ,
\end{equation}
which is used to enforce the constraint using an appropriately defined Lagrange multiplier.

Finally, for the postprocess~\eqref{eq:StokesHDGpostprocess}, the following matrices are required
\begin{align*}
[\mat{A}_{\star\star}]_e := & \nu \sumJ \sumgS \QmatStarSub[j](\bXigS) \QmatStarSub[j]^T\!(\bXigS) \abs{\bJs\!(\bXigS)} \wgS , \\[-0.5ex]
[\mat{A}_{\star\lambda}]_e := & \sumgS \NmatStarSub[\nsd](\bXigS) \abs{\bJs\!(\bXigS)} \wgS , \\[-1.5ex]
[\mat{A}_{\star L}]_e := & - \sqrt{\nu} \sumJ \sumgS \QmatStarSub[j](\bXigS) \NmatStarSub[\msd]^T\!(\bXigS) \abs{\bJs\!(\bXigS)} \wgS .
\end{align*}

\section{An interface with the mesh generator \texttt{Gmsh}}
\label{app:Gmsh}

\begin{figure*}[!tb]
	\centering
	\includegraphics[width=0.9\textwidth]{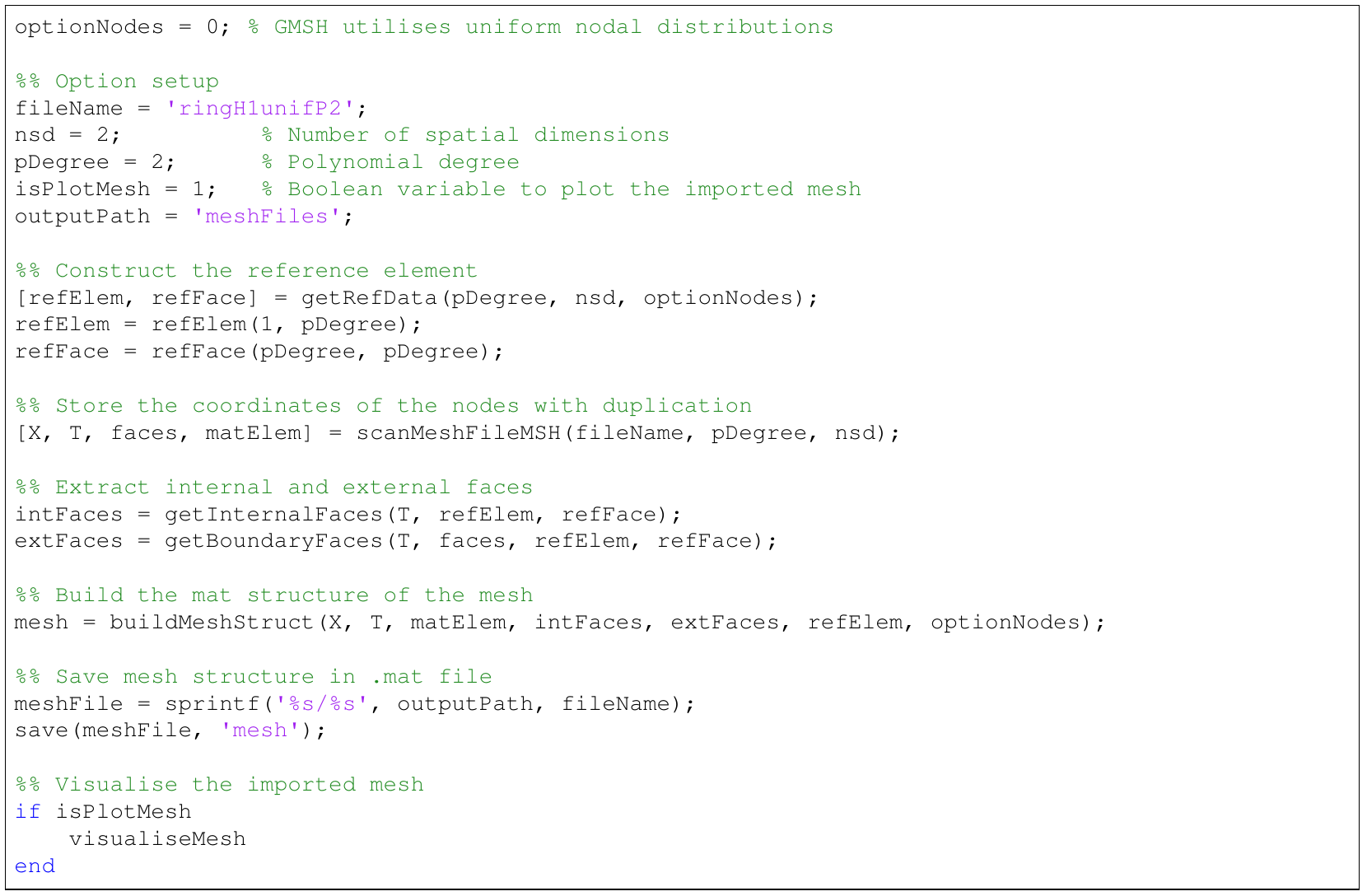}
	\caption{Function \texttt{convertMSHtoMAT} to convert the mesh file from \textit{.msh} to \textit{.mat} format.}
	\label{fig:convertMesh}
\end{figure*}

In this appendix, the interface between the high-order open-source mesh generator \texttt{Gmsh}~\cite{GMSH-09} and \texttt{HDGlab} is described. Starting from the structure of the variable \texttt{mesh} introduced in section~\ref{sc:mesh}, the algorithm to convert a mesh file from the \textit{.msh} to the \textit{.mat} format of \texttt{HDGlab} is presented (Fig.~\ref{fig:convertMesh}).
\begin{remark}
The routines described in this appendix are designed starting from the \textit{.msh} ASCII file format version 2.2.
\end{remark}

First, it is worth recalling that \texttt{HDGlab} offers the feature of utilising either equally-spaced or Fekete points for the approximation. \texttt{Gmsh} provides mesh discretisation based on equally-spaced nodal sets, whence the variable \texttt{optionNodes} is set to 0, see section~\ref{sc:mesh}.

The function \texttt{convertMSHtoMAT} requires the definition of the following data concerning the mesh to be imported:
\begin{itemize}
\item \texttt{fileName}: String that specifies the name of the \textit{.msh} mesh file without extension. The mesh files are archived in the directory \texttt{example}.
\item \texttt{nsd}: Scalar variable defining the number of spatial dimensions (either 2 or 3).
\item \texttt{pDegree}: Scalar variable defining the degree of the polynomial order of the mesh.
\item \texttt{isPlotMesh}: Boolean variable to activate the option for visualising the imported mesh with the nodal distribution.
\item \texttt{outputPath}: String that specifies the location where the imported mesh will be stored. The default location is the directory \texttt{meshFiles}.
\end{itemize}

Given the number of spatial dimensions and the polynomial degree defined above, the \texttt{refElem} and \texttt{refFace} data structures are constructed.

\subsection{Reading the \textit{.msh} file} \label{sc:meshReadFile}

\begin{figure}[!tb]
	\centering
	\includegraphics[width=0.45\textwidth]{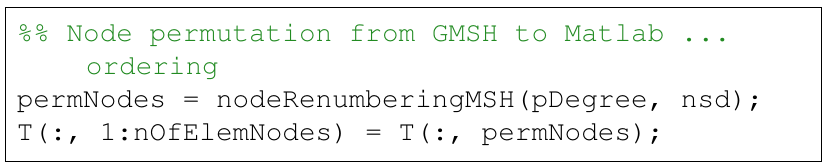}
	\caption{Extract of the \texttt{scanMeshFileMSH} function that constructs the permutation vector for the appropriate renumbering of the mesh nodes.}
	\label{fig:nodePermutation}
\end{figure}

The \textit{.msh} file is read by the function \texttt{scanMeshFileMSH} and the following information is extracted:
\begin{itemize}
\item \texttt{X}: Array containing the coordinates of the mesh nodes.
\item \texttt{T}: Connectivity matrix featuring the identifiers of the nodes associated with each element.
\item \texttt{faces}: Array containing the information on the boundary faces, namely the identifier of the element the face belongs to, the list of vertices, the flags of the boundary and of the type of boundary condition imposed.
\item \texttt{matElem}: Array containing the flag of the region each element belongs to. 
\end{itemize}
A detailed description of the structure of the \textit{.msh} file as well as the notation utilised by the mesh generator to identify element and face types is available in the official \texttt{Gmsh} documentation~\cite{GMSH-DOC}. It is worth noting that the numbering of the nodes in \texttt{HDGlab} differs from the one provided by \texttt{Gmsh}. Hence, an appropriate permutation is performed as reported in figure~\ref{fig:nodePermutation}.

\subsection{Retrieving the information on the faces} \label{sc:meshConstructFace}

As described in section~\ref{sc:mesh}, the interior and exterior faces of the mesh are stored in the data structures \texttt{intFaces} and \texttt{extFaces}, respectively.

\begin{figure*}[!tb]
	\centering
	\includegraphics[width=0.9\textwidth]{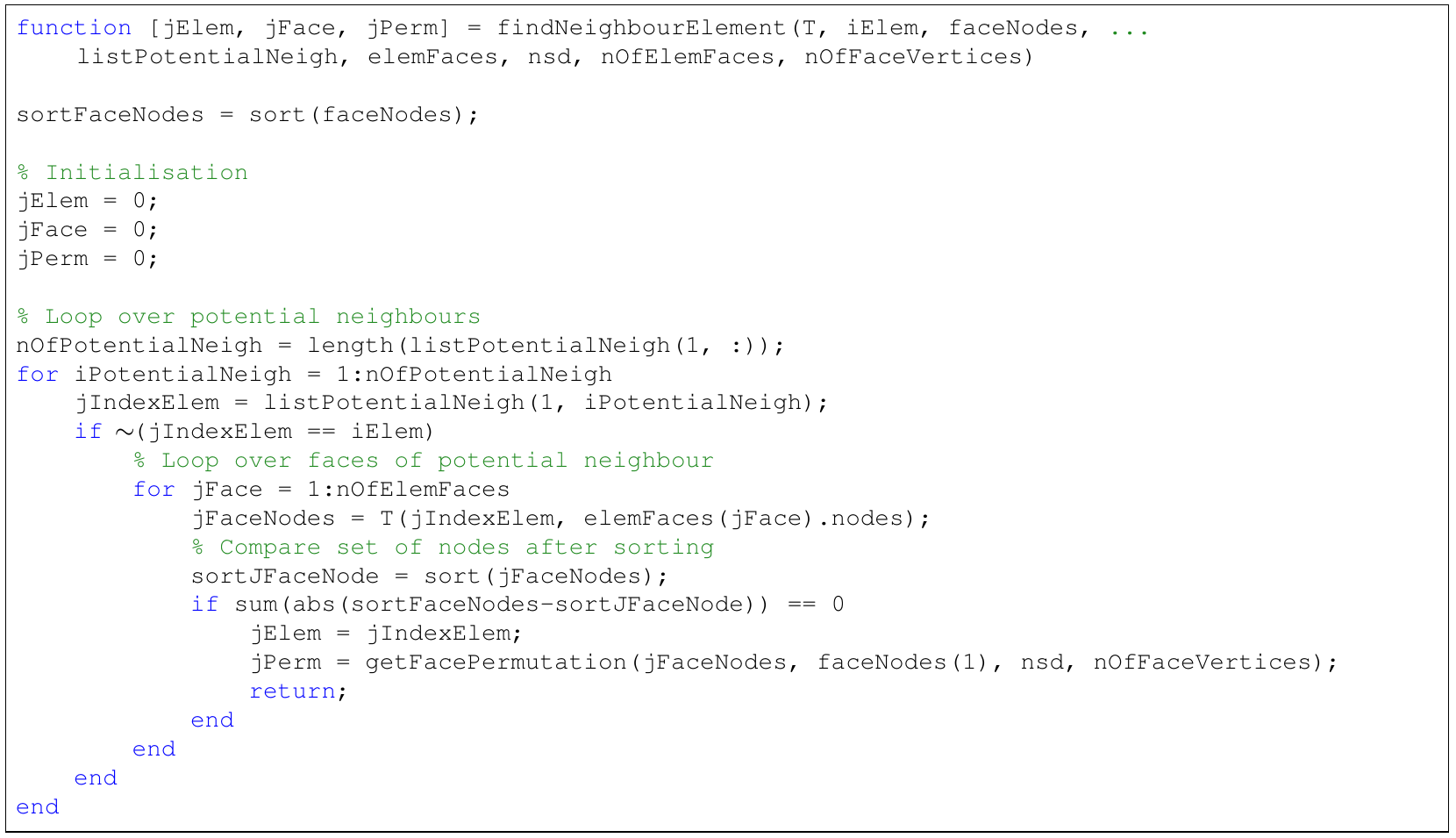}
	\caption{Function \texttt{findNeighbourElement} that identifies the element sharing a face of a given element.}
	\label{fig:findNeighbour}
\end{figure*}

\begin{figure}[!tb]
	\centering
	\includegraphics[width=0.45\textwidth]{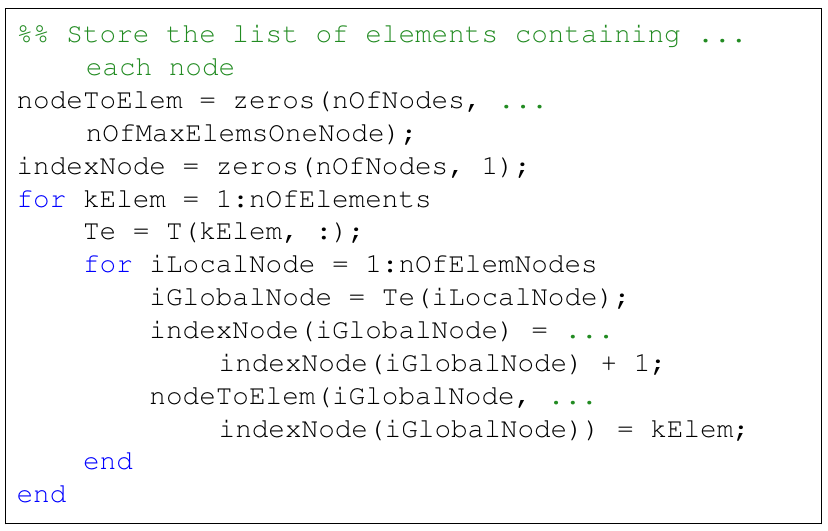}
	\caption{Extract of the \texttt{getInternalFaces} function that constructs the list of elements containing each node.}
	\label{fig:potentialNeighbour}
\end{figure}

To construct the structure \texttt{intFaces}, the \texttt{getInternalFaces} function explores the mesh and, for each face of each element, it identifies the neighbouring element by comparing the list of face nodes, see figure~\ref{fig:findNeighbour}. In order to optimise this operation, a list of potential neighbours is preliminarily stored by identifying the list of  elements containing each node, as detailed by the extract in figure~\ref{fig:potentialNeighbour}.

The function \texttt{getBoundaryFaces}, not reported here for brevity, is responsible for constructing the data structure of the exterior faces. More precisely, the structure \texttt{extFaces} is obtained by rearranging the information previously stored in the data structure \texttt{faces}, according to the rationale described in section~\ref{sc:mesh}.

\subsection{Assemblying the mesh structure} \label{sc:meshConstructMesh}

\begin{figure}[!tb]
	\centering
	\includegraphics[width=0.45\textwidth]{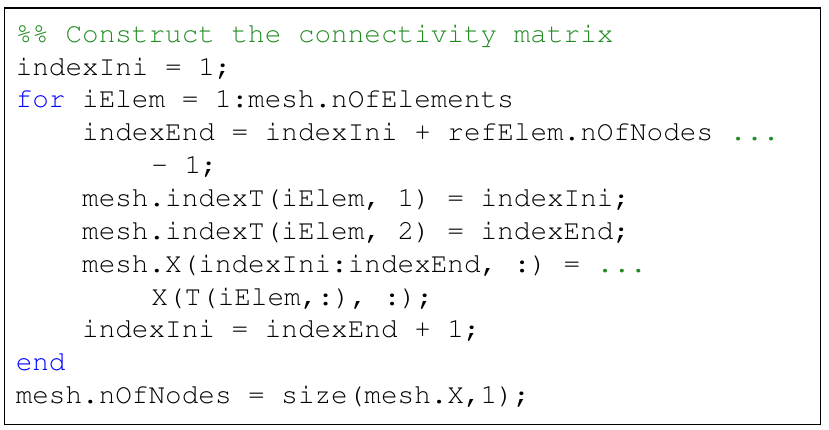}
	\caption{Extract of the \texttt{buildMeshStruct} function that constructs the connectivity matrix.}
	\label{fig:constructConnectivity}
\end{figure}

The structure \texttt{mesh} is finally assembled by the \texttt{buildMeshStruct} function using the information obtained by the mesh generator, namely \texttt{X}, \texttt{T} and \texttt{matElem}, and the structures of the interior and exterior faces, \texttt{intFaces} and \texttt{extFaces}, previously generated. More precisely, the connectivity matrix is constructed via a loop on the mesh elements as reported in figure~\ref{fig:constructConnectivity}.

It is worth noting that the framework provided in \texttt{HDGlab} to import meshes generated using \texttt{Gmsh} can be easily extended to any mesh generator by introducing appropriate functions to construct the \texttt{intFaces} and \texttt{extFaces} data structures, as described above.

\subsection{Some examples of high-order meshes using \texttt{Gmsh}} \label{sc:meshExamples}

\begin{figure*}
\centering
\subfigure[Elasticity: nodal distribution]{\includegraphics[width=0.35\textwidth]{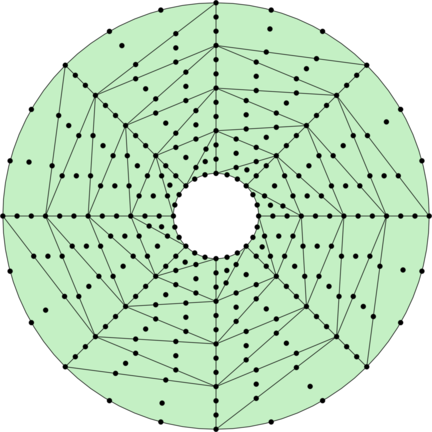}}
\hspace{40pt}
\subfigure[Optimisation: nodal distribution]{\includegraphics[width=0.35\textwidth]{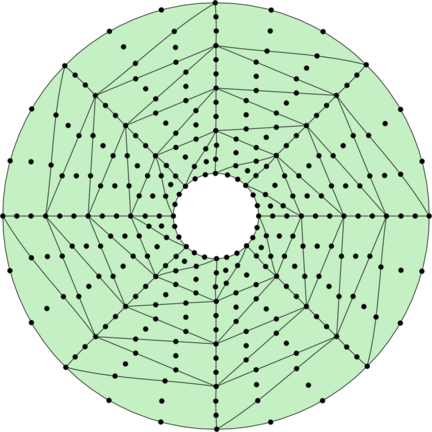}}

\subfigure[Elasticity: $h$-convergence of $p$ and $\bL$]{\includegraphics[width=0.45\textwidth]{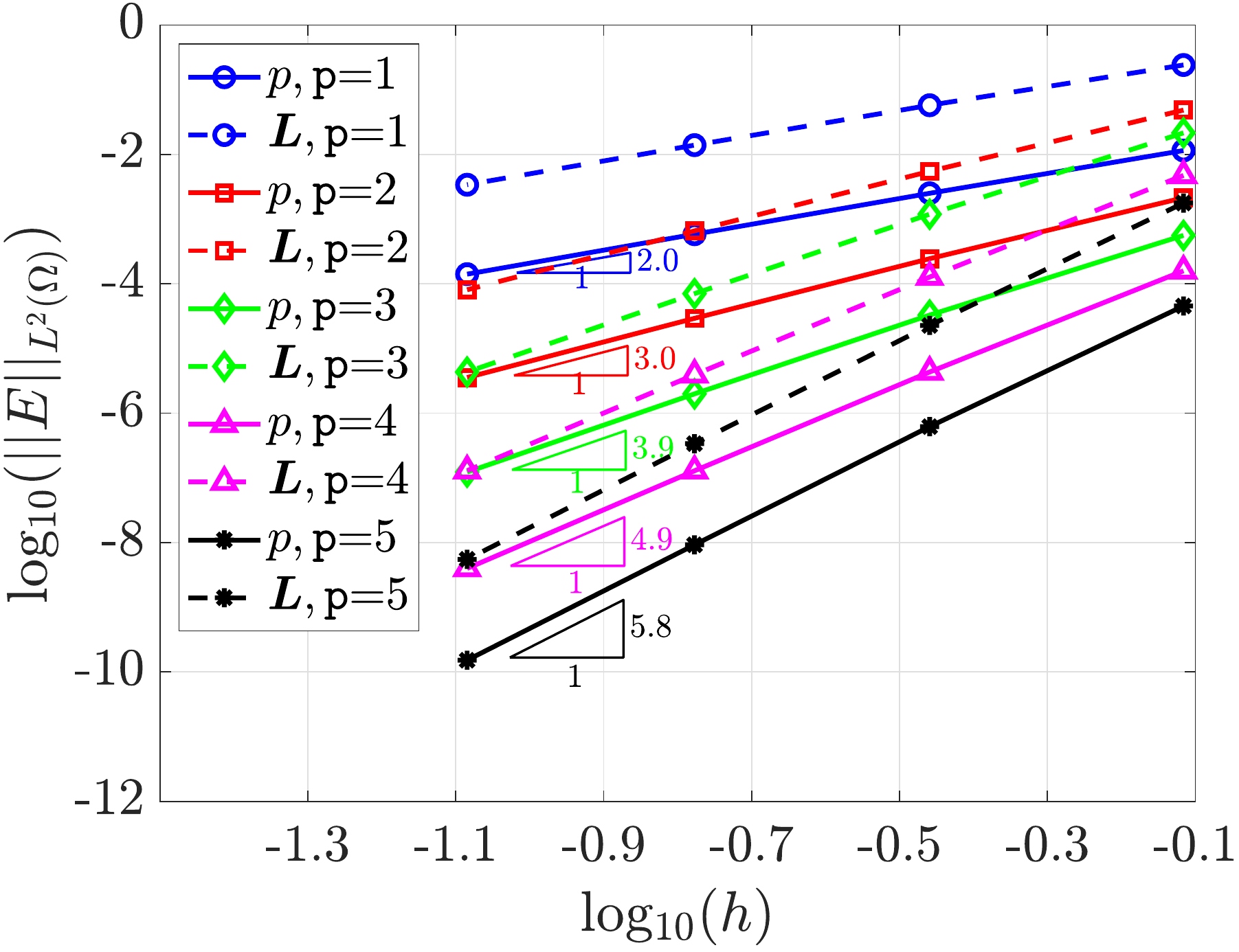}}
\subfigure[Optimisation: $h$-convergence of $p$ and $\bL$]{\includegraphics[width=0.45\textwidth]{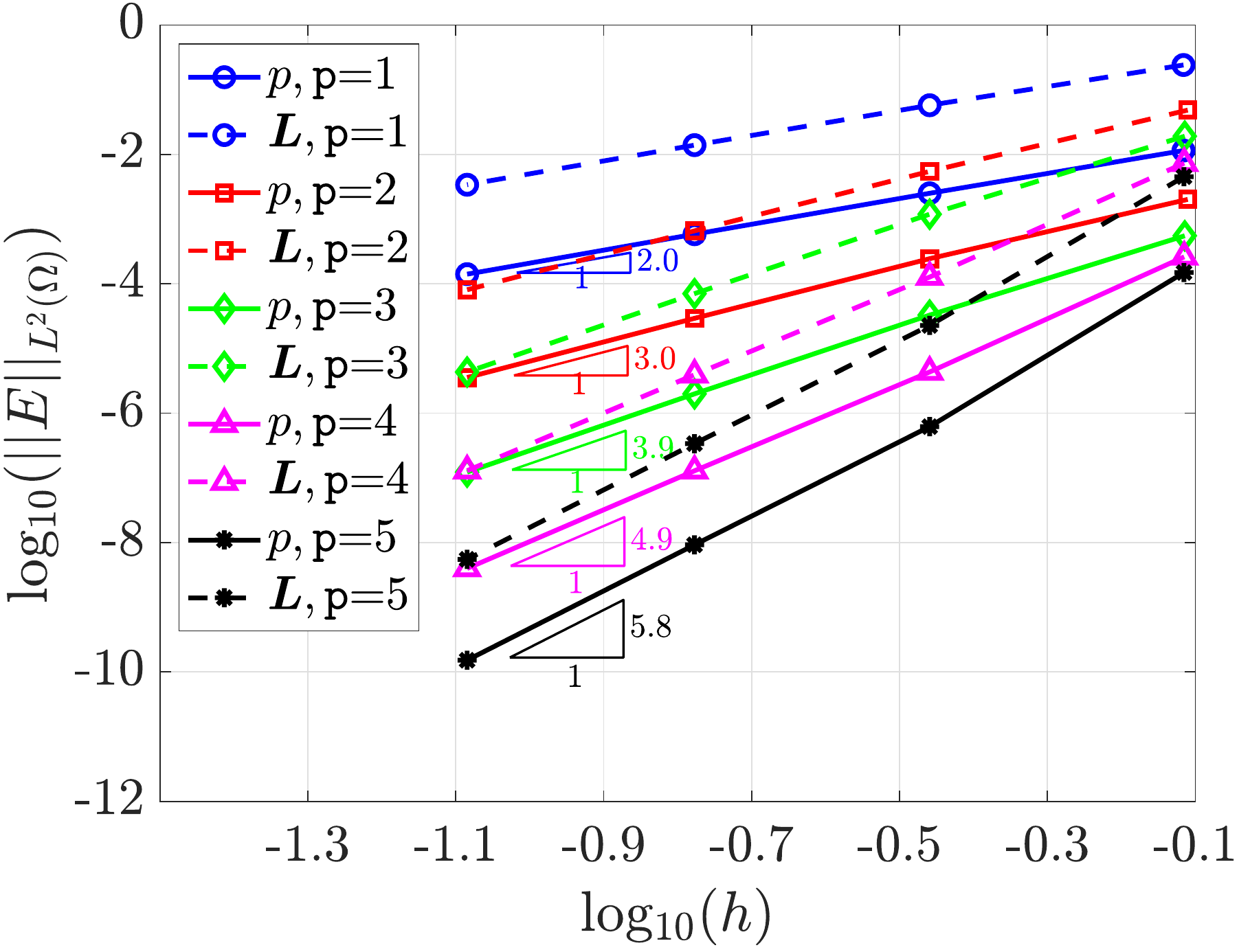}}

\subfigure[Elasticity: $h$-convergence of $\bu$ and $\buS$]{\includegraphics[width=0.45\textwidth]{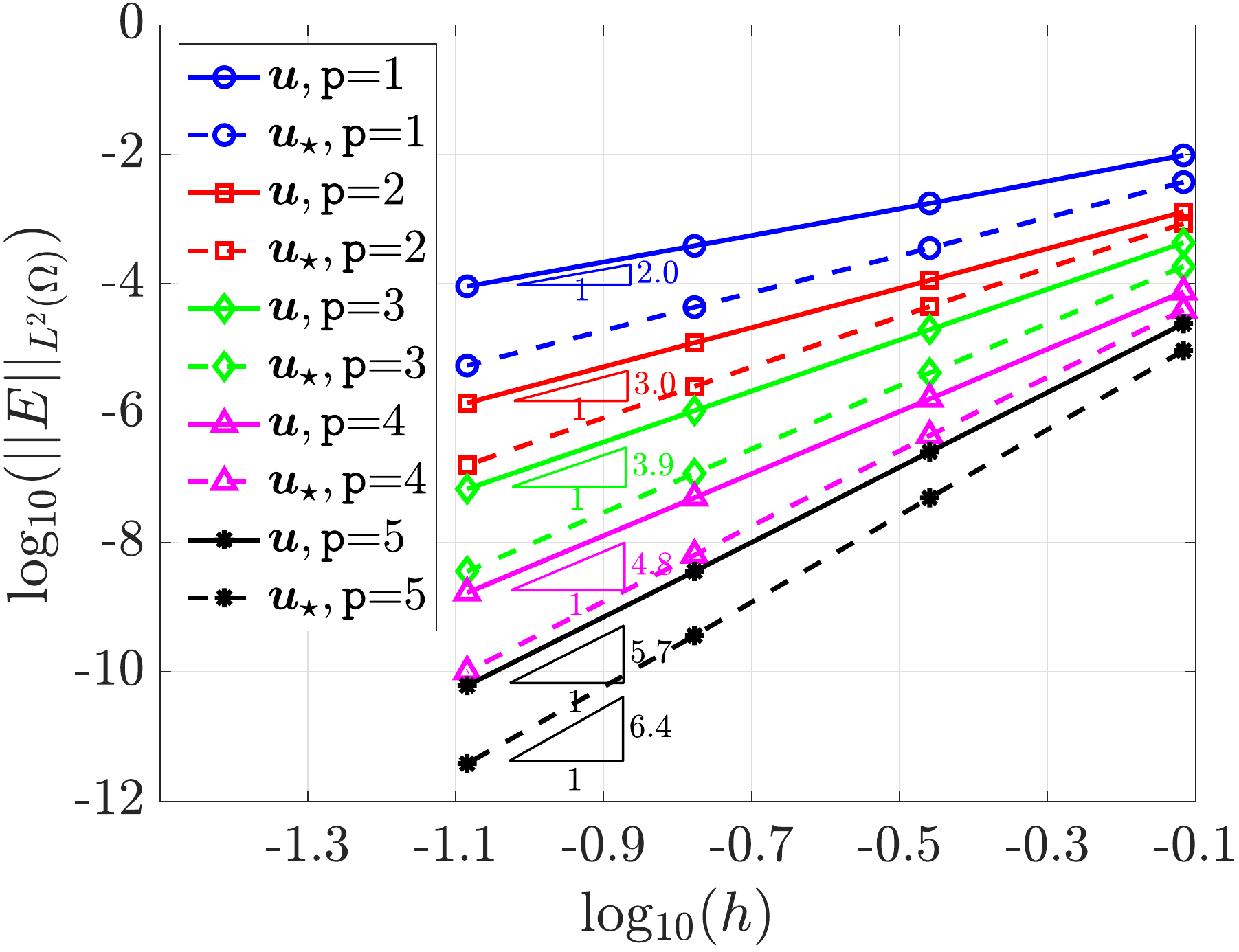}}
\subfigure[Optimisation: $h$-convergence of $\bu$ and $\buS$]{\includegraphics[width=0.45\textwidth]{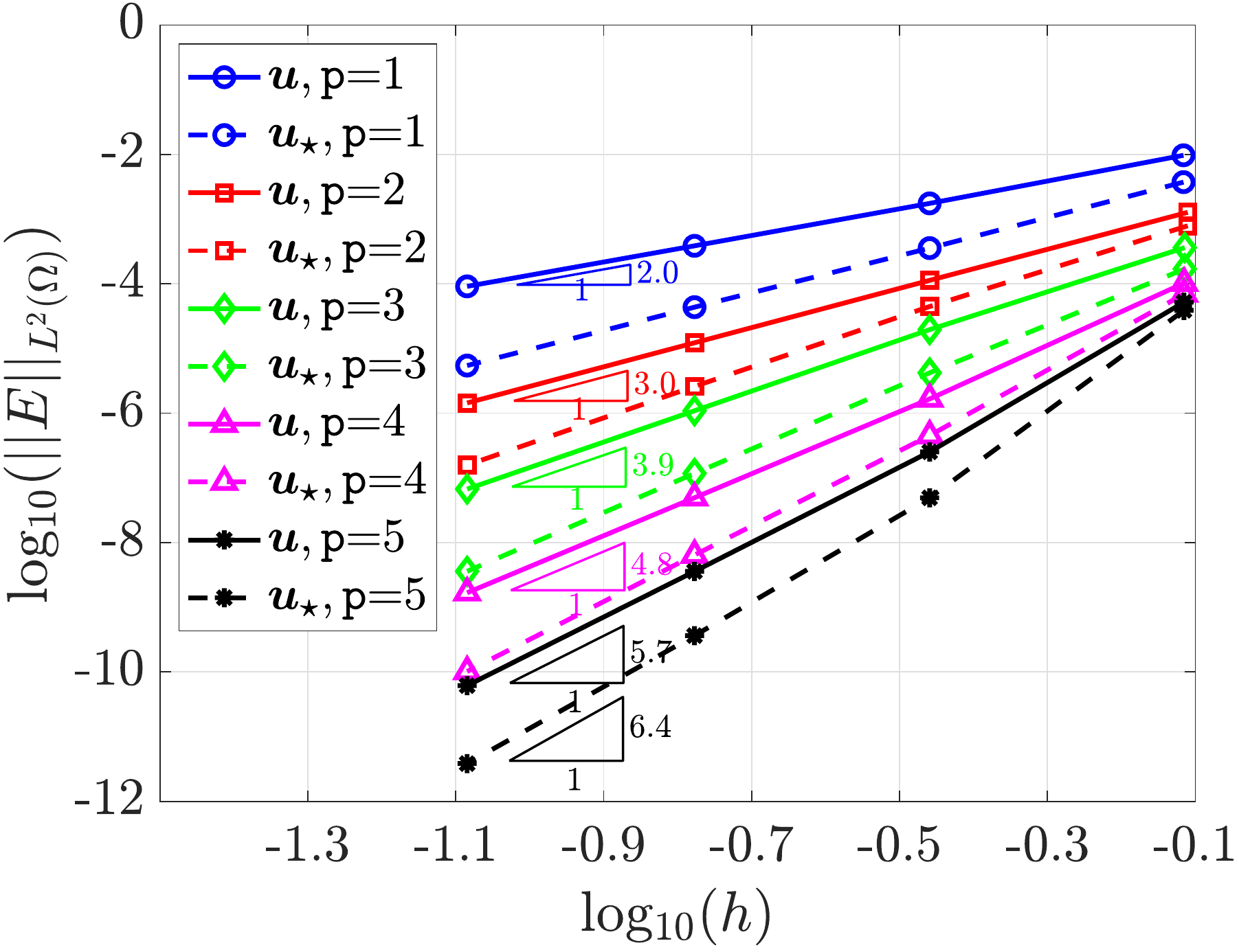}}
\caption{Comparison of high-order meshes generated by \texttt{Gmsh} using the elastic approach (left) and the optimisation algorithm (right) for the Stokes equation in a ring domain. (a-b) Nodal distributions for meshes of degree 3. Convergence of the $\eltwo(\Omega)$ error of (c-d) pressure, $p$, and mixed variable, $\bL$, and (e-f) primal, $\bu$, and postprocessed, $\bu_\star$, velocities as a function of the characteristic mesh size $h$ for polynomial degree of approxiomation $\deg {=} 1, \ldots, 5$.}
\label{fig:couetteGmsh}
\end{figure*}

In this section, several meshes of the ring domain introduced in section~\ref{sc:stokesCouette} are generated using \texttt{Gmsh} and tested to verify the optimal convergence rate of the code when high-order meshes with equally-spaced nodal sets are considered. More precisely, figure~\ref{fig:couetteGmsh} displays the comparison of the third order meshes provided by \texttt{Gmsh} using the \texttt{Mesh.HighOrderOptimize} option either with an elastic approach or an optimisation algorithm~\cite{GMSH-DOC}. It is worth noting that the elastic approach mainly induces the curvature of the boundary of the domain, whereas the optimisation strategy is responsible for curved edges to appear in the interior of the domain as well (Fig.~\ref{fig:couetteGmsh}a-b). In both cases, the nodes are equally-spaced.

In addition, the $\eltwo(\Omega)$ error of the pressure and the gradient of velocity (Fig~\ref{fig:couetteGmsh}c-d) and the primal and postprocessed velocities (Fig~\ref{fig:couetteGmsh}e-f) is reported as a function of the characteristic mesh size. The results diplay the expected optimal convergence of order $\deg {+} 1$ for pressure, velocity and gradient of velocity and the superconvergence of the postprocessed variable $\buS$, showing the capability of \texttt{HDGlab} to work using both equally-spaced and Fekete nodal distributions. It is worth noting that the results obtained using Fekete nodal sets (Fig.~\ref{fig:stokesConvergenceCouette}) provides up to one order of magnitude of extra accuracy for a given mesh size compared to the equally-spaced nodes in figure~\ref{fig:couetteGmsh}.

\section*{Acknowledgements}

This work was partially supported by the Spanish Ministry of Economy and Competitiveness (Grant number: DPI2017-85139-C2-2-R). M.G. and A.H. are also grateful for the support provided by the Spanish Ministry of Economy and Competitiveness through the Severo Ochoa programme for centres of excellence in RTD (Grant number: CEX2018-000797-S) and the Generalitat de Catalunya (Grant number: 2017-SGR-1278). R.S. also acknowledges the support of the Engineering and Physical Sciences Research Council (Grant number: EP/T009071/1).

\section*{Conflict of interest}
The authors declare that they have no conflict of interest.

\bibliographystyle{spmpsci}      
\bibliography{Ref-HDG}   

\begin{thebibliography}{100}
\providecommand{\url}[1]{{#1}}
\providecommand{\urlprefix}{URL }
\expandafter\ifx\csname urlstyle\endcsname\relax
  \providecommand{\doi}[1]{DOI~\discretionary{}{}{}#1}\else
  \providecommand{\doi}{DOI~\discretionary{}{}{}\begingroup
  \urlstyle{rm}\Url}\fi

\bibitem{Aster-DOC}
{\texttt{code\_aster}: Structures and Thermomechanics Analysis for Studies and
  Research}.
\newblock \url{https://code-aster.org/}

\bibitem{Saturne-DOC}
{\texttt{Code\_Saturne}: EDF open-source software to solve computational fluid
  dynamics applications}.
\newblock \url{https://www.code-saturne.org/cms/}

\bibitem{disk-DOC}
{\texttt{DiSk++}: A C++ template library for Discontinuous Skeletal methods}.
\newblock \url{https://github.com/wareHHOuse/diskpp}

\bibitem{Feelpp-DOC}
{\texttt{Feel++}: A poweful, scalable and expressive finite element embedded
  library in C++}.
\newblock \url{http://www.feelpp.org}

\bibitem{FESTUNG-DOC}
{\texttt{FESTUNG}: A MATLAB/GNU Octave toolbox for the discontinuous Galerkin
  method}.
\newblock \url{https://github.com/FESTUNG/FESTUNG}

\bibitem{Firedrake-DOC}
{\texttt{Firedrake}: An automated system for the solution of partial
  differential equations using the finite element method}.
\newblock \url{https://www.firedrakeproject.org}

\bibitem{GetFEM-DOC}
{\texttt{GetFEM}: An open-source finite element library}.
\newblock \url{http://getfem.org}

\bibitem{HArDCore-DOC}
{\texttt{HArDCore}: Hybrid Arbitrary Degree::Core}.
\newblock \url{https://github.com/jdroniou/HArDCore}

\bibitem{HDG3D-DOC}
{\texttt{HDG3D}: Matlab implementation of the Hybridizable Discontinuous
  Galerkin method on general tetrahedrizations of polyhedra in three
  dimensional space}.
\newblock \url{https://github.com/team-pancho/HDG3D}

\bibitem{Nektar-DOC}
{\texttt{Nektar++}: Spectral/hp element framework}.
\newblock \url{https://www.nektar.info}

\bibitem{NGSolve-DOC}
{\texttt{Netgen/NGSolve}: A high performance multiphysics finite element
  software}.
\newblock \url{https://ngsolve.org}

\bibitem{deal-DOC}
\texttt{deal.II}: An open source finite element library.
\newblock \url{https://www.dealii.org}

\bibitem{GMSH-DOC}
\texttt{Gmsh}: A three-dimensional finite element mesh generator with built-in
  pre- and post-processing facilities.
\newblock \url{https://gmsh.info}

\bibitem{MFEM-DOC}
\texttt{MFEM}: Modular finite element methods library.
\newblock \url{https://mfem.org}

\bibitem{Ern-AEP-18}
Abbas, M., Ern, A., Pignet, N.: Hybrid high-order methods for finite
  deformations of hyperelastic materials.
\newblock Computational Mechanics \textbf{62}(4), 909--928 (2018)

\bibitem{Ern-AEP-19b}
Abbas, M., Ern, A., Pignet, N.: A hybrid high-order method for finite
  elastoplastic deformations within a logarithmic strain framework.
\newblock International Journal for Numerical Methods in Engineering
  \textbf{120}(3), 303--327 (2019)

\bibitem{Ern-AEP-19}
Abbas, M., Ern, A., Pignet, N.: A hybrid high-order method for incremental
  associative plasticity with small deformations.
\newblock Computer Methods in Applied Mechanics and Engineering \textbf{346},
  891--912 (2019)

\bibitem{Lanteri-AGGKLM-20}
Agullo, E., Giraud, L., Gob\'{e}, A., Kuhn, M., Lanteri, S., Moya, L.: High
  order {HDG} method and domain decomposition solvers for frequency-domain
  electromagnetics.
\newblock International Journal of Numerical Modelling: Electronic Networks,
  Devices and Fields \textbf{33}(2) (2020)

\bibitem{Fu-AF-18}
Ainsworth, M., Fu, G.: Fully computable a posteriori error bounds for
  hybridizable discontinuous {G}alerkin finite element approximations.
\newblock Journal of Scientific Computing \textbf{77}(1), 443--466 (2018)

\bibitem{MFEM-19}
Anderson, R., Andrej, J., Barker, A., Bramwell, J., Camier, J.S., Cerveny, J.,
  Dobrev, V., Dudouit, Y., Fisher, A., Kolev, T., Pazner, W., Stowell, M.,
  Tomov, V., Dahm, J., Medina, D., Zampini, S.: {MFEM: a modular finite element
  methods library}.
\newblock Tech. rep., arXiv (2019).
\newblock \urlprefix\url{https://arxiv.org/abs/1911.09220}

\bibitem{Monk-ACML-20}
Anderson, T.H., Civiletti, B.J., Monk, P.B., Lakhtakia, A.: Coupled
  optoelectronic simulation and optimization of thin-film photovoltaic solar
  cells.
\newblock Journal of Computational Physics \textbf{407}, 109,242 (2020)

\bibitem{Solano-ASV-19}
Araya, R., Solano, M., Vega, P.: Analysis of an adaptive {HDG} method for the
  {B}rinkman problem.
\newblock IMA Journal of Numerical Analysis \textbf{39}(3), 1502--1528 (2019)

\bibitem{Solano-ASV-19b}
Araya, R., Solano, M., Vega, P.: A posteriori error analysis of an {HDG} method
  for the {O}seen problem.
\newblock Applied Numerical Mathematics \textbf{146}, 291--308 (2019)

\bibitem{Wheeler-APWY-07}
Arbogast, T., Pencheva, G., Wheeler, M.F., Yotov, I.: {A Multiscale Mortar
  Mixed Finite Element Method}.
\newblock Multiscale Modeling \& Simulation \textbf{6}(1), 319--346 (2007)

\bibitem{deal-07}
Bangerth, W., Hartmann, R., Kanschat, G.: {deal.II} -- a general purpose object
  oriented finite element library.
\newblock ACM Transactions on Mathematical Software \textbf{33}(4), 24/1--24/27
  (2007)

\bibitem{Dolean-BBDNT-18}
Barrenechea, G.R., Bosy, M., Dolean, V., Nataf, F., Tournier, P.H.: Hybrid
  discontinuous {G}alerkin discretisation and domain decomposition
  preconditioners for the {S}tokes problem.
\newblock Computational Methods in Applied Mathematics  (2018)

\bibitem{DiPietro-BDGK-18}
Bonaldi, F., Di~Pietro, D.A., Geymonat, G., Krasucki, F.: A hybrid high-order
  method for {Kirchhoff-Love} plate bending problems.
\newblock ESAIM: Mathematical Modelling and Numerical Analysis \textbf{52}(2),
  393--421 (2018)

\bibitem{Lanteri-BCDL-17}
Bonnasse-Gahot, M., Calandra, H., Diaz, J., Lanteri, S.: {Hybridizable
  discontinuous {G}alerkin method for the 2-D frequency-domain elastic wave
  equations}.
\newblock Geophysical Journal International \textbf{213}(1), 637--659 (2017)

\bibitem{DiPietro-BD-18}
Botti, L., Di~Pietro, D.A.: Assessment of hybrid high-order methods on curved
  meshes and comparison with discontinuous {G}alerkin methods.
\newblock Journal of Computational Physics \textbf{370}, 58--84 (2018)

\bibitem{DiPietro-BDD-18}
Botti, L., Di~Pietro, D.A., Droniou, J.: A hybrid high-order discretisation of
  the {B}rinkman problem robust in the {D}arcy and {S}tokes limits.
\newblock Computer Methods in Applied Mechanics and Engineering \textbf{341},
  278--310 (2018)

\bibitem{DiPietro-BDD-19}
Botti, L., Di~Pietro, D.A., Droniou, J.: {A Hybrid High-Order method for the
  incompressible Navier-Stokes equations based on Temam's device}.
\newblock Journal of Computational Physics \textbf{376}, 786--816 (2019)

\bibitem{DiPietro-BDG-19}
Botti, M., Di~Pietro, D.A., Guglielmana, A.: A low-order nonconforming method
  for linear elasticity on general meshes.
\newblock Computer Methods in Applied Mechanics and Engineering \textbf{354},
  96--118 (2019)

\bibitem{DiPietro-BDLS-20}
Botti, M., Di~Pietro, D.A., Le~Ma\^{i}tre, O., Sochala, P.: Numerical
  approximation of poroelasticity with random coefficients using polynomial
  chaos and hybrid high-order methods.
\newblock Computer Methods in Applied Mechanics and Engineering \textbf{361},
  112,736 (2020)

\bibitem{DiPietro-BDS-17}
Botti, M., Di~Pietro, D.A., Sochala, P.: A hybrid high-order method for
  nonlinear elasticity.
\newblock SIAM Journal on Numerical Analysis \textbf{55}(6), 2687--2717 (2017)

\bibitem{DiPietro-BDS-20}
Botti, M., Di~Pietro, D.A., Sochala, P.: A hybrid high-order discretization
  method for nonlinear poroelasticity.
\newblock Computational Methods in Applied Mathematics \textbf{20}(2), 227--249
  (2020)

\bibitem{brezzi1991mixed}
Brezzi, F., Fortin, M.: Mixed and hybrid finite elements methods.
\newblock Springer series in computational mathematics. Springer-Verlag (1991)

\bibitem{BuiThanh-15}
Bui-Thanh, T.: From {G}odunov to a unified hybridized discontinuous {G}alerkin
  framework for partial differential equations.
\newblock Journal of Computational Physics \textbf{295}, 114--146 (2015)

\bibitem{BuiThanh-16}
Bui-Thanh, T.: Construction and analysis of {HDG} methods for linearized
  shallow water equations.
\newblock SIAM Journal on Scientific Computing \textbf{38}(6), A3696--A3719
  (2016)

\bibitem{Burman-BDE-20}
Burman, E., Delay, G., Ern, A.: {An unfitted hybrid high-order method for the
  Stokes interface problem}.
\newblock IMA Journal of Numerical Analysis  (2020)

\bibitem{Burman-BE-18}
Burman, E., Ern, A.: An unfitted hybrid high-order method for elliptic
  interface problems.
\newblock SIAM Journal on Numerical Analysis \textbf{56}(3), 1525--1546 (2018)

\bibitem{Solano-CLOS-20}
Camargo, L., L\'{o}pez-Rodr\'{i}guez, B., Osorio, M., Solano, M.: An {HDG}
  method for {M}axwell's equations in heterogeneous media.
\newblock Computer Methods in Applied Mechanics and Engineering \textbf{368},
  113,178 (2020)

\bibitem{Cangiani2017}
Cangiani, A., Dong, Z., Georgoulis, E., Houston, P.: {$hp$-Version
  Discontinuous Galerkin Methods on Polygonal and Polyhedral Meshes}.
\newblock Springer International Publishing (2017)

\bibitem{Nektar-15}
Cantwell, C.D., Moxey, D., Comerford, A., Bolis, A., Rocco, G., Mengaldo, G.,
  {De Grazia}, D., Yakovlev, S., Lombard, J.E., Ekelschot, D., Jordi, B., Xu,
  H., Mohamied, Y., Eskilsson, C., Nelson, B., Vos, P., Biotto, C., Kirby,
  R.M., Sherwin, S.J.: Nektar++: An open-source spectral/hp element framework.
\newblock Computer Physics Communications \textbf{192}, 205 -- 219 (2015)

\bibitem{Ern-CBCE-18}
Cascavita, K.L., Bleyer, J., Chateau, X., Ern, A.: Hybrid discretization
  methods with adaptive yield surface detection for {B}ingham pipe flows.
\newblock Journal of Scientific Computing \textbf{77}(3), 1424--1443 (2018)

\bibitem{DiPietro-CD-20}
Castanon~Quiroz, D., Di~Pietro, D.A.: {A Hybrid High-Order method for the
  incompressible Navier-Stokes problem robust for large irrotational body
  forces}.
\newblock Computers and Mathematics with Applications \textbf{79}(9),
  2655--2677 (2020)

\bibitem{Castillo-CG-20}
Castillo, P., G\'{o}mez, S.: Conservative super-convergent and hybrid
  discontinuous {G}alerkin methods applied to nonlinear {S}chr\"{o}dinger
  equations.
\newblock Applied Mathematics and Computation \textbf{371}, 124,950 (2020)

\bibitem{Celiker-CCS-10}
Celiker, F., Cockburn, B., Shi, K.: {Hybridizable discontinuous Galerkin
  methods for Timoshenko beams}.
\newblock Journal of Scientific Computing \textbf{44}(1), 1--37 (2010)

\bibitem{Celiker-CCS-11}
Celiker, F., Cockburn, B., Shi, K.: {A projection-based error analysis of HDG
  methods for Timoshenko beams}.
\newblock Mathematics of Computation \textbf{81}(277), 131--151 (2011)

\bibitem{Cesmelioglu-CCNP-13}
Cesmelioglu, A., Cockburn, B., Nguyen, N.C., Peraire, J.: Analysis of {HDG}
  methods for {O}seen equations.
\newblock Journal of Scientific Computing \textbf{55}(2), 392--431 (2013)

\bibitem{Cesmelioglu-CCQ-17}
Cesmelioglu, A., Cockburn, B., Qiu, W.: Analysis of a hybridizable
  discontinuous {G}alerkin method for the steady-state incompressible
  {N}avier-{S}tokes equations.
\newblock Mathematics of Computation \textbf{86}(306), 1643--1670 (2017)

\bibitem{Rhebergen-CRW-20}
Cesmelioglu, A., Rhebergen, S., Wells, G.N.: {An embedded-hybridized
  discontinuous Galerkin method for the coupled Stokes-Darcy system}.
\newblock Journal of Computational and Applied Mathematics \textbf{367},
  112,476 (2020)

\bibitem{DiPietro-CDF-19}
Chave, F., Di~Pietro, D.A., Formaggia, L.: A hybrid high-order method for
  passive transport in fractured porous media.
\newblock GEM - International Journal on Geomathematics \textbf{10}(1) (2019)

\bibitem{DiPietro-CDMP-16}
Chave, F., Di~Pietro, D.A., Marche, F., Pigeonneaux, F.: {A Hybrid High-order
  method for the {C}ahn-{H}illiard problem in mixed form}.
\newblock SIAM Journal on Numerical Analysis \textbf{54}(3), 1873--1898 (2016)

\bibitem{Cockburn-CCSZ-19}
Chen, G., Cockburn, B., Singler, J., Zhang, Y.: {Superconvergent Interpolatory
  {HDG} Methods for Reaction Diffusion Equations I: An {HDG}$_k$ Method}.
\newblock Journal of Scientific Computing \textbf{81}(3), 2188--2212 (2019)

\bibitem{Chen-CCX-19}
Chen, G., Cui, J., Xu, L.: Analysis of a hybridizable discontinuous {G}alerkin
  method for the {M}axwell operator.
\newblock ESAIM: Mathematical Modelling and Numerical Analysis \textbf{53}(1),
  301--324 (2019)

\bibitem{Monk-CMZ-19}
Chen, G., Monk, P., Zhang, Y.: An {HDG} method for the time-dependent
  drift-diffusion model of semiconductor devices.
\newblock Journal of Scientific Computing  (2019)

\bibitem{Qiu-CLQ-14}
Chen, H., Li, J., Qiu, W.: {Robust a posteriori error estimates for HDG method
  for convection-diffusion equations}.
\newblock IMA Journal of Numerical Analysis \textbf{36}(1), 437--462 (2014)

\bibitem{Shi-CQS-18}
Chen, H., Qiu, W., Shi, K.: {A priori and computable a posteriori error
  estimates for an HDG method for the coercive Maxwell equations}.
\newblock Computer Methods in Applied Mechanics and Engineering \textbf{333},
  287--310 (2018)

\bibitem{Shi-CQSS-17}
Chen, H., Qiu, W., Shi, K., Solano, M.: {A Superconvergent HDG Method for the
  Maxwell Equations}.
\newblock Journal of Scientific Computing \textbf{70}(3), 1010--1029 (2017)

\bibitem{Cockburn-CC-12}
Chen, Y., Cockburn, B.: {Analysis of variable-degree HDG methods for
  convection-diffusion equations. Part I: General nonconforming meshes}.
\newblock IMA Journal of Numerical Analysis \textbf{32}(4), 1267--1293 (2012)

\bibitem{Cockburn-CC-14}
Chen, Y., Cockburn, B.: {Analysis of variable-degree HDG methods for
  convection-diffusion equations. Part II: Semimatching nonconforming meshes}.
\newblock Mathematics of Computation \textbf{83}(285), 87--111 (2014)

\bibitem{Chen-CCD-16}
Chen, Y., Cockburn, B., Dong, B.: Superconvergent {HDG} methods for linear,
  stationary, third-order equations in one-space dimension.
\newblock Mathematics of Computation \textbf{85}(302), 2715--2742 (2016)

\bibitem{Chen-CDJ-18}
Chen, Y., Dong, B., Jiang, J.: Optimally convergent hybridizable discontinuous
  {G}alerkin method for fifth-order {K}orteweg-de {V}ries type equations.
\newblock ESAIM: Mathematical Modelling and Numerical Analysis \textbf{52}(6),
  2283--2306 (2018)

\bibitem{childs2010rotating}
Childs, P.R.: Rotating flow.
\newblock Elsevier (2010)

\bibitem{Moon-CM-20}
Cho, K., Moon, M.: {Multiscale hybridizable discontinuous Galerkin method for
  elliptic problems in perforated domains}.
\newblock Journal of Computational and Applied Mathematics \textbf{365},
  112,346 (2020)

\bibitem{Ern-CEP-20}
Chouly, F., Ern, A., Pignet, N.: {A Hybrid High-Order discretization combined
  with Nitsche's method for contact and Tresca friction in small strain
  elasticity}.
\newblock {SIAM Journal on Scientific Computing} \textbf{42}(4), A2300--A2324
  (2020)

\bibitem{Lanteri-CDL-18}
Christophe, A., Descombes, S., Lanteri, S.: An implicit hybridized
  discontinuous {G}alerkin method for the {3D} time-domain {M}axwell equations.
\newblock Applied Mathematics and Computation \textbf{319}, 395--408 (2018)

\bibitem{Chung-CCF-14}
Chung, E., Cockburn, B., Fu, G.: {The staggered DG method is the limit of a
  hybridizable DG method}.
\newblock SIAM Journal on Numerical Analysis \textbf{52}(2), 915--932 (2014)

\bibitem{Chung-CCF-16}
Chung, E., Cockburn, B., Fu, G.: {The Staggered DG Method is the Limit of a
  Hybridizable DG Method. Part II: The Stokes Flow}.
\newblock Journal of Scientific Computing \textbf{66}(2), 870--887 (2016)

\bibitem{Efendiev-CEL-19}
Chung, E., Efendiev, Y., Leung, W.T.: Generalized multiscale finite element
  methods with energy minimizing oversampling.
\newblock International Journal for Numerical Methods in Engineering
  \textbf{117}(3), 316--343 (2019)

\bibitem{disk-18}
Cicuttin, M., {Di Pietro}, D., Ern, A.: Implementation of discontinuous
  skeletal methods on arbitrary-dimensional, polytopal meshes using generic
  programming.
\newblock Journal of Computational and Applied Mathematics \textbf{344}, 852 --
  874 (2018)

\bibitem{Ern-CEG-20}
Cicuttin, M., Ern, A., Gudi, T.: Hybrid high-order methods for the elliptic
  obstacle problem.
\newblock Journal of Scientific Computing \textbf{83}(1), 8 (2020)

\bibitem{Ern-CEL-18}
Cicuttin, M., Ern, A., Lemaire, S.: A hybrid high-order method for highly
  oscillatory elliptic problems.
\newblock Computational Methods in Applied Mathematics  (2018)

\bibitem{Peraire-CFCNP-20}
Ciuc\v{a}, C., Fernandez, P., Christophe, A., Nguyen, N.C., Peraire, J.:
  Implicit hybridized discontinuous {G}alerkin methods for compressible
  magnetohydrodynamics.
\newblock Journal of Computational Physics: X \textbf{5}, 100,042 (2020)

\bibitem{Cockburn-16}
Cockburn, B.: Static condensation, hybridization, and the devising of the {HDG}
  methods.
\newblock In: G.R. Barrenechea, F.~Brezzi, A.~Cangiani, E.~Georgoulis (eds.)
  Building Bridges: Connections and Challenges in Modern Approaches to
  Numerical Partial Differential Equations, pp. 129--177. Springer
  International Publishing, Cham (2016)

\bibitem{Cockburn-CM-15}
Cockburn, B.., Mustapha, K.: {A hybridizable discontinuous Galerkin method for
  fractional diffusion problems}.
\newblock Numerische Mathematik \textbf{130}(2), 293--314 (2015)

\bibitem{Cockburn-CDG:08}
Cockburn, B., Dong, B., Guzm{{\'a}}n, J.: A superconvergent {LDG}-hybridizable
  {G}alerkin method for second-order elliptic problems.
\newblock Mathematics of Computation \textbf{77}(264), 1887--1916 (2008)

\bibitem{Cockburn-CDG-09}
Cockburn, B., Dong, B., Guzm\'{a}n, J.: {A hybridizable and superconvergent
  discontinuous Galerkin method for biharmonic problems}.
\newblock Journal of Scientific Computing \textbf{40}(1-3), 141--187 (2009)

\bibitem{Sacco-CDGRS-09}
Cockburn, B., Dong, B., Guzm\'{a}n, J., Restelli, M., Sacco, R.: {A
  hybridizable discontinuous Galerkin method for steady-state
  convection-diffusion-reaction problems}.
\newblock SIAM Journal on Scientific Computing \textbf{31}(5), 3827--3846
  (2009)

\bibitem{Jay-CDGT-14}
Cockburn, B., Dubois, O., Gopalakrishnan, J., Tan, S.: {Multigrid for an HDG
  method}.
\newblock IMA Journal of Numerical Analysis \textbf{34}(4), 1386--1425 (2014)

\bibitem{Cockburn-CF-17}
Cockburn, B., Fu, G.: {Devising superconvergent {HDG} methods with symmetric
  approximate stresses for linear elasticity by $M$-decompositions}.
\newblock IMA Journal of Numerical Analysis \textbf{38}(2), 566--604 (2017)

\bibitem{Cockburn-CF-17-II}
Cockburn, B., Fu, G.: {Superconvergence by $M$-decompositions. Part II:
  Construction of two-dimensional finite elements}.
\newblock ESAIM: Mathematical Modelling and Numerical Analysis \textbf{51}(1),
  165--186 (2017)

\bibitem{Cockburn-CF-17-III}
Cockburn, B., Fu, G.: {Superconvergence by $M$-decompositions. Part III:
  Construction of three-dimensional finite elements}.
\newblock ESAIM: Mathematical Modelling and Numerical Analysis \textbf{51}(1),
  365--398 (2017)

\bibitem{Cockburn-CFQ-17}
Cockburn, B., Fu, G., Qiu, W.: A note on the devising of superconvergent {HDG}
  methods for {S}tokes flow by {$M$}-decompositions.
\newblock IMA Journal of Numerical Analysis \textbf{37}(2), 730--749 (2017)

\bibitem{Cockburn-CF-17-I}
Cockburn, B., Fu, G., Sayas, F.J.: {Superconvergence by $M$-decompositions.
  Part I: General theory for HDG methods for diffusion}.
\newblock Mathematics of Computation \textbf{86}(306), 1609--1641 (2017)

\bibitem{Cockburn-CFHJSS-18}
Cockburn, B., Fu, Z., Hungria, A., Ji, L., S\'{a}nchez, M.A., Sayas, F.J.:
  S{tormer-Numerov HDG Methods for Acoustic Waves}.
\newblock Journal of Scientific Computing \textbf{75}(2), 597--624 (2018)

\bibitem{cockburn2004characterization}
Cockburn, B., Gopalakrishnan, J.: A characterization of hybridized mixed
  methods for second order elliptic problems.
\newblock SIAM Journal on Numerical Analysis \textbf{42}(1), 283--301 (2004)

\bibitem{Jay-CG:09}
Cockburn, B., Gopalakrishnan, J.: The derivation of hybridizable discontinuous
  {G}alerkin methods for {S}tokes flow.
\newblock SIAM Journal on Numerical Analysis \textbf{47}(2), 1092--1125 (2009)

\bibitem{Jay-CGL:09}
Cockburn, B., Gopalakrishnan, J., Lazarov, R.: Unified hybridization of
  discontinuous {G}alerkin, mixed, and continuous {G}alerkin methods for second
  order elliptic problems.
\newblock SIAM Journal on Numerical Analysis \textbf{47}(2), 1319--1365 (2009)

\bibitem{Jay-CGNPS-11}
Cockburn, B., Gopalakrishnan, J., Nguyen, N.C., Peraire, J., Sayas, F.J.:
  Analysis of {HDG} methods for {S}tokes flow.
\newblock Mathematics of Computation \textbf{80}(274), 723--760 (2011)

\bibitem{Karniadakis-CKS-00}
Cockburn, B., Karniadakis, G.E., Shu, C.W.: The development of discontinuous
  {G}alerkin methods.
\newblock In: Discontinuous {G}alerkin methods ({N}ewport, {RI}, 1999),
  \emph{Lect. Notes Comput. Sci. Eng.}, vol.~11, pp. 3--50. Springer, Berlin
  (2000)

\bibitem{Nguyen-CNP:10}
Cockburn, B., Nguyen, N.C., Peraire, J.: A comparison of {HDG} methods for
  {S}tokes flow.
\newblock Journal of Scientific Computing \textbf{45}(1-3), 215--237 (2010)

\bibitem{Cockburn-CQ-14}
Cockburn, B., Quenneville-B\'{e}lair, V.: {Uniform-in-time superconvergence of
  the HDG methods for the acoustic wave equation}.
\newblock Mathematics of Computation \textbf{83}(285), 65--85 (2014)

\bibitem{Cockburn-CS-14}
Cockburn, B., Sayas, F.J.: {Divergence-conforming HDG methods for Stokes
  flows}.
\newblock Mathematics of Computation \textbf{83}(288), 1571--1598 (2014)

\bibitem{Cockburn-CS-16}
Cockburn, B., Shen, J.: A hybridizable discontinuous {G}alerkin method for the
  $p$-{L}aplacian.
\newblock SIAM Journal on Scientific Computing \textbf{38}(1), A545--A566
  (2016)

\bibitem{Cockburn-CS-19}
Cockburn, B., Shen, J.: An algorithm for stabilizing hybridizable discontinuous
  {G}alerkin methods for nonlinear elasticity.
\newblock Results in Applied Mathematics \textbf{1} (2019)

\bibitem{Shi-CS-13}
Cockburn, B., Shi, K.: {Superconvergent HDG methods for linear elasticity with
  weakly symmetric stresses}.
\newblock IMA Journal of Numerical Analysis \textbf{33}(3), 747--770 (2013)

\bibitem{Cockburn-CS:14}
Cockburn, B., Shi, K.: Devising {HDG} methods for {S}tokes flow: an overview.
\newblock Computers \& Fluids \textbf{98}, 221--229 (2014)

\bibitem{Cockburn-CS-98}
Cockburn, B., Shu, C.W.: The local discontinuous {G}alerkin method for
  time-dependent convection-diffusion systems.
\newblock SIAM Journal on Numerical Analysis \textbf{35}(6), 2440--2463 (1998)

\bibitem{Cockburn-CSZ-19}
Cockburn, B., Singler, J.R., Zhang, Y.: {Interpolatory HDG Method for Parabolic
  Semilinear PDEs}.
\newblock Journal of Scientific Computing \textbf{79}(3), 1777--1800 (2019)

\bibitem{Solano-CS-12}
Cockburn, B., Solano, M.: Solving {D}irichlet boundary-value problems on curved
  domains by extensions from subdomains.
\newblock SIAM Journal on Scientific Computing \textbf{34}(1), A497--A519
  (2012)

\bibitem{Solano-CS-14}
Cockburn, B., Solano, M.: Solving convection-diffusion problems on curved
  domains by extensions from subdomains.
\newblock Journal of Scientific Computing \textbf{59}(2), 512--543 (2014)

\bibitem{Cockburn-CZ-17}
Cockburn, B., Wang, Z.: Adjoint-based, superconvergent {G}alerkin
  approximations of linear functionals.
\newblock Journal of Scientific Computing \textbf{73}(2-3), 644--666 (2017)

\bibitem{Cockburn-CZ-12}
Cockburn, B., Zhang, W.: {A posteriori error estimates for HDG methods}.
\newblock Journal of Scientific Computing \textbf{51}(3), 582--607 (2012)

\bibitem{Cockburn-CZ-13}
Cockburn, B., Zhang, W.: A posteriori error analysis for hybridizable
  discontinuous {G}alerkin methods for second order elliptic problems.
\newblock SIAM Journal on Numerical Analysis \textbf{51}(1), 676--693 (2013)

\bibitem{Ern2016-CDPE}
{Cockburn, B.}, {Di Pietro, D. A.}, {Ern, A.}: Bridging the hybrid high-order
  and hybridizable discontinuous {G}alerkin methods.
\newblock ESAIM: Mathematical Modelling and Numerical Analysis \textbf{50}(3),
  635--650 (2016)

\bibitem{CostaSole-CRS-19}
Costa-Sol\'{e}, A., Ruiz-Giron\'{e}s, E., Sarrate, J.: An {HDG} formulation for
  incompressible and immiscible two-phase porous media flow problems.
\newblock International Journal of Computational Fluid Dynamics \textbf{33}(4),
  137--148 (2019)

\bibitem{ErnBook}
Di~Pietro, D., Ern, A.: Mathematical aspects of discontinuous {G}alerkin
  methods, vol.~69.
\newblock Springer, Heidelberg (2012)

\bibitem{Ern-DPE-15}
Di~Pietro, D., Ern, A.: A hybrid high-order locking-free method for linear
  elasticity on general meshes.
\newblock Computer Methods in Applied Mechanics and Engineering \textbf{283},
  1--21 (2015)

\bibitem{DiPietro-DPEL-14}
Di~Pietro, D., Ern, A., Lemaire, S.: An arbitrary-order and compact-stencil
  discretization of diffusion on general meshes based on local reconstruction
  operators.
\newblock Computational Methods in Applied Mathematics \textbf{14}(4), 461--472
  (2014)

\bibitem{DiPietro-DD-17}
Di~Pietro, D.A., Droniou, J.: {A Hybrid High-Order method for Leray-Lions
  elliptic equations on general meshes}.
\newblock Mathematics of Computation \textbf{86}(307), 2159--2191 (2017)

\bibitem{HHO-book}
Di~Pietro, D.A., Droniou, J.: {The Hybrid High-Order Method for Polytopal
  Meshes}.
\newblock Modeling, Simulation and Applications series. {Springer International
  Publishing} (2020)

\bibitem{Ern-DDE-15}
Di~Pietro, D.A., Droniou, J., Ern, A.: A discontinuous-skeletal method for
  advection-diffusion-reaction on general meshes.
\newblock SIAM Journal on Numerical Analysis \textbf{53}(5), 2135--2157 (2015)

\bibitem{DiPietro-DDM-18}
Di~Pietro, D.A., Droniou, J., Manzini, G.: {Discontinuous Skeletal Gradient
  Discretisation methods on polytopal meshes}.
\newblock Journal of Computational Physics \textbf{355}, 397--425 (2018)

\bibitem{Ern-DE-15}
Di~Pietro, D.A., Ern, A.: Hybrid high-order methods for variable-diffusion
  problems on general meshes.
\newblock Comptes Rendus Mathematique \textbf{353}(1), 31--34 (2015)

\bibitem{Ern-DELS-16}
Di~Pietro, D.A., Ern, A., Linke, A., Schieweck, F.: {A discontinuous skeletal
  method for the viscosity-dependent Stokes problem}.
\newblock Computer Methods in Applied Mechanics and Engineering \textbf{306},
  175--195 (2016)

\bibitem{DiPietro-DK-18}
Di~Pietro, D.A., Krell, S.: A hybrid high-order method for the steady
  incompressible {N}avier-{S}tokes problem.
\newblock Journal of Scientific Computing \textbf{74}(3), 1677--1705 (2018)

\bibitem{diskin2011comparison}
Diskin, B., Thomas, J.L.: Comparison of node-centered and cell-centered
  unstructured finite-volume discretizations: inviscid fluxes.
\newblock AIAA journal \textbf{49}(4), 836--854 (2011)

\bibitem{diskin2010comparison}
Diskin, B., Thomas, J.L., Nielsen, E.J., Nishikawa, H., White, J.A.: Comparison
  of node-centered and cell-centered unstructured finite-volume
  discretizations: viscous fluxes.
\newblock AIAA journal \textbf{48}(7), 1326 (2010)

\bibitem{donea2003finite}
Donea, J., Huerta, A.: Finite Element Methods for Flow Problems.
\newblock Finite Element Methods for Flow Problems. John Wiley \& Sons (2003)

\bibitem{Dong-DWXW-16}
Dong, H., Wang, B., Xie, Z., Wang, L.L.: {An unfitted hybridizable
  discontinuous {G}alerkin method for the {P}oisson interface problem and its
  error analysis}.
\newblock IMA Journal of Numerical Analysis \textbf{37}(1), 444--476 (2016)

\bibitem{Sayas-DS-20}
Du, S., Sayas, F.J.: A unified error analysis of hybridizable discontinuous
  {G}alerkin methods for the static {M}axwell equations.
\newblock SIAM Journal on Numerical Analysis \textbf{58}(2), 1367--1391 (2020)

\bibitem{Shi-ELMS-15}
Efendiev, Y., Lazarov, R., Moon, M., Shi, K.: {A spectral multiscale
  hybridizable discontinuous Galerkin method for second order elliptic
  problems}.
\newblock Computer Methods in Applied Mechanics and Engineering \textbf{292},
  243 -- 256 (2015).
\newblock Special Issue on Advances in Simulations of Subsurface Flow and
  Transport (Honoring Professor Mary F. Wheeler)

\bibitem{Egger-ES-09}
Egger, H., Sch\"{o}berl, J.: {A hybrid mixed discontinuous Galerkin
  finite-element method for convection-diffusion problems}.
\newblock IMA Journal of Numerical Analysis \textbf{30}(4), 1206--1234 (2009)

\bibitem{Egger-EW-12}
Egger, H., Waluga, C.: {$hp$-analysis of a hybrid DG method for Stokes flow}.
\newblock IMA Journal of Numerical Analysis \textbf{33}(2), 687--721 (2012)

\bibitem{Egger-EW-12b}
Egger, H., Waluga, C.: A hybrid mortar method for incompressible flow.
\newblock International Journal of Numerical Analysis and Modeling
  \textbf{9}(4), 793--812 (2012)

\bibitem{Fabien-20}
Fabien, M.S.: A {GPU}-accelerated hybridizable discontinuous {G}alerkin method
  for linear elasticity.
\newblock Communications in Computational Physics \textbf{27}(2), 513--545
  (2020)

\bibitem{Fabien-20-shallow}
Fabien, M.S.: A high-order implicit {HDG} method for the
  {B}enjamin-{B}ona-{M}ahony equation.
\newblock International Journal for Numerical Methods in Fluids  (2020)

\bibitem{Riviere-FKMR-19}
Fabien, M.S., Knepley, M.G., Mills, R.T., Riviere, B.M.: Manycore parallel
  computing for a hybridizable discontinuous {G}alerkin nested multigrid
  method.
\newblock SIAM Journal on Scientific Computing \textbf{41}(2), C73--C96 (2019)

\bibitem{Riviere-FKR-18}
Fabien, M.S., Knepley, M.G., Rivi\`{e}re, B.M.: A hybridizable discontinuous
  {G}alerkin method for two-phase flow in heterogeneous porous media.
\newblock International Journal for Numerical Methods in Engineering
  \textbf{116}(3), 161--177 (2018)

\bibitem{Farahinia-FZ-19}
Farahinia, A., Zhang, W.J.: Numerical investigation into the mixing performance
  of micro {T}-mixers with different patterns of obstacles.
\newblock Journal of the Brazilian Society of Mechanical Sciences and
  Engineering \textbf{41}(11), 491 (2019)

\bibitem{Peraire-FCTNP-18}
Fernandez, P., Christophe, A., Terrana, S., Nguyen, N.C., Peraire, J.:
  Hybridized discontinuous {G}alerkin methods for wave propagation.
\newblock Journal of Scientific Computing \textbf{77}(3), 1566--1604 (2018)

\bibitem{Peraire-FNP-17}
Fernandez, P., Nguyen, N.C., Peraire, J.: The hybridized discontinuous
  {G}alerkin method for implicit large-eddy simulation of transitional
  turbulent flows.
\newblock Journal of Computational Physics \textbf{336}, 308 -- 329 (2017)

\bibitem{Fidkowski-16}
Fidkowski, K.J.: A hybridized discontinuous {G}alerkin method on mapped
  deforming domains.
\newblock Computers \& Fluids \textbf{139}, 80 -- 91 (2016)

\bibitem{Fidkowski-19}
Fidkowski, K.J.: Comparison of hybrid and standard discontinuous {G}alerkin
  methods in a mesh-optimisation setting.
\newblock International Journal of Computational Fluid Dynamics
  \textbf{33}(1-2), 34--42 (2019)

\bibitem{Fidkowski-FC-20}
Fidkowski, K.J., Chen, G.: Output-based mesh optimization for hybridized and
  embedded discontinuous {G}alerkin methods.
\newblock International Journal for Numerical Methods in Engineering
  \textbf{121}(5), 867--887 (2020)

\bibitem{Saturne-11}
Fournier, Y., Bonelle, J., Moulinec, C., Shang, Z., Sunderland, A.G., Uribe,
  J.C.: {Optimizing Code\_Saturne computations on Petascale systems}.
\newblock Computers \& Fluids \textbf{45}(1), 103 -- 108 (2011)

\bibitem{Fraeijs-65}
{Fraeijs de Veubeke}, B.: Displacement and equilibrium models in the finite
  element method.
\newblock In: {O.C. Zienkiewicz and G.S. Holister} (ed.) Stress Analysis, pp.
  145--197. {John Wiley \& Sons} (1965)

\bibitem{Crivellini-FFC-20}
Franciolini, M., Fidkowski, K.J., Crivellini, A.: Efficient discontinuous
  {G}alerkin implementations and preconditioners for implicit unsteady
  compressible flow simulations.
\newblock Computers \& Fluids \textbf{203}, 104,542 (2020)

\bibitem{Fu-20}
Fu, G.: {Arbitrary Lagrangian-Eulerian hybridizable discontinuous Galerkin
  methods for incompressible flow with moving boundaries and interfaces}.
\newblock Computer Methods in Applied Mechanics and Engineering \textbf{367},
  113,158 (2020)

\bibitem{Fu-FCS-15}
Fu, G., Cockburn, B., Stolarski, H.: Analysis of an {HDG} method for linear
  elasticity.
\newblock International Journal for Numerical Methods in Engineering
  \textbf{102}(3-4), 551--575 (2015)

\bibitem{Fu-FJQ-19}
Fu, G., Jin, Y., Qiu, W.: {Parameter-free superconvergent H(div)-conforming HDG
  methods for the Brinkman equations}.
\newblock IMA Journal of Numerical Analysis \textbf{39}(2), 957--982 (2018)

\bibitem{HDG3D-15}
Fu, Z., Gatica, L.F., Sayas, F.J.: {Algorithm 949: MATLAB Tools for HDG in
  Three Dimensions}.
\newblock ACM Transactions on Mathematical Software \textbf{41}(3) (2015)

\bibitem{Gander-GH-18}
Gander, M.J., Hajian, S.: Analysis of {S}chwarz methods for a hybridizable
  discontinuous {G}alerkin discretization: the many-subdomain case.
\newblock Mathematics of Computation \textbf{87}(312), 1635--1657 (2018)

\bibitem{Gatica-GS-15}
Gatica, G.N., Sequeira, F.A.: Analysis of an augmented {HDG} method for a class
  of quasi-{N}ewtonian {S}tokes flows.
\newblock Journal of Scientific Computing \textbf{65}(3), 1270--1308 (2015)

\bibitem{GMSH-09}
Geuzaine, C., Remacle, J.F.: {Gmsh: A 3-D finite element mesh generator with
  built-in pre- and post-processing facilities}.
\newblock International Journal for Numerical Methods in Engineering
  \textbf{79}(11), 1309--1331 (2009)

\bibitem{MG-GBSH-20}
Giacomini, M., Borchini, L., Sevilla, R., Huerta, A.: Separated response
  surfaces for flows in parametrised domains: comparison of a priori and a
  posteriori {PGD} algorithms.
\newblock Tech. rep., arXiv (2020).
\newblock \urlprefix\url{https://arxiv.org/abs/2009.02176}.
\newblock Submitted.

\bibitem{MG-GKSH-18}
Giacomini, M., Karkoulias, A., Sevilla, R., Huerta, A.: A superconvergent {HDG}
  method for {S}tokes flow with strongly enforced symmetry of the stress
  tensor.
\newblock Journal of Scientific Computing \textbf{77}(3), 1679--1702 (2018)

\bibitem{MG-GS:19}
Giacomini, M., Sevilla, R.: Discontinuous {G}alerkin approximations in
  computational mechanics: hybridization, exact geometry and degree adaptivity.
\newblock SN Applied Sciences \textbf{1}, 1047 (2019)

\bibitem{MG-GS-20}
Giacomini, M., Sevilla, R.: A second-order face-centred finite volume method on
  general meshes with automatic mesh adaptation.
\newblock International Journal for Numerical Methods in Engineering  (2020)

\bibitem{MG-GSH-20}
Giacomini, M., Sevilla, R., Huerta, A.: Tutorial on {H}ybridizable
  {D}iscontinuous {G}alerkin ({HDG}) {F}ormulation for {I}ncompressible {F}low
  {P}roblems.
\newblock In: L.D. Lorenzis, A.~D\"{u}ster (eds.) Modeling in Engineering Using
  Innovative Numerical Methods for Solids and Fluids, \emph{CISM International
  Centre for Mechanical Sciences}, vol. 599, pp. 163--201. Springer
  International Publishing (2020)

\bibitem{Giorgiani-GFH-13}
Giorgiani, G., Fern{\'a}ndez-M{\'e}ndez, S., Huerta, A.: Hybridizable
  discontinuous {G}alerkin $p$-adaptivity for wave propagation problems.
\newblock International Journal for Numerical Methods in Fluids
  \textbf{72}(12), 1244--1262 (2013)

\bibitem{Giorgiani-GFH-14}
Giorgiani, G., Fern{\'a}ndez-M{\'e}ndez, S., Huerta, A.: Hybridizable
  discontinuous {G}alerkin with degree adaptivity for the incompressible
  {N}avier--{S}tokes equations.
\newblock Computers \& Fluids \textbf{98}, 196--208 (2014)

\bibitem{Gurkan-GKF-17}
G{\"u}rkan, C., Kronbichler, M., Fern{\'a}ndez-M{\'e}ndez, S.: e{X}tended
  {H}ybridizable {D}iscontinuous {G}alerkin with {H}eaviside enrichment for
  heat bimaterial problems.
\newblock Journal of Scientific Computing \textbf{72}(2), 542--567 (2017)

\bibitem{Gurkan-GKF-19}
G\"{u}rkan, C., Kronbichler, M., Fern\'{a}ndez-M\'{e}ndez, S.: e{X}tended
  hybridizable discontinuous {G}alerkin for incompressible flow problems with
  unfitted meshes and interfaces.
\newblock International Journal for Numerical Methods in Engineering
  \textbf{117}(7), 756--777 (2019)

\bibitem{Gurkan-GSKF-16}
G{\"u}rkan, C., Sala-Lardies, E., Kronbichler, M., Fern{\'a}ndez-M{\'e}ndez,
  S.: e{X}tended {H}ybridizable {D}iscontinous {G}alerkin ({X-HDG}) for void
  problems.
\newblock Journal of Scientific Computing \textbf{66}(3), 1313--1333 (2016)

\bibitem{Guyan-65}
Guyan, R.: Reduction of stiffness and mass matrices.
\newblock AIAA Journal \textbf{3}(2), 380--380 (1965)

\bibitem{Hesthaven-HW-02}
Hesthaven, J., Warburton, T.: {Nodal High-Order Methods on Unstructured Grids:
  I. Time-Domain Solution of {M}axwell's Equations}.
\newblock Journal of Computational Physics \textbf{181}(1), 186--221 (2002)

\bibitem{Kronbichler-HBKPCW-18}
Hoermann, J.M., Bertoglio, C., Kronbichler, M., Pfaller, M.R., Chabiniok, R.,
  Wall, W.A.: An adaptive hybridizable discontinuous {G}alerkin approach for
  cardiac electrophysiology.
\newblock International Journal for Numerical Methods in Biomedical Engineering
  \textbf{34}(5) (2018)

\bibitem{Rhebergen-HR-19}
Horv\'{a}th, T.L., Rhebergen, S.: {A locally conservative and energy-stable
  finite-element method for the Navier-Stokes problem on time-dependent
  domains}.
\newblock International Journal for Numerical Methods in Fluids
  \textbf{89}(12), 519--532 (2019)

\bibitem{Rhebergen-HR-20}
Horv\'{a}th, T.L., Rhebergen, S.: An exactly mass conserving space-time
  embedded-hybridized discontinuous {G}alerkin method for the {N}avier-{S}tokes
  equations on moving domains.
\newblock Journal of Computational Physics \textbf{417}, 109,577 (2020)

\bibitem{Huang-HH-19}
Huang, J., Huang, X.: A hybridizable discontinuous {G}alerkin method for
  {K}irchhoff plates.
\newblock Journal of Scientific Computing \textbf{78}(1), 290--320 (2019)

\bibitem{AA-HARP:13}
Huerta, A., Angeloski, A., Roca, X., Peraire, J.: Efficiency of high-order
  elements for continuous and discontinuous {G}alerkin methods.
\newblock International Journal for Numerical Methods in Engineering
  \textbf{96}(9), 529--560 (2013)

\bibitem{Sayas-HPS-17}
Hungria, A., Prada, D., Sayas, F.J.: {HDG} methods for elastodynamics.
\newblock Computers \& Mathematics with Applications \textbf{74}(11), 2671 --
  2690 (2017)

\bibitem{Peraire-HNPK-13}
Huynh, L.N.T., Nguyen, N.C., Peraire, J., Khoo, B.C.: A high-order hybridizable
  discontinuous {G}alerkin method for elliptic interface problems.
\newblock International Journal for Numerical Methods in Engineering
  \textbf{93}(2), 183--200 (2013)

\bibitem{FESTUNG-18}
Jaust, A., Reuter, B., Aizinger, V., Sch\"{u}tz, J., Knabner, P.: {FESTUNG: A
  MATLAB/GNU Octave toolbox for the discontinuous Galerkin method. Part III:
  Hybridized discontinuous Galerkin (HDG) formulation}.
\newblock Computers \& Mathematics with Applications \textbf{75}(12), 4505 --
  4533 (2018)

\bibitem{Lew-KLC-15}
Kabaria, H., Lew, A.J., Cockburn, B.: A hybridizable discontinuous {G}alerkin
  formulation for non-linear elasticity.
\newblock Computer Methods in Applied Mechanics and Engineering \textbf{283},
  303--329 (2015)

\bibitem{BuiThanh-KBA-19}
Kang, S., Bui-Thanh, T., Arbogast, T.: {A hybridized discontinuous Galerkin
  method for a linear degenerate elliptic equation arising from two-phase
  mixtures}.
\newblock Computer Methods in Applied Mechanics and Engineering \textbf{350},
  315--336 (2019)

\bibitem{BuiThanh-KGB-20}
Kang, S., Giraldo, F.X., Bui-Thanh, T.: {IMEX HDG-DG}: A coupled implicit
  hybridized discontinuous {G}alerkin and explicit discontinuous {G}alerkin
  approach for shallow water systems.
\newblock Journal of Computational Physics \textbf{401}, 109,010 (2020)

\bibitem{Cockburn-KSC:11}
Kirby, R., Sherwin, S.J., Cockburn, B.: To {CG} or to {HDG}: A comparative
  study.
\newblock Journal of Scientific Computing \textbf{51}(1), 183--212 (2011)

\bibitem{Rhebergen-KR-19}
Kirk, K.L.A., Rhebergen, S.: Analysis of a pressure-robust hybridized
  discontinuous {G}alerkin method for the stationary {N}avier-{S}tokes
  equations.
\newblock Journal of Scientific Computing \textbf{81}(2), 881--897 (2019)

\bibitem{Sevilla-KSH-20}
Komala-Sheshachala, S., Sevilla, R., Hassan, O.: A coupled {HDG-FV} scheme for
  the simulation of transient inviscid compressible flows.
\newblock Computers \& Fluids \textbf{202}, 104,495 (2020)

\bibitem{Kronbichler-KSMW-16}
Kronbichler, M., Schoeder, S., M\"{u}ller, C., Wall, W.A.: Comparison of
  implicit and explicit hybridizable discontinuous {G}alerkin methods for the
  acoustic wave equation.
\newblock International Journal for Numerical Methods in Engineering
  \textbf{106}(9), 712--739 (2016)

\bibitem{Kronbichler-KW-18}
Kronbichler, M., Wall, W.A.: A performance comparison of continuous and
  discontinuous {G}alerkin methods with fast multigrid solvers.
\newblock SIAM Journal on Scientific Computing \textbf{40}(5), A3423--A3448
  (2018)

\bibitem{MG-SpinaGH-20}
{La Spina}, A., Giacomini, M., Huerta, A.: {Hybrid coupling of CG and HDG
  discretizations based on Nitsche's method}.
\newblock Computational Mechanics \textbf{65}(2), 311--330 (2020)

\bibitem{LaSpina-SKGWH-20}
{La Spina}, A., Kronbichler, M., Giacomini, M., Wall, W., Huerta, A.: A weakly
  compressible hybridizable discontinuous {G}alerkin formulation for
  fluid-structure interaction problems.
\newblock Computer Methods in Applied Mechanics and Engineering \textbf{372},
  113,392 (2020)

\bibitem{Lederer-LLS-18}
Lederer, P.L., Lehrenfeld, C., Sch\"{o}berl, J.: {Hybrid discontinuous Galerkin
  methods with relaxed ${H}$(div)-conformity for incompressible flows. Part I}.
\newblock SIAM Journal on Numerical Analysis \textbf{56}(4), 2070--2094 (2018)

\bibitem{Lederer-LLS-19}
Lederer, P.L., Lehrenfeld, C., Sch\"{o}berl, J.: {Hybrid discontinuous Galerkin
  methods with relaxed ${H}$(div)-conformity for incompressible flows. Part
  II}.
\newblock ESAIM: Mathematical Modelling and Numerical Analysis \textbf{53}(2),
  503--522 (2019)

\bibitem{Lederer-LLS-20}
Lederer, P.L., Lehrenfeld, C., Sch\"{o}berl, J.: Divergence-free tangential
  finite element methods for incompressible flows on surfaces.
\newblock International Journal for Numerical Methods in Engineering
  \textbf{121}(11), 2503--2533 (2020)

\bibitem{BuiThanh-LSBS-19}
Lee, J.J., Shannon, S.J., Bui-Thanh, T., Shadid, J.N.: Analysis of an {HDG}
  method for linearized incompressible resistive {MHD} equations.
\newblock SIAM Journal on Numerical Analysis \textbf{57}(4), 1697--1722 (2019)

\bibitem{Schoberl-LS-16}
Lehrenfeld, C., Sch\"{o}berl, J.: High order exactly divergence-free hybrid
  discontinuous {G}alerkin methods for unsteady incompressible flows.
\newblock Computer Methods in Applied Mechanics and Engineering \textbf{307},
  339--361 (2016)

\bibitem{Chen-LC-20}
Leng, H., Chen, Y.: Adaptive hybridizable discontinuous {G}alerkin methods for
  nonstationary convection-diffusion problems.
\newblock Advances in Computational Mathematics \textbf{46}(4) (2020)

\bibitem{Shi-LS-18}
Li, G., Shi, K.: Upscaled {HDG} methods for {B}rinkman equations with
  high-contrast heterogeneous coefficient.
\newblock Journal of Scientific Computing \textbf{77}(3), 1780--1800 (2018)

\bibitem{Lanteri-LLMW-17}
Li, L., Lanteri, S., Mortensen, N.A., Wubs, M.: {A hybridizable discontinuous
  Galerkin method for solving nonlocal optical response models}.
\newblock Computer Physics Communications \textbf{219}, 99--107 (2017)

\bibitem{Lanteri-LLP-14}
Li, L., Lanteri, S., Perrussel, R.: {A hybridizable discontinuous Galerkin
  method combined to a Schwarz algorithm for the solution of 3D time-harmonic
  Maxwell's equation}.
\newblock Journal of Computational Physics \textbf{256}, 563--581 (2014)

\bibitem{Lanteri-LLP-15}
Li, L., Lanteri, S., Perrussel, R.: {A class of locally well-posed hybridizable
  discontinuous Galerkin methods for the solution of time-harmonic Maxwell's
  equations}.
\newblock Computer Physics Communications \textbf{192}, 23 -- 31 (2015)

\bibitem{liu2009fast}
Liu, Y.: Fast multipole boundary element method: theory and applications in
  engineering.
\newblock Cambridge university press (2009)

\bibitem{loseille2018vizir}
Loseille, A., Feuillet, R.: Vizir: High-order mesh and solution visualization
  using {OpenGL} 4.0 graphic pipeline.
\newblock In: 2018 AIAA Aerospace Sciences Meeting, p. 1174 (2018)

\bibitem{Qiu-LCQ-17}
Lu, P.., Chen, H., Qiu, W.: {An absolutely stable hp-HDG method for the
  time-harmonic Maxwell equations with high wave number}.
\newblock Mathematics of Computation \textbf{86}(306), 1553--1577 (2017)

\bibitem{Stern-MS-20}
McLachlan, R.I., Stern, A.: Multisymplecticity of hybridizable discontinuous
  {G}alerkin methods.
\newblock Foundations of Computational Mathematics \textbf{20}(1), 35--69
  (2020)

\bibitem{AdM-MFH:08}
Montlaur, A., Fern\'andez-M\'endez, S., Huerta, A.: Discontinuous {G}alerkin
  methods for the {S}tokes equations using divergence-free approximations.
\newblock International Journal for Numerical Methods in Fluids \textbf{57}(9),
  1071--1092 (2008)

\bibitem{Lazarov-MLJ-19}
Moon, M., Lazarov, R., Jun, H.K.: Multiscale {HDG} model reduction method for
  flows in heterogeneous porous media.
\newblock Applied Numerical Mathematics \textbf{140}, 115--133 (2019)

\bibitem{Peraire-MNP-11}
Moro, D., Nguyen, N.C., Peraire, J.: Navier-{S}tokes solution using
  hybridizable discontinuous {G}alerkin methods.
\newblock In: 20th AIAA Computational Fluid Dynamics Conference. AIAA (2011)

\bibitem{Muixi-MRF-20}
Muix\'{i}, A., Rodr\'{i}guez-Ferran, A., Fern\'{a}ndez-M\'{e}ndez, S.: A
  hybridizable discontinuous {G}alerkin phase-field model for brittle fracture
  with adaptive refinement.
\newblock International Journal for Numerical Methods in Engineering
  \textbf{121}(6), 1147--1169 (2020)

\bibitem{BuiThanh-MBS-20}
Muralikrishnan, S., Bui-Thanh, T., Shadid, J.N.: A multilevel approach for
  trace system in {HDG} discretizations.
\newblock Journal of Computational Physics \textbf{407}, 109,240 (2020)

\bibitem{BuiThanh-MTB-18}
Muralikrishnan, S., Tran, M., Bui-Thanh, T.: An improved iterative {HDG}
  approach for partial differential equations.
\newblock Journal of Computational Physics \textbf{367}, 295 -- 321 (2018)

\bibitem{Cockburn-MNC-16}
Mustapha, K., Nour, M., Cockburn, B.: {Convergence and superconvergence
  analyses of HDG methods for time fractional diffusion problems}.
\newblock Advances in Computational Mathematics \textbf{42}(2), 377--393 (2016)

\bibitem{nelson2012elvis}
Nelson, B., Liu, E., Kirby, R.M., Haimes, R.: Elvis: A system for the accurate
  and interactive visualization of high-order finite element solutions.
\newblock IEEE transactions on visualization and computer graphics
  \textbf{18}(12), 2325--2334 (2012)

\bibitem{Nguyen-NPC:10}
Nguyen, N., Peraire, J., Cockburn, B.: A hybridizable discontinuous {G}alerkin
  method for {S}tokes flow.
\newblock Computer Methods in Applied Mechanics and Engineering
  \textbf{199}(9-12), 582--597 (2010)

\bibitem{Nguyen-NPC:09}
Nguyen, N.C., Peraire, J., Cockburn, B.: An implicit high-order hybridizable
  discontinuous {G}alerkin method for linear convection-diffusion equations.
\newblock Journal of Computational Physics \textbf{228}(9), 3232--3254 (2009)

\bibitem{Nguyen-NPC:09b}
Nguyen, N.C., Peraire, J., Cockburn, B.: An implicit high-order hybridizable
  discontinuous {G}alerkin method for nonlinear convection-diffusion equations.
\newblock Journal of Computational Physics \textbf{228}(23), 8841--8855 (2009)

\bibitem{nguyen2011high}
Nguyen, N.C., Peraire, J., Cockburn, B.: High-order implicit hybridizable
  discontinuous {G}alerkin methods for acoustics and elastodynamics.
\newblock Journal of Computational Physics \textbf{230}(10), 3695--3718 (2011)

\bibitem{nguyen2011hybridizable}
Nguyen, N.C., Peraire, J., Cockburn, B.: Hybridizable discontinuous {G}alerkin
  methods for the time-harmonic {M}axwell's equations.
\newblock Journal of Computational Physics \textbf{230}(19), 7151--7175 (2011)

\bibitem{Nguyen-NPC:11}
Nguyen, N.C., Peraire, J., Cockburn, B.: An implicit high-order hybridizable
  discontinuous {G}alerkin method for the incompressible {N}avier-{S}tokes
  equations.
\newblock Journal of Computational Physics \textbf{230}(4), 1147--1170 (2011)

\bibitem{Peraire-NPC-15}
Nguyen, N.C., Peraire, J., Cockburn, B.: {A class of embedded discontinuous
  Galerkin methods for computational fluid dynamics}.
\newblock Journal of Computational Physics \textbf{302}, 674--692 (2015)

\bibitem{oikawa2015hybridized}
Oikawa, I.: A hybridized discontinuous {G}alerkin method with reduced
  stabilization.
\newblock Journal of Scientific Computing \textbf{65}(1), 327--340 (2015)

\bibitem{oikawa2016analysis}
Oikawa, I.: Analysis of a reduced-order {HDG} method for the {S}tokes
  equations.
\newblock Journal of Scientific Computing \textbf{67}(2), 475--492 (2016)

\bibitem{FernandezMendez-PTF-19}
Paipuri, M., Tiago, C., Fern{\'a}ndez-M{\'e}ndez, S.: Coupling of continuous
  and hybridizable discontinuous {G}alerkin methods: Application to conjugate
  heat transfer problem.
\newblock Journal of Scientific Computing \textbf{78}(1), 321--350 (2019)

\bibitem{peraire2010hybridizable}
Peraire, J., Nguyen, N.C., Cockburn, B.: A hybridizable discontinuous
  {G}alerkin method for the compressible {E}uler and {N}avier-{S}tokes
  equations.
\newblock AIAA paper \textbf{363}, 2010 (2010)

\bibitem{Evans-PE-19}
Peters, E., Evans, J.: A divergence-conforming hybridized discontinuous
  {G}alerkin method for the incompressible {R}eynolds-averaged
  {N}avier-{S}tokes equations.
\newblock International Journal for Numerical Methods in Fluids \textbf{91},
  112--133 (2019)

\bibitem{Aster-19}
Pignet, N.: {Hybrid High-Order methods for nonlinear solid mechanics}.
\newblock {PhD thesis}, {Universit{\'e} Paris-Est Marne la Vall{\'e}e} (2019).
\newblock TEL 02318157

\bibitem{poya2016unified}
Poya, R., Sevilla, R., Gil, A.J.: A unified approach for a posteriori
  high-order curved mesh generation using solid mechanics.
\newblock Computational Mechanics \textbf{58}(3), 457--490 (2016)

\bibitem{Feelpp-06}
Prud'homme, C.: {A Domain Specific Embedded Language in C++ for Automatic
  Differentiation, Projection, Integration and Variational Formulations}.
\newblock Scientific Programming \textbf{14}, 150,736 (2006)

\bibitem{Shi-QSS-18}
Qiu, W., Shen, J., Shi, K.: An {HDG} method for linear elasticity with strong
  symmetric stresses.
\newblock Mathematics of Computation \textbf{87}(309), 69--93 (2018)

\bibitem{Shi-QS-16}
Qiu, W., Shi, K.: A superconvergent {HDG} method for the incompressible
  {N}avier-{S}tokes equations on general polyhedral meshes.
\newblock IMA Journal of Numerical Analysis \textbf{36}(4), 1943--1967 (2016)

\bibitem{Shi-QS-19}
Qiu, W., Shi, K.: Analysis on an {HDG} method for the $p$-{L}aplacian
  equations.
\newblock Journal of Scientific Computing \textbf{80}(2), 1019--1032 (2019)

\bibitem{Solano-QSV-16}
Qiu, W., Solano, M., Vega, P.: A high order {HDG} method for curved-interface
  problems via approximations from straight triangulations.
\newblock Journal of Scientific Computing \textbf{69}(3), 1384--1407 (2016)

\bibitem{Quarteroni-book}
Quarteroni, A.: Numerical models for differential problems, \emph{{MS\&A.
  Modeling, Simulation and Applications}}, vol.~16.
\newblock Springer, Cham (2017)

\bibitem{Firedrake-16}
Rathgeber, F., Ham, D.A., Mitchell, L., Lange, M., Luporini, F., Mcrae, A.T.T.,
  Bercea, G.T., Markall, G.R., Kelly, P.H.J.: Firedrake: Automating the finite
  element method by composing abstractions.
\newblock ACM Transactions on Mathematical Software \textbf{43}(3) (2016)

\bibitem{remacle2007efficient}
Remacle, J.F., Chevaugeon, N., Marchandise, E., Geuzaine, C.: Efficient
  visualization of high-order finite elements.
\newblock International Journal for Numerical Methods in Engineering
  \textbf{69}(4), 750--771 (2007)

\bibitem{GetFEM-20}
Renard, Y., Poulios, K.: {GetFEM: Automated FE modeling of multiphysics
  problems based on a generic weak form language}.
\newblock Tech. rep., HAL (2020).
\newblock \urlprefix\url{https://hal.archives-ouvertes.fr/hal-02532422}

\bibitem{Cockburn-RC-12}
Rhebergen, S., Cockburn, B.: {A space-time hybridizable discontinuous Galerkin
  method for incompressible flows on deforming domains}.
\newblock Journal of Computational Physics \textbf{231}(11), 4185--4204 (2012)

\bibitem{Wells-RW-18b}
Rhebergen, S., Wells, G.: A hybridizable discontinuous {G}alerkin method for
  the {N}avier-{S}tokes equations with pointwise divergence-free velocity
  field.
\newblock Journal of Scientific Computing \textbf{76}(3), 1484--1501 (2018)

\bibitem{Wells-RW-18}
Rhebergen, S., Wells, G.: Preconditioning of a hybridized discontinuous
  {G}alerkin finite element method for the {S}tokes equations.
\newblock Journal of Scientific Computing \textbf{77}(3), 1936--1952 (2018)

\bibitem{Rhebergen-RW-20}
Rhebergen, S., Wells, G.N.: {An embedded-hybridized discontinuous Galerkin
  finite element method for the Stokes equations}.
\newblock Computer Methods in Applied Mechanics and Engineering \textbf{358},
  112,619 (2020)

\bibitem{Riviere2008}
Rivi\`{e}re, B.: {Discontinuous Galerkin Methods for Solving Elliptic and
  Parabolic Equations}.
\newblock Society for Industrial and Applied Mathematics (2008)

\bibitem{Loula-RDIL-20}
Rocha, B.M., {dos Santos, R. W.}, Igreja, I., Loula, A.F.D.: Stabilized hybrid
  discontinuous {G}alerkin finite element method for the cardiac monodomain
  equation.
\newblock International Journal for Numerical Methods in Biomedical Engineering
  \textbf{36}(7) (2020)

\bibitem{Dawson-SD-18}
Samii, A., Dawson, C.: An explicit hybridized discontinuous {G}alerkin method
  for {S}erre-{G}reen-{N}aghdi wave model.
\newblock Computer Methods in Applied Mechanics and Engineering \textbf{330},
  447 -- 470 (2018)

\bibitem{Dawson-SKMD-19}
Samii, A., Kazhyken, K., Michoski, C., Dawson, C.: A comparison of the explicit
  and implicit hybridizable discontinuous {G}alerkin methods for nonlinear
  shallow water equations.
\newblock Journal of Scientific Computing \textbf{80}(3), 1936--1956 (2019)

\bibitem{Dawson-SMD-16}
Samii, A., Michoski, C., Dawson, C.: A parallel and adaptive hybridized
  discontinuous {G}alerkin method for anisotropic nonhomogeneous diffusion.
\newblock Computer Methods in Applied Mechanics and Engineering \textbf{304},
  118 -- 139 (2016)

\bibitem{Dawson-SPMD-16}
Samii, A., Panda, N., Michoski, C., Dawson, C.: A hybridized discontinuous
  {G}alerkin method for the nonlinear {K}orteweg-de {V}ries equation.
\newblock Journal of Scientific Computing \textbf{68}(1), 191--212 (2016)

\bibitem{Peraire-SCNPC-17}
S\'{a}nchez, M.A., Ciuca, C., Nguyen, N.C., Peraire, J., Cockburn, B.:
  {Symplectic Hamiltonian HDG methods for wave propagation phenomena}.
\newblock Journal of Computational Physics \textbf{350}, 951--973 (2017)

\bibitem{SanchezVizuet-SS-19}
S\'{a}nchez-Vizuet, T., Solano, M.E.: A hybridizable discontinuous {G}alerkin
  solver for the {G}rad-{S}hafranov equation.
\newblock Computer Physics Communications \textbf{235}, 120--132 (2019)

\bibitem{SanchezVizuet-SSC-20}
S\'{a}nchez-Vizuet, T., Solano, M.E., Cerfon, A.J.: {Adaptive Hybridizable
  Discontinuous Galerkin discretization of the Grad-Shafranov equation by
  extension from polygonal subdomains}.
\newblock Computer Physics Communications \textbf{255}, 107,239 (2020)

\bibitem{NGSolve-14}
Sch\"{o}berl, J.: {C++11 Implementation of Finite Elements in NGSolve}.
\newblock Tech. Rep. {ASC-30/2014}, {Institute for Analysis and Scientific
  Computing - TU Wien} (2014).
\newblock
  \urlprefix\url{https://www.asc.tuwien.ac.at/~schoeberl/wiki/publications/ngs-cpp11.pdf}

\bibitem{Kronbichler-SKW-18}
Schoeder, S., Kronbichler, M., Wall, W.A.: Arbitrary high-order explicit
  hybridizable discontinuous {G}alerkin methods for the acoustic wave equation.
\newblock Journal of Scientific Computing \textbf{76}(2), 969--1006 (2018)

\bibitem{Kronbichler-SSKK-20}
Schoeder, S., Sticko, S., Kreiss, G., Kronbichler, M.: High-order cut
  discontinuous {G}alerkin methods with local time stepping for acoustics.
\newblock International Journal for Numerical Methods in Engineering
  \textbf{121}(13), 2979--3003 (2020)

\bibitem{Schutz-SA-17}
Sch\"{u}tz, J., Aizinger, V.: A hierarchical scale separation approach for the
  hybridized discontinuous {G}alerkin method.
\newblock Journal of Computational and Applied Mathematics \textbf{317}, 500 --
  509 (2017)

\bibitem{RS-19}
Sevilla, R.: {HDG-NEFEM} for two dimensional linear elasticity.
\newblock Computers \& Structures \textbf{220}, 69--80 (2019)

\bibitem{RS-SBGH-20}
Sevilla, R., Borchini, L., Giacomini, M., Huerta, A.: Hybridisable
  discontinuous {G}alerkin solution of geometrically parametrised {S}tokes
  flows.
\newblock Computer Methods in Applied Mechanics and Engineering \textbf{372},
  113,397 (2020)

\bibitem{Sevilla-SFH-08}
Sevilla, R., Fern{{\'a}}ndez-M{{\'e}}ndez, S., Huerta, A.: N{URBS}-enhanced
  finite element method ({NEFEM}).
\newblock International Journal for Numerical Methods in Engineering
  \textbf{76}(1), 56--83 (2008)

\bibitem{Sevilla-SFH-11}
Sevilla, R., Fern\'{a}ndez-M\'{e}ndez, S., Huerta, A.: {3D NURBS-enhanced
  finite element method (NEFEM)}.
\newblock International Journal for Numerical Methods in Engineering
  \textbf{88}(2), 103--125 (2011)

\bibitem{RS-SGH-18}
Sevilla, R., Giacomini, M., Huerta, A.: A face-centred finite volume method for
  second-order elliptic problems.
\newblock International Journal for Numerical Methods in Engineering
  \textbf{115}(8), 986--1014 (2018)

\bibitem{RS-SGH-19}
Sevilla, R., Giacomini, M., Huerta, A.: A locking-free face-centred finite
  volume ({FCFV}) method for linear elastostatics.
\newblock Computers \& Structures \textbf{212}, 43--57 (2019)

\bibitem{RS-SGKH-18}
Sevilla, R., Giacomini, M., Karkoulias, A., Huerta, A.: A superconvergent
  hybridisable discontinuous {G}alerkin method for linear elasticity.
\newblock International Journal for Numerical Methods in Engineering
  \textbf{116}(2), 91--116 (2018)

\bibitem{RS-SH:16}
Sevilla, R., Huerta, A.: Tutorial on {H}ybridizable {D}iscontinuous {G}alerkin
  ({HDG}) for second-order elliptic problems.
\newblock In: J.~Schr{\"o}der, P.~Wriggers (eds.) Advanced Finite Element
  Technologies, \emph{CISM International Centre for Mechanical Sciences}, vol.
  566, pp. 105--129. Springer International Publishing (2016)

\bibitem{RS-SH-18}
Sevilla, R., Huerta, A.: {HDG-NEFEM} with degree adaptivity for {S}tokes flows.
\newblock Journal of Scientific Computing \textbf{77}(3), 1953--1980 (2018)

\bibitem{Pitt-SMP-16}
Sheldon, J.P., Miller, S.T., Pitt, J.S.: A hybridizable discontinuous
  {G}alerkin method for modeling fluid-structure interaction.
\newblock Journal of Computational Physics \textbf{326}, 91 -- 114 (2016)

\bibitem{Shen-SSZ-19}
Shen, J., Singler, J.R., Zhang, Y.: {HDG-POD} reduced order model of the heat
  equation.
\newblock Journal of Computational and Applied Mathematics \textbf{362},
  663--679 (2019)

\bibitem{Solano-SV-19}
Solano, M., Vargas, F.: A high order {HDG} method for {S}tokes flow in curved
  domains.
\newblock Journal of Scientific Computing \textbf{79}(3), 1505--1533 (2019)

\bibitem{soon2009hybridizable}
Soon, S.C., Cockburn, B., Stolarski, H.K.: A hybridizable discontinuous
  {G}alerkin method for linear elasticity.
\newblock International Journal for Numerical Methods in Engineering
  \textbf{80}(8), 1058--1092 (2009)

\bibitem{Peraire-SNPC-16}
Stanglmeier, M., Nguyen, N.C., Peraire, J., Cockburn, B.: An explicit
  hybridizable discontinuous {G}alerkin method for the acoustic wave equation.
\newblock Computer Methods in Applied Mechanics and Engineering \textbf{300},
  748--769 (2016)

\bibitem{Stenberg-90-Stokes}
Stenberg, R.: Some new families of finite elements for the {S}tokes equations.
\newblock Numerische Mathematik \textbf{56}(8), 827--838 (1990)

\bibitem{Su-SWZW-19}
Su, W., Wang, P., Zhang, Y., Wu, L.: A high-order hybridizable discontinuous
  {G}alerkin method with fast convergence to steady-state solutions of the gas
  kinetic equation.
\newblock Journal of Computational Physics \textbf{376}, 973--991 (2019)

\bibitem{Peraire-TNBP-19}
Terrana, S., Nguyen, N.C., Bonet, J., Peraire, J.: A hybridizable discontinuous
  {G}alerkin method for both thin and 3{D} nonlinear elastic structures.
\newblock Computer Methods in Applied Mechanics and Engineering \textbf{352},
  561 -- 585 (2019)

\bibitem{Terrana-TVG-17}
Terrana, S., Vilotte, J., Guillot, L.: {A spectral hybridizable discontinuous
  Galerkin method for elastic-acoustic wave propagation}.
\newblock Geophysical Journal International \textbf{213}(1), 574--602 (2017)

\bibitem{Peraire-VMCYNOP-20}
Vidal-Codina, F., Mart\'{i}n-Moreno, L., Cirac\`{i}, C., Yoo, D., Nguyen, N.C.,
  Oh, S.H., Peraire, J.: Terahertz and infrared nonlocality and field
  saturation in extreme-scale nanoslits.
\newblock Optics Express \textbf{28}(6), 8701--8715 (2020)

\bibitem{Peraire-VCNOP-18}
Vidal-Codina, F., Nguyen, N., Oh, S.H., Peraire, J.: A hybridizable
  discontinuous {G}alerkin method for computing nonlocal electromagnetic
  effects in three-dimensional metallic nanostructures.
\newblock Journal of Computational Physics \textbf{355}, 548 -- 565 (2018)

\bibitem{Peraire-VNP-18}
Vidal-Codina, F., Nguyen, N., Peraire, J.: Computing parametrized solutions for
  plasmonic nanogap structures.
\newblock Journal of Computational Physics \textbf{366}, 89 -- 106 (2018)

\bibitem{Peraire-VNGP-15}
Vidal-Codina, F., Nguyen, N.C., Giles, M.B., Peraire, J.: A model and variance
  reduction method for computing statistical outputs of stochastic elliptic
  partial differential equations.
\newblock Journal of Computational Physics \textbf{297}, 700--720 (2015)

\bibitem{Peraire-VNGP-16}
Vidal-Codina, F., Nguyen, N.C., Giles, M.B., Peraire, J.: An empirical
  interpolation and model-variance reduction method for computing statistical
  outputs of parametrized stochastic partial differential equations.
\newblock SIAM-ASA Journal on Uncertainty Quantification \textbf{4}(1),
  244--265 (2016)

\bibitem{Vieira-VGSH-20}
Vieira, L.M., Giacomini, M., Sevilla, R., Huerta, A.: A second-order
  face-centred finite volume method for elliptic problems.
\newblock Computer Methods in Applied Mechanics and Engineering \textbf{358},
  112,655 (2020)

\bibitem{Vila-VGSH-review}
Vila-P\'{e}rez, J., Giacomini, M., Sevilla, R., Huerta, A.: Hybridisable
  discontinuous {G}alerkin formulation of compressible flows.
\newblock Tech. rep., arXiv (2020).
\newblock Submitted.

\bibitem{wang1991exact}
Wang, C.Y.: Exact solutions of the steady-state {N}avier-{S}tokes equations.
\newblock Annual Review of Fluid Mechanics \textbf{23}(1), 159--177 (1991)

\bibitem{BuiThanh-WMB-19}
Wildey, T., Muralikrishnan, S., Bui-Thanh, T.: Unified geometric multigrid
  algorithm for hybridized high-order finite element methods.
\newblock SIAM Journal on Scientific Computing \textbf{41}(5), S172--S195
  (2019)

\bibitem{Williams-18}
Williams, D.M.: An entropy stable, hybridizable discontinuous {G}alerkin method
  for the compressible {N}avier-{S}tokes equations.
\newblock Mathematics of Computation \textbf{87}(309), 95--121 (2018)

\bibitem{May-WBMS-14}
Woopen, M., Balan, A., May, G., Sch\"{u}tz, J.: A comparison of hybridized and
  standard {DG} methods for target-based $hp$-adaptive simulation of
  compressible flow.
\newblock Computers \& Fluids \textbf{98}, 3 -- 16 (2014)

\bibitem{May-WMS-14}
Woopen, M., May, G., Sch\"{u}tz, J.: Adjoint-based error estimation and mesh
  adaptation for hybridized discontinuous {G}alerkin methods.
\newblock International Journal for Numerical Methods in Fluids
  \textbf{76}(11), 811--834 (2014)

\bibitem{xie2013generation}
Xie, Z.Q., Sevilla, R., Hassan, O., Morgan, K.: The generation of arbitrary
  order curved meshes for {3D} finite element analysis.
\newblock Computational Mechanics \textbf{51}, 361--374 (2013)

\bibitem{Shi-YSF-19}
Yang, Y., Shi, K., Fu, S.: Multiscale hybridizable discontinuous {G}alerkin
  method for flow simulations in highly heterogeneous media.
\newblock Journal of Scientific Computing \textbf{81}(3), 1712--1731 (2019)

\bibitem{Peraire-YVCNSPO-19}
Yoo, D., Vidal-Codina, F., Cirac\`{i}, C., Nguyen, N.C., Smith, D.R., Peraire,
  J., Oh, S.H.: Modeling and observation of mid-infrared nonlocality in
  effective epsilon-near-zero ultranarrow coaxial apertures.
\newblock Nature Communications \textbf{10}(1) (2019)

\end{thebibliography}

\end{document}